\documentclass[preprint,12pt]{elsarticle}

\usepackage{a4}
\usepackage{amsmath}
\usepackage{amsfonts}
\usepackage{theorem}
\usepackage{delarray}
\usepackage{makeidx}
\usepackage{amssymb}
\usepackage{indentfirst}
\usepackage{graphicx}
\usepackage[T1]{fontenc}
\usepackage{graphics,epsfig,fancyhdr}
\usepackage{subfigure}
\usepackage[usenames]{color}
\usepackage{alltt}
\usepackage{float}
\usepackage[T1]{fontenc}

\journal{Journal of Computational Physics}

\begin{document}

\begin{frontmatter}



\title{Particle-in-cell modelling of relativistic laser-plasma interaction with the adjustable-damping,
direct implicit method}

\author[cea]{M. Drouin\corref{cor1}}
\ead{mathieu.drouin@cea.fr}
\author[cea]{L. Gremillet}
\ead{laurent.gremillet@cea.fr}
\author[cpht]{J.-C. Adam}
\author[cpht]{A. H\'eron}

\cortext[cor1]{Corresponding author}

\address[cea]{CEA, DAM, DIF, F-91297 Arpajon Cedex, France}
\address[cpht]{Centre de Physique Th\' eorique, UMR 7644, \'Ecole Polytechnique, CNRS, 91128 Palaiseau, France}

\begin{abstract}

Implicit particle-in-cell codes offer advantages over their explicit counterparts in that they suffer
weaker stability constraints on the need to resolve the higher frequency modes of the system. This feature
may prove particularly valuable for modeling the interaction of high-intensity laser pulses with overcritical
plasmas, in the case where the electrostatic modes in the denser regions are of negligible influence on the physical
processes under study. To this goal, we have developed the new two-dimensional electromagnetic code ELIXIRS
(standing for ELectromagnetic Implicit X-dimensional Iterative Relativistic Solver) based on the relativistic extension of
the so-called Direct Implicit Method [D. Hewett and A. B. Langdon, J. Comp. Phys. \textbf{72}, 121(1987)].
Dissipation-free propagation of light waves into vacuum is achieved by an adjustable-damping electromagnetic solver.
In the high-density case where the Debye length is not resolved, satisfactory energy conservation is ensured by the
use of high-order weight factors. In this paper, we first present an original derivation of the electromagnetic direct
implicit method within a Newton iterative scheme. Its linear properties are then investigated through numerically solving the
relation dispersions obtained for both light and plasma waves, accounting for finite space and time steps.
Finally, our code is successfully benchmarked against explicit particle-in-cell simulations for two kinds of physical problems:
plasma expansion into vacuum and relativistic laser-plasma interaction. In both cases, we will demonstrate the robustness
of the implicit solver for crude discretizations, as well as the gains in efficiency which can be realized
over standard explicit simulations.
\end{abstract}


\end{frontmatter}

\section{Introduction}

Particle-in-cell (PIC) codes have become widely used plasma simulation tools owing to their ability to mimic real
plasma behavior. Yet the standard PIC algorithm employs an explicit time-differencing, and hence suffers from strict stability
constraints on the time step, which needs to resolve the highest-frequency modes of the system  \cite{bird85}.
Furthermore, the mesh size must be comparable to the Debye length $\lambda_D$ in order to prevent the finite-grid instability \cite{bird85}.
As a consequence, explicit PIC codes may find it difficult to cope with the large spatial and temporal scales associated with
a number of physical scenarios, thus requiring massively parallel computing facilities \cite{bowersalbright2008}. Several
alternatives have been developed over the past decades to relax these constraints so that the choice of the space and time
steps can be dictated by physical accuracy rather than stability conditions. The simplest way to do so is to suppress
high-frequency processes within the mathematical model itself. Codes based on the Darwin-field approximation \cite{hewe94,tagu04},
gyrokinetic equations \cite{cand03} or hybrid particle-fluid models \cite{mason1980,lipa02,davi97,grem02,liljo2008} rely
precisely on such an approach. The shortcoming inherent in these codes is the somewhat uncertain domain of validity of
their basic assumptions. A second, more involved numerically, possibility retains a fully kinetic and electromagnetic description
by using an implicit scheme for the entire Vlasov-Maxwell set of equations. This is the approach dealt with in this work.

The main feature, and difficulty, of a fully implicit PIC scheme is the prediction of the future particles' charge and current densities
as functions of the future electromagnetic fields. Two main techniques have been designed to this goal. The first one to be
published, the so-called moment method, makes use of the fluid equations to predict future source terms
\cite{dena81,maso81,brac82,maso87,vu92,lape06}. and has been recently extended to the relativistic regime \cite{nogu07}.
The present article will focus on the alternate approach, referred to as the direct implicit method, which is based on a direct linearization
of the Lorentz equations \cite{langdon1982,langdon1983, hewett1987, cohenlangdon1989}. Most implementations of the direct
implicit method start with the so-called $D_1$ discretization of the Lorentz equation, first presented in Ref. \cite{friedman1981}. The
relativistic formulation, originally derived in Ref. \cite{langdon1987}, was  implemented, albeit in a simplified form, in the LSP code
\cite{welchrose2001, welchrose2004,camp04,evans2006, weisolodov2008}.

The direct implicit method proceeds as follows. First, particles' momenta and positions are advanced to an intermediate time level using known
fields, yielding predicted charge and current densities. Second, by linearizing the latter quantities around the predicted momenta and positions,
we can express correction terms as functions of the future fields and thus derive an implicit wave equation. Once this equation is solved,
the particles' quantities are updated.  Here we will show that the direct method can be derived as a simplified Newton scheme.

Our main motivation is the simulation of the interaction of an ultra-intense laser pulse with solid-density plasma slabs.
The energetic particle beams originating from this interaction stir great interest in many fields spanning
inertial confinement fusion \cite{taba94,camp04,maso06,honr06,atze08,chri08}, high energy density physics
\cite{pate03,mart06,akli08,nils09}, nuclear physics \cite{cowa00,ledi03} or medical physics \cite{bula02}. For
the high plasma densities considered, the electron plasma frequency $\omega_p$ largely exceeds the laser frequency. Using an
explicit PIC code, the space and time steps should resolve the high-frequency electron plasma modes of the plasma
bulk. However, these modes are of no interest for the problem since they do not affect the laser-plasma
interaction nor other potentially important related processes as the subsequent, fast electron-driven ion expansion. By contrast,
resorting to an implicit scheme would allow a significantly increased time step, that is, determined only by the need to resolve the incoming
laser wave. In this respect, one should realize that the strong wave damping inherent with implicit methods may be harmful in the context of
laser-plasma interaction, for which light waves have to travel over many wavelengths. This prompted us to develop an electromagnetic
solver with adjustable damping, based on a generalization of the scheme initially proposed by Friedman \cite{friedman1990} for the Lorentz
equation. We will demonstrate that our adjustable damping scheme tolerates abrupt spatial jumps in the controlling parameter.
Our code therefore allows for dissipation-free laser propagation into vacuum, along with strong damping of undesirable plasma waves
into the densest part of the target.

As explicit codes, implicit codes suffer from the artificial heating arising from a crude discretization of the Debye length, as is
commonplace when handling large-scale, high-density plasmas. This detrimental effect is generally attributed to the so-called
grid-instability \cite{bird85}. To keep it at an acceptable level, we will exploit the well-known mitigating influence of high-order
weight factors \cite{abe1986,sento08} by using quadratic weight factors. We will also take advantage of the stabilizing effect of
the large time steps allowed by the implicit scheme.

The paper is organized as follows. In Sec. \ref{sec:direct_implicit_method}, we recall the basic principles of the PIC technique, give
the implicit time-discretized equations to solve, and derive within a simplified Newton formalism the relativistic direct implicit method.
In Sec. \ref{sec:numerical_resolution}, we outline the numerical resolution of the wave equation as implemented in our newly developed,
2Dx-3Dv code ELIXIRS (ELectromagnetic Implicit X-dimensionnal Iterative Relativistic Solver). The introduction of implicit injecting/outgoing
boundary conditions for the electromagnetic field is also discussed. Sec. \ref{sec:dispersion_relation} is devoted to the linear properties
of the direct implicit method through the resolution of the electromagnetic and electrostatic dispersion relations. The effects of finite
space and time steps, adjustable damping and high-order weight factors will be accounted for. Finally, in Sec. \ref{sec:applications},
our code is benchmarked against explicit simulations for two kinds of physical problems: the expansion of a plasma slab in vacuum,
and the interaction of an ultra-intense laser pulse with an overcritical plasma target.
The sensitivity of the simulation results to the damping parameter and the number of macro-particules will be addressed.

\section{The relativistic direct implicit method as a simplified Newton scheme}
\label{sec:direct_implicit_method}

In contrast to Ref. \cite{langdon1987}, we present here a derivation of the electromagnetic direct implicit method for the relativistic case within a Newton
iterative scheme and a weak formulation of Maxwell's equations. Anticipating our need of a dissipation-free propagation of light waves inside the vacuum
region of the simulation domain, we introduce a generalization of the adjustable damping scheme proposed and used in the electrostatic regime by
Friedman \cite{friedman1990}.

\subsection{Basic equations}

Consider Maxwell's equations
\begin{align}
\label{bp001}
	\mathbf{\nabla}\times \mathbf{E} &=-\frac{\partial \mathbf{B}}{\partial t}  \, , \\
 	\mathbf{\nabla}\times \mathbf{B}&=\mu_0 \mathbf{j} +\frac{1}{c^2}\frac{\partial \mathbf{E}}{\partial t} \,,
\end{align}
and the collisionless Vlasov equation for the distribution function $ f_s(\mathbf{x},\mathbf{u},t)$ of the $s$th particle species
\begin{equation}
\label{bp01}
	\frac{\partial f_s}{\partial t} + \frac{\mathbf{u}}{\gamma}\frac{\partial f_s}{\partial \mathbf{x}}+
	\frac{q_s}{m_s} \left(  \mathbf{E} + \frac{\mathbf{u}}{\gamma}\times \mathbf{B} \right)\cdot
	\frac{\partial f_s}{\partial\mathbf{u}}=0 \, .
\end{equation}
Here $q_s$ and $m_s$ are the charge and the rest mass  of the $s$th particle species, respectively. $\mathbf{u}$ denotes the
relativistic momentum normalized by $m_s$. The relativistic factor then writes
$\gamma=\left( 1 + u^2/{c^2} \right)^{1/2}$.
The particle method consists in describing the distribution function $f_s$ as an ensemble of macro-particles in the form
\begin{equation}
	f_s(\mathbf{x},\mathbf{u},t)=\sum_{p=1}^{N_s} S(\mathbf{x}-\mathbf{X}_p(t))\delta (\mathbf{u}-\mathbf{U}_p(t))  \, ,
\end{equation}
where $S$ is the shape function \cite{bird85}, $N_s$ the total number of particles of the $s$th species, and $\delta$ the Dirac distribution.
The relativistic motion of each macro-particle obeys the following equations:
\begin{align}
\frac{d\mathbf{X}_p(t)}{dt} & =  \mathbf{V}_p(t) = \frac{\mathbf{U}_p(t)}{\gamma_p(t)} \, , \\
\frac{d\mathbf{U}_p(t)}{dt} & =  \frac{q_s}{m_s}\left\{ \mathbf{E} \left[ \mathbf{X}_p(t),t \right] +
\frac{\mathbf{U}_p(t)}{\gamma_p(t)} \times \mathbf{B} \left[ \mathbf{X}_p(t),t\right]\right\} \, .
\end{align}

We now make use of the implicit scheme with adjustable damping proposed by Friedman \cite{friedman1990} for an electrostatic
problem, which generalizes the so-called $D_1$-scheme of Langdon \emph{et al.} \cite{langdon1982,langdon1983,hewett1987,langdon1987}.
The equations of motion are discretised as
\begin{align}
\label{bp02a}
\bold{X}_{n+1} &= \bold{X}_n + \Delta t \frac{\bold{U}_{n+1/2}}{\gamma_{n+1/2}} \, ,  \\
\label{bp02b}
\bold{U}_{n+1/2} & = \bold{U}_{n-1/2} + \frac{\Delta t}{2}(\bold{a}_{n+1}+\bold{\bar{A}}_{n-1})
 + \frac{q_s \Delta t}{2 m_s} \left(\frac{\bold{U}_{n+1/2}+\bold{U}_{n-1/2}}{\gamma_n}\right)
 \times \bold{B}_n(\bold{X}_n) \, ,\\
\label{bp02c}
\bold{\bar{A}}_{n-1}  &= \frac{\theta_f}{2}\bold{a}_n + \left(1-\frac{\theta_f}{2}\right)\bold{\bar{a}}_{n-2} \, ,\\
\label{bp02d}
\bold{\bar{a}}_{n-1} & =  \left(1-\frac{\theta_f}{2}\right)\bold{a}_n + \frac{\theta_f}{2}\bold{\bar{a}}_{n-2} \, ,
\end{align}
where the index $n$ denotes the time step index and we have defined
\begin{align}
\label{bp02e}
&\mathbf{a}_n =\frac{q_s}{m_s}\mathbf{E}_n \, ,\\
\label{bp02f}
&\gamma_n = \left\{1+\frac{1}{c^2}\left[ \mathbf{U}_{n-1/2}+\frac{\Delta t}{4}
\left( \mathbf{a}_{n+1} + \bar{\mathbf{A}}_{n-1}\right) \right]^2 \right\}^{1/2}  \, , \\
\label{bp02g}
&\gamma_{n+1/2} = \left(1+\frac{\mathbf{U}_{n+1/2}^2}{c^2} \right)^{1/2}  \, .
\end{align}
Friedman's scheme can be readily applied to Maxwell's equations, which yields
\begin{align}
\label{bp06}
&\bold{E}_{n+1} = \bold{E}_n + c^2\Delta{t}\bold{\nabla} \times \bold{B}_{n+1/2} -
\frac{\Delta t}{\epsilon_0} \bold{j}_{n+1/2} \, ,\\
\label{bp07}
&\bold{B}_{n+1/2} =  \bold{B}_{n-1/2} - \frac{\Delta t}{2} \bold{\nabla} \times \big(\bold{E}_{n+1} +
\bold{\bar{\bar{E}}}_{n-1}\big) \, , \\
&\mathbf{B}_n = \mathbf{B}_{n-1/2} -  \frac{\Delta t}{2} \bold{\nabla} \times \bold{E}_{n}  \, , \\
\label{bp08}
&\bold{\bar{\bar{E}}}_{n-1} =  \frac{\theta_f}{2}\bold{E}_n+ \left( 1-\frac{\theta_f}{2} \right) \bold{\bar{E}}_{n-2} \, , \\
\label{bp09}
&\bold{\bar{E}}_{n-1} =  \left(1-\frac{\theta_f}{2}\right)\bold{E}_{n}+\frac{\theta_f}{2}\bold{\bar{E}}_{n-2} \, .
\end{align}
where $\mathbf{j}$ denotes the current density.

As will be demonstrated in Sec. \ref{sec:dispersion_relation}, this scheme allows, via the parameter $\theta_f$, a flexible control of the damping of the high-frequency (electrostatic and
electromagnetic) waves of the system. This property is of major interest for applications such as laser-plasma interaction involving a traveling electromagnetic wave into vacuum, for which
the numerical damping associated with the standard $D_1$ method may prove too severe. It is worth noting that, even though referred to uniquely as $\theta_f$, the damping parameters
involved in the electromagnetic scheme and the particle pusher may assume distinct values. The next sections will be devoted to the solution of the set of Eqs. (\ref{bp02a})-(\ref{bp09})
within a Newton iterative scheme. We will show that for a proper choice of the initial conditions, this scheme reduces to the direct implicit method developed in Refs. \cite{hewett1987,langdon1987}.


\subsection{Weak formulation of the electric field equation}

By replacing Eq. (\ref{bp07}) into Eq. (\ref{bp06}), one obtains the following wave equation
\begin{equation}
\label{bp10}
\mathbf{E}_{n+1} + \frac{c^2\Delta t^2}{2}\mathbf{\nabla}\times\mathbf{\nabla}\times\mathbf{E}_{n+1} +
\frac{\Delta t}{\epsilon_0}\mathbf{j}_{n+1/2} = \mathbf{Q}'  \, ,
\end{equation}
with the (known) source term
\begin{equation}
\label{bp11}
\mathbf{Q}' = \mathbf{E}_n + c^2\Delta t \mathbf{\nabla}\times\mathbf{B}_{n-1/2} -
\frac{c^2 \Delta t^2}{2} \mathbf{\nabla}\times\mathbf{\nabla}\times\bar{\bar{\mathbf{E}}}_{n-1}  \, .
\end{equation}
For any test function $\psi$, we assume the following weak formulation of the current density
\begin{align}
\label{bp12}
&\int \mathbf{j}_{n+1/2}(\mathbf{x})\psi(\mathbf{x}) d\mathbf{x} \notag \\
&= \sum_s \frac{q_s}{2} \int f_{s,0}(\mathbf{x},\mathbf{u})
\mathbf{V}_{n+1/2}(\mathbf{x},\mathbf{u})
\left[ \psi \left(\mathbf{X}_{n+1}(\mathbf{x},\mathbf{u}) \right)+\psi \left(\mathbf{X}_n(\mathbf{x},\mathbf{u})\right) \right] \, d\mathbf{x}d\mathbf{u}  \, ,
\end{align}
where $f_{s,0}=f_s(\mathbf{x},\mathbf{u},0)$ is the initial particle distribution function and $\mathbf{V}_{n+1/2}=\mathbf{U}_{n+1/2}/\gamma_{n+1/2}$.

The problem then consists in finding $(\mathbf{E}_{n+1},\mathbf{X}_{n+1},\mathbf{U}_{n+1/2})$ which solve
\begin{align}
\int \bold{E}_{n+1}(\bold{x})\psi(\bold{x})d\bold{x} + \frac{c^2\Delta t^2}{2}\int \bold{\nabla} \times \bold{\nabla}
\times \bold{E}_{n+1}(\bold{x}) \psi(\bold{x})d\bold{x} \notag \\
+ \frac{\Delta t}{\epsilon_0} \int \bold{j}_{n+1/2}(\bold{x})\psi(\bold{x}) d\bold{x}
= \int \bold{Q}'(\bold{x}) \psi(\bold{x})d\bold{x} \label{bp13} \,
\end{align}
together with Eqs. (\ref{bp02a})-(\ref{bp02g}). We employ the Newton method to solve this system: for each quantity
of interest $Y$, we introduce the ansatz
\begin{equation}
Y^{(k+1)}_{n+\alpha} = Y^{(k)}_{n+\alpha} + \delta Y^{(k)}_{n+\alpha} \quad k = 0,1,\dots
\label{eq:newton}
\end{equation}
where $\alpha=(1/2, 1)$ depending on whether $Y$ is centered at full or half time steps.
The subscript $n+1$ will be hereafter omitted for clarity. Substituting the above ansatz into Eq. (\ref{bp11}) yields
\begin{align}
&\int \left[ \bold{E}^{(k)}(\bold{x}) + \delta \bold{E}^{(k)}(\bold{x}) \right] \psi(\bold{x})  d\bold{x}
+ \frac{c^2 \Delta t^2}{2}\int \bold{\nabla} \times
\bold{\nabla} \times \left[\bold{E}^{(k)}(\bold{x}) +\delta\bold{E}^{(k)}(\bold{x}) \right]\psi(\bold{x})  d\bold{x} \notag \\
&+ \frac{\Delta t}{\epsilon_0} \int \bold{j}^{(k+1)}(\bold{x}) \psi(\bold{x}) d\bold{x} =
\int \bold{Q'}(\bold{x}) \psi(\bold{x})  d\bold{x} \, .
\label{bp14}
\end{align}
The term involving $\mathbf{j}^{(k+1)}$ is calculated with positions $\mathbf{X}^{(k+1)}$ and velocities $\mathbf{V}^{(k+1)}$
\begin{align}
&\int \bold{j}^{(k+1)} \psi(\bold{x})d\bold{x} =
\sum_s \frac{q_s}{2} \int f_{s,0}(\bold{x}, \bold{u}) \bold{V}^{(k)}\left[\psi(\bold{X}^{(k)})+\psi(\bold{X}_n)\right]d\bold{x}d\bold{u}  \notag \\
&+ \sum_s \frac{q_s}{2}\int f_{s,0}(\bold{x},\bold{u}) \delta\bold{V}^{(k)}\left[\psi(\bold{X}^{(k)})+\psi(\bold{X}_n)\right]d\bold{x}d\bold{u} \notag \\
&+ \sum_s \frac{q_s}{2}\int f_{s,0}(\bold{x},\bold{u}) \bold{V}^{(k)} \left[ \bold{\nabla}\psi(\bold{X}^{(k)} ) \cdot \delta \bold{X}^{(k)} \right] d\bold{x}d\bold{u} \, .
\label{bp15}
\end{align}

To obtain the equation solved for the electric field, we need to express the terms $\mathbf{X}^{(k)}$, $\delta \mathbf{X}^{(k)}$, $\mathbf{V}^{(k)}$ and
$\delta \mathbf{V}^{(k)}$ as functions of the electric field. Before proceeding, let us first define the following quantities
\begin{align}
&\gamma^{(k)} =  \left(1 + \frac{\mathbf{U}^{(k)2}}{c^2} \right)^{1/2} \, , \\
&\Gamma^{(k)} = \left \{ 1+\frac{1}{c^2} \left[ \mathbf{U}_{n-1/2}
+ \frac{\Delta t}{4} \left( \frac{q_s}{m_s} \mathbf{E}^{(k)} (\mathbf{X}^{(k)})
+ \mathbf{\bar{A}}_{n-1} \right ) \right]^2 \right \}^{1/2}\, , \\
& \theta (\bold{X}_n) = \frac{q_s \Delta t }{2 m_s \Gamma^{(k)}} \bold{B}_n(\bold{X}_n) \, , \\
& \bold{R} (\bold{X}_n) = \frac{2}{1+\theta^2}(\bold{I}+\bold{\theta}  \otimes  \bold{\theta} -\bold{\theta} \times\bold{I}) -\bold{I}  \, , \\
&\mathbf{M}(\mathbf{U}^{(k)}) = \frac{1}{\gamma^{(k)}}
\left( \mathbf{I} - \frac{\mathbf{U}^{(k)} \otimes \mathbf{U}^{(k)}}{\gamma^{(k)2} c^2} \right) \, , \\
&\mathbf{N}\left(\bold{E}^{(k)}(\bold{X}^{(k)}),\bold{U}^{(k)}\right) = \frac{q_s \Delta t}{4 m_s c^2}
\left[ \frac{\bold{U}_{n-1/2}+\bold{U}^{(k)}}{\Gamma^{(k)3}}  \times \bold{B}_n(\bold{X}_n) \right]
 \notag \\
 & \otimes \left[\bold{U}_{n-1/2} + \frac{\Delta t}{4}\left( \frac{q_s}{m_s}
\bold{E}^{(k)}(\bold{X}^{(k)}) + \bold{\bar{A}}_{n-1} \right) \right] \, ,
\end{align}
with $\bold{I}$ the identity matrix. Straightforward calculations then yield
\begin{align}
\bold{X}^{(k)} &=  \bold{X}_n + \frac{\Delta t \bold{U}^{(k)}}{\gamma^{(k)}} \, , \\
\delta \bold{X}^{(k)} & = \Delta t \bold{M}\delta \bold{U}^{(k)} \, , \\
\bold{V}^{(k)} & = \frac{\bold{U}^{(k)}}{\gamma^{(k)}} \, , \\
\delta \bold{V}^{(k)} & = \bold{M}\delta \bold{U}^{(k)} \, ,
\label{bp16}
\end{align}
Using the above expressions and the Newton ansatz (\ref{eq:newton}), the Lorentz equation becomes
\begin{align}
\bold{U}^{(k)}+\delta \bold{U}^{(k)} & = \bold{U}_{n-1/2} +
\frac{q_s \Delta t}{2 m_s} \left[ \bold{E}^{(k)}(\bold{X}^{(k)}) +
\bold{\nabla} \bold{E}^{(k)}(\bold{X}^{(k)})\delta \bold{X}^{(k)}
+ \delta \bold{E}^{(k)}(\bold{X}^{(k)}) \right]  \notag\\
& + \frac{\Delta t}{2} \bold{\bar{A}}_{n-1} + \frac{q_s\Delta t}{2 m_s} \left( \frac{\bold{U}^{(k)}+
\delta\bold{U}^{(k)}+\bold{U}_{n-1/2}}{\Gamma^{(k)}} \right) \times \bold{B}_n(\bold{X}_n) \notag \\
& - \frac{q_s \Delta t}{2 m_s}\bold{N}(\bold{E}^{(k)}(\bold{X}^{(k)}),\bold{U}^{(k)})
\bold{\nabla}\bold{E}^{(k)}(\bold{X}^{(k)})\delta\bold{X}^{(k)} \notag \\
& - \frac{q_s \Delta t}{2 m_s}\bold{N}(\bold{E}^{(k)}(\bold{X}^{(k)}),\bold{U}^{(k)})
\delta \bold{E}^{(k)}(\bold{X}^{(k)}) \, ,
\label{bp17}
\end{align}
where we have dropped second-order terms. Assuming the electric field gradient term is negligible, this equation further simplifies as
\begin{align}
&\mathbf{U}^{(k)} + \delta\bold{U}^{(k)} = \mathbf{U}_{n-1/2}
+ \frac{\Delta t}{4} \left[ \mathbf{I}+\mathbf{R} (\mathbf{X}_n) \right]  \left[ \mathbf{\bar{A}}_{n-1} + \frac{q_s}{m_s} \bold{E}^{(k)}(\mathbf{X}^{(k)})  \right]  \notag \\
& + \frac{\Delta t q_s}{4 m_s} \left[ \mathbf{I}+\mathbf{R} (\mathbf{X}_n)Ê\right]
\left[ \mathbf{I} - \mathbf{N} \left(\mathbf{E}^{(k)} (\mathbf{X}^{(k)}),\mathbf{U}^{(k)} \right) \right] \delta \mathbf{E}^{(k)}(\mathbf{X}^{(k)}) \, .
\label{bp18}
\end{align}
The set of equations (\ref{bp13})-(\ref{bp20}) constitutes the weak formulation of the problem. We will now show how to recover the direct implicit method
as a simplified Newton algorithm.

\subsubsection{The direct implicit method}

The simplest scheme consists in considering only one iteration in the above system and choosing the following initial values
\begin{equation}
\begin{array}{ll}
\label{bp20}
\left \{ \begin{array}{ll}
\bold{X}^{(0)} & = \bold{\widetilde{X}}_{n+1} \\
\bold{U}^{(0)} & = \bold{\widetilde{U}}_{n+1/2} \\
\bold{E}^{(0)} & = 0
\end{array}
\right.
&
\left \{ \begin{array}{ll}
\delta \bold{X}^{(0)} & = \delta \bold{X} \\
\delta \bold{U}^{(0)} & = \delta \bold{U} \\
\delta \bold{E}^{(0)} & = \mathbf{E}^{(1)} = \bold{E}_{n+1}  \, ,
\end{array}
\right.
\end{array}
\end{equation}
where we have introduced the predicted position and momentum $\mathbf{\widetilde{X}}_{n+1}$ and $\mathbf{\widetilde{V}}_{n+\frac{1}{2}}$ computed from the
known fields $\mathbf{\bar{A}}_{n-1}$ and $\mathbf{B}_n$. We have
\begin{align}
\label{bp21}
\widetilde{\bold{X}}_{n+1} & = \bold{X}_n +
\Delta t \frac{\widetilde{\bold{U}}_{n+1/2}}{\widetilde{\gamma}_{n+1/2}} \, , \\
\widetilde{\bold{U}}_{n+1/2} & = \bold{R}(\bold{X}_n)\bold{U}_{n-1/2} +
\frac{\Delta t}{4} \left[ \bold{I}+\bold{R}(\bold{X}_n) \right] \bar{\bold{A}}_{n-1} \, .
\end{align}
with $\widetilde{\gamma}_{n+1/2} = \gamma^{(0)}$. The correction terms then write
\begin{align}
\delta \bold{U} &=  \frac{q_s \Delta t }{4 m_s} [ \bold{I} + \bold{R}(\bold{X}_n) ]
[ \bold{I} - \bold{N}(\widetilde{\bold{U}}_{n+1/2}) ] \bold{E}_{n+1} (\widetilde{\mathbf{X}}_{n+1}) \, , \label{eq:dU} \\
\delta \bold{V} & = \bold{M} \delta \bold{U} \, , \label{eq:dV} \\
\delta \bold{X} & = \Delta t \bold{M} \delta \bold{U} \, ,
\label{eq:dX}
\end{align}
where we have defined
\begin{align}
&\mu(\widetilde{\bold{U}}_{n+1/2}) = \mathbf{N}(0, \widetilde{\bold{U}}_{n+1/2}) \notag \\
&= \frac{q_s \Delta t}{4 m_s c^2}
\left[ \frac{ \bold{U}_{n-1/2}+\widetilde{\bold{U}}_{n+1/2} } {\widetilde{\gamma}_n^3}
\times \bold{B}_n(\bold{X}_n) \right]
 \otimes \left(\bold{U}_{n-1/2} + \frac{\Delta t}{4}\bar{\bold{A}}_{n-1} \right) \, ,
\end{align}
and $\widetilde{\gamma}_n = \Gamma^{(0)}$. After substituting the above equations into (\ref{bp15}), using
$\mathbf{X}_n=\widetilde{\mathbf{X}}_{n+1}-\Delta t \widetilde{\mathbf{V}}_{n+1/2}$ and replacing the resulting expression into (\ref{bp14}), we obtain
\begin{align}
\label{bp23}
&\int \bold{E}_{n+1}(\bold{x}) \psi (\bold{x})d\bold{x} \notag
+ \frac{c^2\Delta t^2}{2} \int \bold{\nabla}\times\bold{\nabla}\times\bold{E}_{n+1}(\bold{x})\psi(\bold{x}) d\bold{x} \notag \\
& + \sum_s \frac{q_s \Delta t}{2 \epsilon_0}\int f_{s,0}(\bold{x}, \bold{u}) \bold{\widetilde{V}}_{n+1/2}(\bold{x},\bold{u})
\left[ \psi(\bold{\widetilde{X}}_{n+1}\left(\bold{x},\bold{u})\right)+\psi(\bold{X}_{n}\left(\bold{x},\bold{u})\right) \right] d\bold{x}d\bold{u} \notag  \\
& + \sum_s \frac{q_s \Delta t}{\epsilon_0} \int f_{s,0}(\bold{x},\bold{u})\delta \bold{V} (\bold{x},\bold{u})
\psi(\bold{\widetilde{X}}_{n+1}(\bold{x},\bold{u})) d\bold{x}d\bold{u} \notag \\
& + \sum_s \frac{q_s \Delta t}{2\epsilon_0} \int f_{s,0}(\bold{x},\bold{u}) \left[ \delta \bold{X} \otimes \bold{\widetilde{V}}_{n+1/2}
 - \widetilde{\bold{V}}_{n+1/2} \otimes \delta \bold{X} \right]  \bold{\nabla} \psi(\bold{\widetilde{X}}_{n+1}) d\mathbf{x}d\mathbf{u}  \notag \\
&= \int \bold{Q'}(\bold{x})\psi(\bold{x}) d\bold{x} \, .
\end{align}
From Eq. (\ref{bp12}), we identify
\begin{align}
&\sum_s \frac{q_s \Delta t}{2 \epsilon_0} \int f_{s,0}(\bold{x}, \bold{u}) \bold{\widetilde{V}}_{n+1/2}(\bold{x},\bold{u})
\left[ \psi(\bold{\widetilde{X}}_{n+1}\left(\bold{x},\bold{u})\right)+\psi(\bold{X}_{n}\left(\bold{x},\bold{u})\right) \right] d\bold{x}d\bold{u} \notag \\
&= \frac{\Delta t}{\epsilon_0} \int  \widetilde{\mathbf{j}}_{n+1/2}(\mathbf{x}) \psi (\mathbf{x}) d\mathbf{x} \, . \label{eq:I1}
\end{align}
To reduce the next integral, it is convenient to introduce the weak formulation of the predicted charge density
\begin{equation}
\label{bp24}
\int \widetilde{\rho}_{s}(\mathbf{x}) \psi(\mathbf{x}) d\mathbf{x} \notag
= q_s \int f_{s,0}(\mathbf{x},\mathbf{u}) \psi \left(\widetilde{\mathbf{X}}_{n+1} (\mathbf{x},\mathbf{u})\right) d\mathbf{x}d\mathbf{u}
\, .
\end{equation}
Approximating $\mathbf{R} (\mathbf{X}_n) \approx \mathbf{R} (\widetilde{\mathbf{X}}_{n+1})$, we obtain
\begin{align}
&\frac{q_s \Delta t}{\epsilon_0} \int f_{s,0} \delta \bold{V} \psi(\bold{\widetilde{X}}_{n+1}) d\bold{x}d\bold{u}  \notag \\
&= \frac{q_s \Delta t^2}{4 m_s \epsilon_0} \int \widetilde{\mathbf{\rho}} (\mathbf{x}) \bold{M}(\bold{x}) (\bold{I}+\bold{R}(\bold{x}))
\left[ \bold{I} - \bold{N}(\bold{x}) \right] \mathbf{E}_{n+1} (\bold{x}) d\bold{x}  \, .
\end{align}
Defining the implicit susceptibility $\chi$ as
\begin{equation}
\chi(\bold{x}) = \sum_s \frac{q_s \Delta t^2}{4 m_s \epsilon_0} \bold{M}(\bold{x}) (\bold{I}+\bold{R}_{s,n}(\bold{x}))
\left[ \bold{I} - \bold{N}(\bold{x}) \right] \widetilde{\rho}_s(\bold{x}) \, ,
\label{bp25}
\end{equation}
we have
\begin{equation}
\sum_s \frac{q_s \Delta t}{\epsilon_0} \int f_{s,0}(\bold{x},\bold{u})\delta \bold{V} (\bold{x},\bold{u})
\psi(\bold{\widetilde{X}}_{n+1}(\bold{x},\bold{u})) d\bold{x}d\bold{u}
=  \int   \psi (\mathbf{x}) \chi (\bold{x}) \mathbf{E}_{n+1} (\mathbf{x})  d\mathbf{x} \, . \label{eq:I2}
\end{equation}
We treat the remaining integral by introducing the modified current $\widetilde{\mathbf{j}}_{s}^{+}$
\begin{equation}
\int \widetilde{\mathbf{j}}_{s}^{+}(\mathbf{x}) \psi(\mathbf{x}) d\mathbf{x} \notag \\
= q_s \int f_{s,0}(\mathbf{x},\mathbf{u}) \widetilde{\mathbf{V}}_{n+1/2}
\left( \mathbf{x},\mathbf{u}\right) \psi \left(\widetilde{\mathbf{X}}_{n+1} (\mathbf{x},\mathbf{u})\right) d\mathbf{x}d\mathbf{u} \, .
\end{equation}
We then have
\begin{align}
&\frac{q_s \Delta t}{2\epsilon_0} \int f_{s,0}(\bold{x},\bold{u}) \left[ \delta \mathbf{X} \otimes \bold{\widetilde{V}}_{n+1/2}
 - \bold{\widetilde{V}}_{n+1/2} \otimes \delta \mathbf{X} \right]  \bold{\nabla} \psi(\bold{\widetilde{X}}_{n+1}) d\mathbf{x}d\mathbf{u} \notag \\
& = - \frac{q_s\Delta t^3}{8 m_s \epsilon_0} \int \nabla \times \left\{  \left[  \widetilde{\mathbf{j}}_s^{+}(\mathbf{x})
  \times \mathbf{M}(\mathbf{x}) \left[ \mathbf{I} + \mathbf{R}(\mathbf{x}) \right] \left[ \mathbf{I}-\mathbf{N}(\mathbf{x}) \right]  \right]\mathbf{E}_{n+1}(\mathbf{x}) \right\}
  \psi(\mathbf{x}) d\mathbf{x} \notag \\
  & = - \frac{q_s\Delta t^3}{8 m_s \epsilon_0} \int \nabla \times \left\{  \left[ \frac{ \widetilde{\mathbf{j}}_s^{+}(\mathbf{x})}{\widetilde{\gamma}_{n+1/2} (\mathbf{x})}
  \times \left[ \mathbf{I} + \mathbf{R}(\mathbf{x}) \right] \left[\mathbf{I} -\mathbf{N}(\mathbf{x}) \right] \right]\mathbf{E}_{n+1}(\mathbf{x}) \right\}
  \psi(\mathbf{x}) d\mathbf{x}
\end{align}
where use has been made of the identity $\mathbf{U} \times \mathbf{U} \otimes \mathbf{U} = 0$. We are then led to define the tensor $\zeta$ as
\begin{equation}
\zeta (\mathbf{x}) = \frac{\Delta t^2}{8\epsilon_0} \sum_s  \frac{q_s}{m_s} \frac{\widetilde{\mathbf{j}}_s^{+}}{\widetilde{\gamma}_{n+1/2}}
 \times \left[ \mathbf{I}+\mathbf{R}(\mathbf{x}) \right] \left[ \mathbf{I}-\mathbf{N}(\mathbf{x}) \right] \, .
\end{equation}
There follows
\begin{equation}
\frac{q_s \Delta t}{2\epsilon_0} \int f_{s,0} \left( \delta \mathbf{X} \otimes \bold{\widetilde{V}}_{n+1/2}
 - \widetilde{\bold{V}}_{n+1/2} \otimes \delta \mathbf{X} \right) \nabla \psi d\mathbf{x}d\mathbf{u}
= -\Delta t \int \nabla \times (\zeta \mathbf{E}_{n+1}) d\mathbf{x} \label{eq:I3} \, .
\end{equation}
Equation (\ref{bp15}) supplemented by Eqs. (\ref{eq:I1}), (\ref{eq:I2}) and (\ref{eq:I3}) should be satisfied for any test function $\psi$.
As a result, we have to solve the local field equation
\begin{equation}
\label{bp26}
\bold{E}_{n+1} + \frac{c^2 \Delta t^2}{2} \bold{\nabla}\times\bold{\nabla}\times \bold{E}_{n+1}
+ \bold{\chi} \bold{E}_{n+1} - \Delta t \bold{\nabla} \times \big( \bold{\zeta} \bold{E}_{n+1}\big) = \bold{Q} \, ,
\end{equation}
where the source term now reads
\begin{equation}
\label{bp27}
\bold{Q} = \bold{E}_n - \frac{\Delta t}{\epsilon_0}\widetilde{\mathbf{j}}_{n+1/2}+ c^2\Delta t \bold{\nabla} \times \bold{B}_{n-1/2} -
\frac{c^2\Delta t^2}{2}\bold{\nabla} \times \bold{\nabla} \times \bold{\bar{\bar{E}}}_{n-1} \, .
\end{equation}
We have thus recovered the relativistic implicit method based on the $D_1$ scheme which was presented in Ref. \cite{langdon1987},
with the only difference that the source term now involves the time-averaged field $\bold{\bar{\bar{E}}}_{n-1}$. It then appears that
the direct implicit method can be derived as a one-iteration Newton method with the starting values
$\mathbf{X}^{(0)}=\widetilde{\mathbf{X}}_{n+1}$, $\mathbf{U}^{(0)}=\widetilde{\mathbf{U}}_{n+1/2}$ and $\mathbf{E}^{(0)}=0$.

\section{Numerical resolution}
\label{sec:numerical_resolution}

\subsection{Resolution of the field equation}

In this section, we sketch the numerical procedure used to solve Eq. (\ref{bp26}) in the case of a 2Dx-3Dv phase space with periodic boundary conditions along the
transverse $y$ axis. We have first to evaluate the implicit susceptibilities. These terms are computed for each macroparticle, yielding $\chi(\mathbf{X_p},\mathbf{U}_p)$
and $\zeta(\mathbf{X_p},\mathbf{U}_p)$, before being projected onto the $(x,y)$ grid through the usual formulas:
\begin{align}
\chi(\mathbf{x})&=\sum_{s} \sum_{p} S(\mathbf{X}_p-\mathbf{x}) \chi(\mathbf{X}_p,\mathbf{U}_p) \, ,\\
\zeta(\mathbf{x})&=\sum_{s} \sum_{p} S(\mathbf{X}_p-\mathbf{x}) \zeta(\mathbf{X}_p,\mathbf{U}_p) \, .
\end{align}
We then apply the iterative method of Concus and Golub \cite{golub1973} to solve the elliptic system defined by Eq. (\ref{bp26}), which reads in the present case
\begin{equation}
\label{bp28}
\mathbf{E}^{(m+1)}+\frac{c^2\Delta t^2}{2}\mathbf{\nabla}\times\mathbf{\nabla}\times \mathbf{E}^{(m+1)}
+ \mathbf{\chi} \mathbf{E}^{(m+1)} -\Delta t \mathbf{\nabla}\times\left( \mathbf{\zeta}\mathbf{E}^{(m+1)}\right)
= \widetilde{\mathbf{Q}}^{(m)}
\end{equation}
The right-hand side of Eq. (\ref{bp28}) is given by
\begin{equation}
\widetilde{\mathbf{Q}}^{(m)}=\mathbf{Q}-(\mathbf{\chi}-\mathbf{\chi}^{0})\mathbf{E}^{(m)}
+\Delta t\mathbf{\nabla}\times \left[ \left( \mathbf{\zeta}-\mathbf{\zeta}^0 \right)\mathbf{E}^{(m)} \right]
\end{equation}
where $m$ is the iteration index and $\chi^0$ and $\zeta^0$ denote the $y$-averaged susceptibilities.
The fast convergence of the scheme implies, in principle, slow variations of the field quantities in the $y$ direction, but this has not
proved particularly constraining for the physical situations we have considered.

As is usual in electromagnetic PIC codes, two interleaved meshes are used for the spatial differencing of the grid quantities. The fields
are discretized as follows: $\rho_{i,j}$, $J_{z,i,j}$, $E_{z,i,j}$, $J_{x,i+1/2,j}$, $E_{x,i+1/2,j}$, $B_{y,i+1/2,j}$, $J_{y,i, j+1/2}$,
$E_{y,i,j+1/2}$, $B_{x,i,j+1/2}$ and $B_{z,i+1/2, j+1/2}$. The  $\chi$ and $\zeta$ are stored at $(i,j)$ except for $\chi_{11}$,
$\zeta_{11}$, $\zeta_{21}$, $\zeta_{31}$, which are located at $(i+1/2,j)$, and $\chi_{22}$, $\zeta_{12}$, $\zeta_{22}$, $\zeta_{32}$,
located at $(i,j+1/2)$. Once space-discretized, the above equations are Fourier transformed along the $y$ direction.
Considering $N_y$ grid cells, we obtain $N_y$ one-dimensional equations to solve.
Considering $N_x$ grid cells in the $x$ direction, each equation gives a $6N_x$ system of equations.
These systems have a band-diagonal structure and are solved by a standard LU technique, using routines
\textsf{bandec} and \textsf{banbks} of the numerical recipes library \cite{numrecipf90}.
Details on spatial discretisations and Fourier transformations used to solve Eq. (\ref{bp28})
are given in Appendix \ref{app:field_equation}.

\subsection{Charge correction}

Our method to accumulate charge and current densities [Eqs. (\ref{bp12}) and (\ref{bp24})] does not satisfy charge conservation,
which results into the violation of Poisson's equation. This is a common flaw of early electromagnetic PIC codes \cite{bird85} which
may be corrected by a more sophisticated projection scheme \cite{villasenor92,esirkepov01}. A well-known alternative approach, which
will be implemented here, is to correct the electrostatic part of the electric field $\mathbf{E}_{n+1}$ solution of Eq. (\ref{bp26}) so
that it fulfills Poisson's equation \cite{bird85}. Using normalized quantities, our best statement of Gauss's law is
\begin{equation}
\mathbf{\nabla}\cdot\mathbf{E}_{n+1}^*=\rho_{n+1} \, ,
\end{equation}
where $\mathbf{E}_{n+1}^*$ represents the sought-for electric field. Using
$\rho_{n+1}=\widetilde{\rho}_{n+1}-\mathbf{\nabla}\cdot\left(\chi\mathbf{E}_{n+1}^* \right)$, this can be reformulated as
\begin{equation}
\label{eqcharge00}
\mathbf{\nabla}\cdot\left[ (1+\chi)\mathbf{E}_{n+1}^*\right]=\widetilde{\rho}_{n+1} \, .
\end{equation}
Now, taking the divergence of Eq. (\ref{bp26}) yields
\begin{equation}
\label{eqcharge01}
\mathbf{\nabla}\cdot\left[ (1+\chi)\mathbf{E}_{n+1}\right]= \mathbf{\nabla}\cdot \mathbf{Q}
\end{equation}
with generally $\mathbf{\nabla}\cdot \mathbf{Q}\ne \widetilde{\rho}_{n+1}$. We may first think of introducing a potential $\psi$
such that $\mathbf{Q}^*=\mathbf{Q}-\mathbf{\nabla}\psi$ fulfills $\mathbf{\nabla}\cdot \mathbf{Q}^*=\widetilde{\rho}_{n+1}$,
but this correction has been shown to cause spurious effects \cite{hewett1987}.
A proper correction makes use of the following form \cite{hewett1987}
\begin{equation}
\label{eqcharge02}
\mathbf{Q}^*=\mathbf{Q}-(\mathbf{I}+\mathbf{\chi})\mathbf{\nabla}\psi \, ,
\end{equation}
There follows
\begin{equation}
\mathbf{\nabla}\cdot\left[ (1+\chi)\mathbf{\nabla}\psi\right]=\mathbf{\nabla}\cdot\mathbf{Q}
-\widetilde{\rho}_{n+1} \, ,
\end{equation}
which is equivalent to
\begin{equation}
\label{eqcharge03}
\mathbf{\nabla}\cdot\left[ (1+\chi)\mathbf{\nabla}\psi\right] = \mathbf{\nabla}\cdot\left[ (1+\chi)\mathbf{E}_{n+1}\right]
-\widetilde{\rho}_{n+1} \, ,
\end{equation}
where the only unknown is the scalar field $\psi$.
Eventually, the corrected field $\mathbf{E}^{*}_{n+1}$ ensuring Eq. (\ref{eqcharge00}) is given by
$\mathbf{E}^{*}_{n+1}=\mathbf{E}_{n+1}-\mathbf{\nabla}\psi$.
Details on the numerical resolution of Eq. (\ref{eqcharge03}) are given in Appendix \ref{app:charge_correction}.

\subsection{Electromagnetic boundary conditions}
\label{subsec:boundary_conditions}

In this section we describe the implementation of injecting/outgoing boundary conditions on both sides of the simulation box. Incident and scattered
electromagnetic waves are assumed linearly polarized and depending on the phase term $\mathbf{k}\cdot \mathbf{x}-\omega t$ only.
Waves polarized in the $(x,y)$ plane then verify
\begin{align}
E_y^{inc} & =  B_z^{inc} \cos \theta_i \, ,\\
E_y^{scat} & = - B_z^{scat} \cos \theta_s \, ,
\end{align}
where $\theta_i$ and $\theta_s$ denote respectively the incident and scattered angles. The total field becomes
\begin{align}
E_y^{tot} & = E_y^{scat} + E_y^{inc} \\
& = - B_z^{tot} \cos \theta_s + \frac{E_y^{inc}}{\cos \theta_i}
\left( \cos \theta_i + \cos \theta_s\right)
\end{align}
Discretizing with centered finite differences in space and time gives
\begin{align}
&\frac{1}{4}\big( E_{y,1,j+1/2}^{n+1} + E_{y,0,j+1/2}^{n+1} + E_{y,1,j+1/2}^{n} + E_{y,0,j+1/2}^{n} \big)
 = - B_{z,1/2,j+1/2}^{n+1/2}\cos\theta_s \notag \\
& + E_{y,1/2,j+1/2}^{inc,n+1/2}
 \frac{(\cos\theta_i + \cos\theta_s)}{\cos\theta_i} \,.
\end{align}
Using Maxwell-Faraday's equation, we can express $E_{y,0,j+1/2}^{n+1}$ as a function of the field values at inner grid points and previous
time steps. We have
\begin{align}
\label{bp35}
& E_{y,0,j+1/2}^{n+1} = A E_{y,1,j+1/2}^{n+1} \left( \frac{2\ \Delta t}{\Delta x}\cos\theta_s -1 \right)
 - \frac{2A \Delta t}{\Delta y}\cos\theta_s \left( E_{x,1/2,j+1}^{n+1} - E_{x,1/2,j}^{n+1}\right) \notag \\
&- 4A\cos\theta_s B_{z,1/2,j+1/2}^{n-1/2} + \frac{2A \Delta t}{\Delta x} \cos\theta_s
 \left( \bar{\bar{E}}_{y,1,j+1/2}^{n-1} - \bar{\bar{E}}_{y,0,j+1/2}^{n-1} \right) \notag \\
&- \frac{2A \Delta t}{\Delta y}\cos\theta_s\left( \bar{\bar{E}}_{x,1/2,j+1}^{n-1} - \bar{\bar{E}}_{x,1/2,j}^{n-1}\right)
 + \frac{4A}{\cos\theta_i}\left( \cos\theta_i + \cos\theta_s\right)E_{y,1/2,j+1/2}^{inc,n+1/2}  \notag \\
&-A \left( E_{y,1,j+1/2}^n + E_{y,0,j+1/2}^n \right) \, ,
\end{align}
where the coefficient $A$ is given by
\begin{equation}
A= \left(1+2\frac{\Delta t}{\Delta x}\cos\theta_s\right)^{-1}  \, .
\end{equation}
A similar equation can be established for $z$-polarized waves, which reads
\begin{align}
\label{bp36}
&E_{z,0,j}^{n+1} = B E_{z,1,j}^{n+1}\left( \frac{2\Delta t}{\Delta x \cos \theta_s} -1 \right) \notag \\
&- B(E_{z,0,j}^n + E_{z,1,j}^n) + \frac{4B}{\cos \theta_s} B_{y,1/2,j}^{n-1/2}
+ \frac{2B\Delta t}{\Delta x \cos \theta_s}
\left( \bar{\bar{E}}_{z,1,j}^{n-1} - \bar{\bar{E}}_{z,0,j}^{n-1} \right) \notag \\
&+4B E_{z,1/2,j}^{inc,n+1/2} \left( 1+ \frac{\cos \theta_s}{\cos \theta_i}\right) \, ,
\end{align}
where we have defined the coefficient $B$ as
\begin{align}
 B = \left( 1 + \frac{2\Delta t}{\Delta x \cos \theta_s }\right)^{-1} \, .
\end{align}

Note that the above equations only apply in vacuum. This is realized in practice by imposing boundary conditions on particles a few grid cells away from the outer boundaries of the
computational domain.

\section{Numerical analysis of the adjustable-damping, direct implicit method}
\label{sec:dispersion_relation}

\subsection{Dispersion relation of electromagnetic waves in vacuum}
\label{subsec:electromagnetic_waves}

Our aim here is to quantify the error in phase velocity and the damping associated with electromagnetic waves as functions
of the space and time steps. In particular, we will demonstrate the possibility to control the wave damping by adjusting the
parameter $\theta_f$.

Combining Maxwell-Amp\`ere's (\ref{bp06}) and Maxwell-Faraday's (\ref{bp07}) equations and assuming propagation in vacuum yield the
wave equation
\begin{equation}
\label{bp40}
\mathbf{E}_{n+1}=2\mathbf{E}_{n}-\mathbf{E}_{n-1}-\frac{c^2\Delta
t^2}{2}\nabla \times \nabla
\times\left(\mathbf{E}_{n+1}+\bar{\bar{\mathbf{E}}}_{n-1}\right) \, .
\end{equation}
The time-filtered term involves the adjustable damping parameter $\theta_f$ [Eq. (\ref{bp08})] and can be expanded as
\begin{align}
\label{bp41}
&\mathbf{E}_{n+1}+\bar{\bar{\mathbf{E}}}_{n-1} = \mathbf{E}_{n+1}+
\frac{\theta_f}{2}\mathbf{E}_{n}+\left(1-\frac{\theta_f}{2}\right)^2\mathbf{E}_{n-1}
+\left(1-\frac{\theta_f}{2}\right)^2 \frac{\theta_f}{2}\mathbf{E}_{n-2} \notag \\
&+\left(1-\frac{\theta_f}{2}\right)^2
\left(\frac{\theta_f}{2}\right)^2\mathbf{E}_{n-3} + \dots
\end{align}
In a 2-D geometry, taking the electric field in the form $\mathbf{E}_n = \mathbf{E}_{0} \Phi(x,y) z^n$ with $z=\exp(-i\omega\Delta
t)$  and $i=\sqrt{-1}$, Eq. (\ref{bp41}) becomes
\begin{align}
\label{bp42}
\mathbf{E}_{n+1} + \bar{\bar{\mathbf{E}}}_{n-1} & =
\mathbf{E}_0 \Phi(x,y) \Bigg\{ z^{-1} \left[ \left( 1-\frac{\theta_f}{2} \right)^2+\frac{\theta_f}{2}z + z^2 \right]  \notag \\
& + \left( 1-\frac{\theta_f}{2} \right)^2 \frac{\theta_f}{2}z^{-2} \left[1+\frac{\theta_f}{2}z^{-1}
+\left(\frac{\theta_f}{2} \right)^2 z^{-2} + \dots \right] \Bigg\} z^n \, .
\end{align}
where the adjustable damping parameter $\theta_f \in [0,1]$. Simplifying the series in the right-hand side of Eq. (\ref{bp42}) yields
\begin{align}
\label{bp43}
\mathbf{E}_{n+1}+\bar{\bar{\mathbf{E}}}_{n-1} &=
\mathbf{E}_{0}\Phi(x,y) \Bigg\{ z^{-1} \left[ \left( 1-\frac{\theta_f}{2} \right)^2 + \frac{\theta_f}{2}z + z^2 \right] \notag \\
&+\left(1-\frac{\theta_f}{2}\right)^2\frac{\theta_f}{2}\frac{2z^{-1}}{2z-\theta_f} \Bigg\} z^n \, .
\end{align}
The electromagnetic wave is assumed polarized in the $(x,y)$ plane with a harmonic dependence $\Phi(x,y)=\exp\left[ i(k_x x +k_y y)\right]$.
Substituting Eq. (\ref{bp43}) into Eq.(\ref{bp40}) and space-differencing the Laplacian leads, we get after some straightforward algrebra
the following third degree polynomial equation
\begin{equation}
\label{bp44}
z^2=2z-1
-\Bigg\{ \bigg[ \left( 1-\frac{\theta_f}{2} \right)^2 +\frac{\theta_f}{2}z + z^2 \bigg]
+ \left( 1-\frac{\theta_f}{2} \right)^2 \frac{\theta_f}{2z-\theta_f} \Bigg\} \frac{\Omega^2}{2} \, ,
\end{equation}
where we have introduced
\begin{equation}
\Omega^2 = 4\left\{ \frac{c^2\Delta t^2}{\Delta x^2}\sin^2\left(\frac{k_x\Delta x}{2}\right)
+ \frac{c^2\Delta t^2}{\Delta y^2}\sin^2\left(\frac{k_y\Delta y}{2}\right)\right\} \, .
\end{equation}
Equation (\ref{bp44}) simplifies as
\begin{equation}
\label{bp45}
z^3(2+\Omega^2)-z^2(4+\theta_f)+z \left[ 2+\Omega^2(1-\theta_f)+2\theta_f \right] - \theta_f=0 \, .
\end{equation}

\begin{figure}[htbp]
\includegraphics[width=7cm]{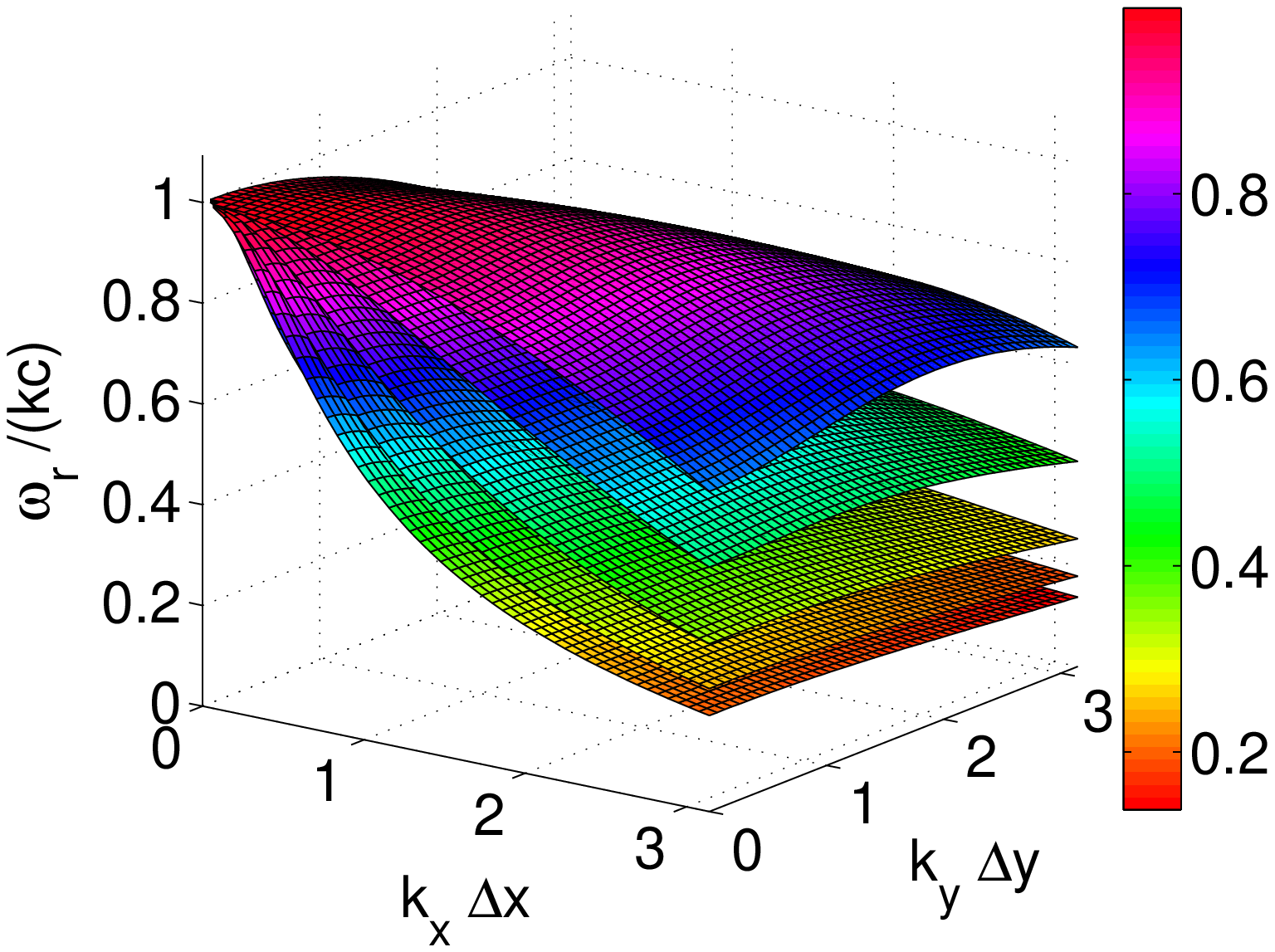}
\includegraphics[width=7cm]{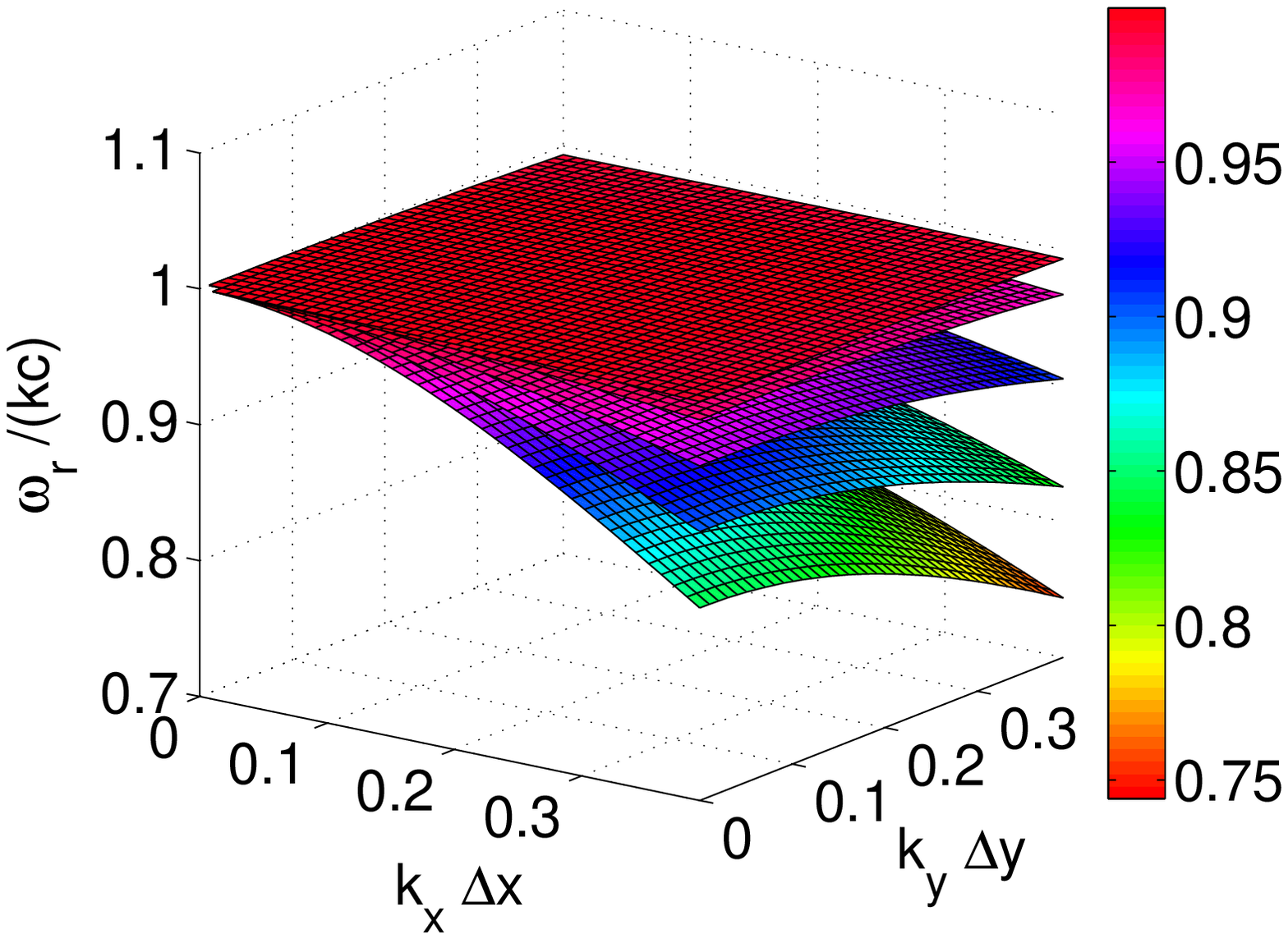}
\caption{Phase velocity of the least damped root of Eq. (\ref{bp45}) as a function of
 $(k_x \Delta x, k_y \Delta y)$,
 for  different values of $c\Delta t/\Delta x = c\Delta t/\Delta y \in \{0.05, 0.66, 1.28, 1.9, 2.5\}$
 (from top to bottom) and $\theta_f=0$. A narrower $(k_x \Delta x, k_y \Delta y)$ range is represented on the right.}
\label{fig3010}
\end{figure}

Let us first examine the special case $\theta_f = 0$. The roots of interest are solutions of
\begin{equation}
z^2(2+\Omega^2)-4z+(2+\Omega^2)=0
\end{equation}
The discriminant $\Delta=4-(2+\Omega^2)^2$ being always negative, we get the roots $z_{\pm}=(2 \pm i\sqrt{-\Delta})/(2+\Omega^2)$,
which statisfy $|z_+|=|z_-|=1$. We have therefore demonstrated the absence of damping when $\theta_f=0$.
Figure \ref{fig3010} plots the normalized phase velocity $v_{\phi}=\frac{\Re \omega}{kc}$ (where $k=\sqrt{k_x^2 + k_y^2}$) 
for different values of 
$c\Delta t/\Delta x = c\Delta t/\Delta y$. The phase velocity error grows for increasing $\Delta x$ and 
$\Delta t/\Delta x$.
A value $c \Delta t/\Delta x > 1$,
that is, violating the stability constraint of the standard explicit scheme, therefore implies a moderate spatial step $k_x \Delta x
\lesssim 0.38$ ($c\Delta t/\Delta x = 1.27$) so as to avoid excessive ($> 5\%$) phase velocity error, 
which, in presence of relativistic particles, may cause unphysical Cerenkov radiation \cite{greenwood04}.

\begin{figure}[htbp]
\includegraphics[width=7cm]{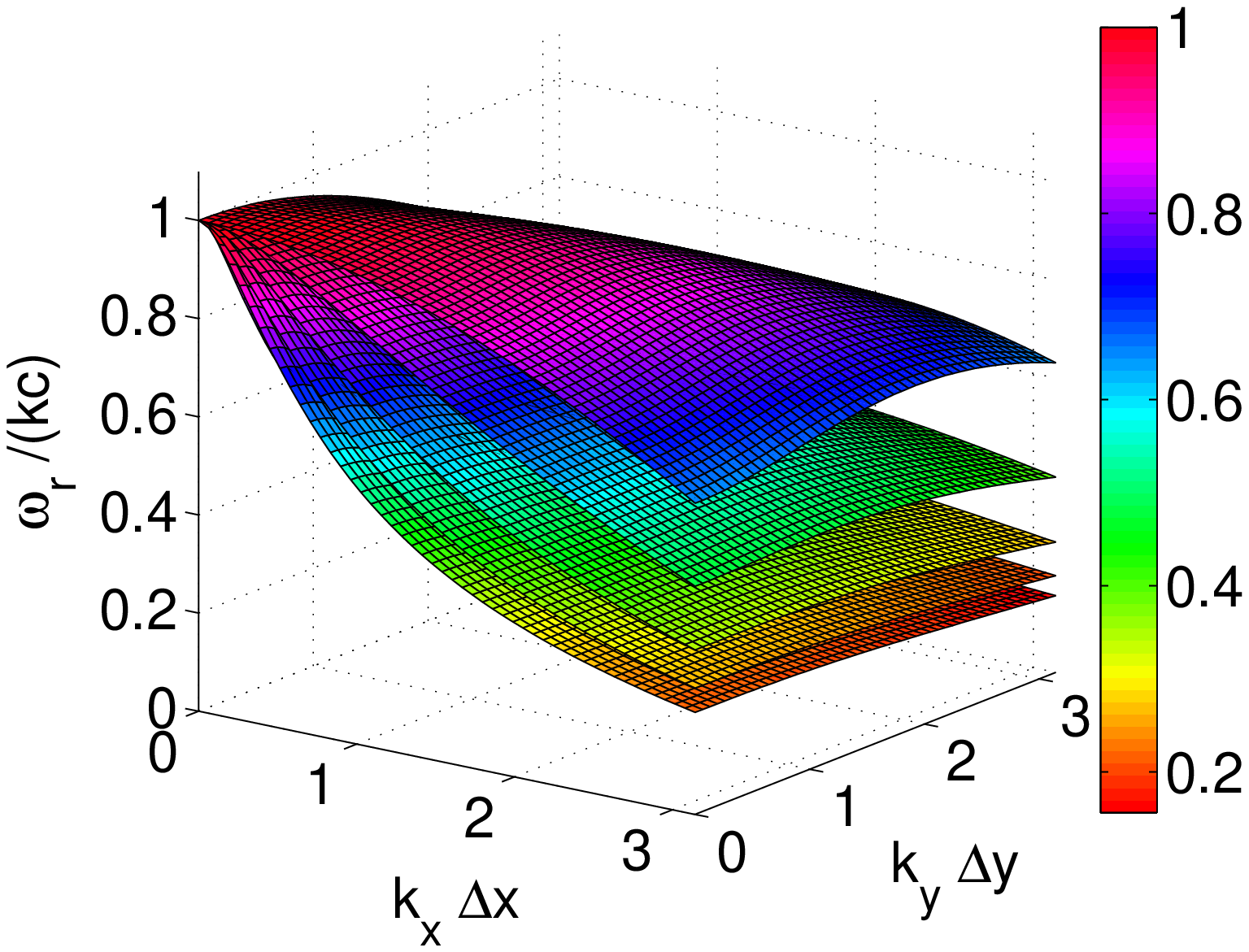}
\includegraphics[width=7cm]{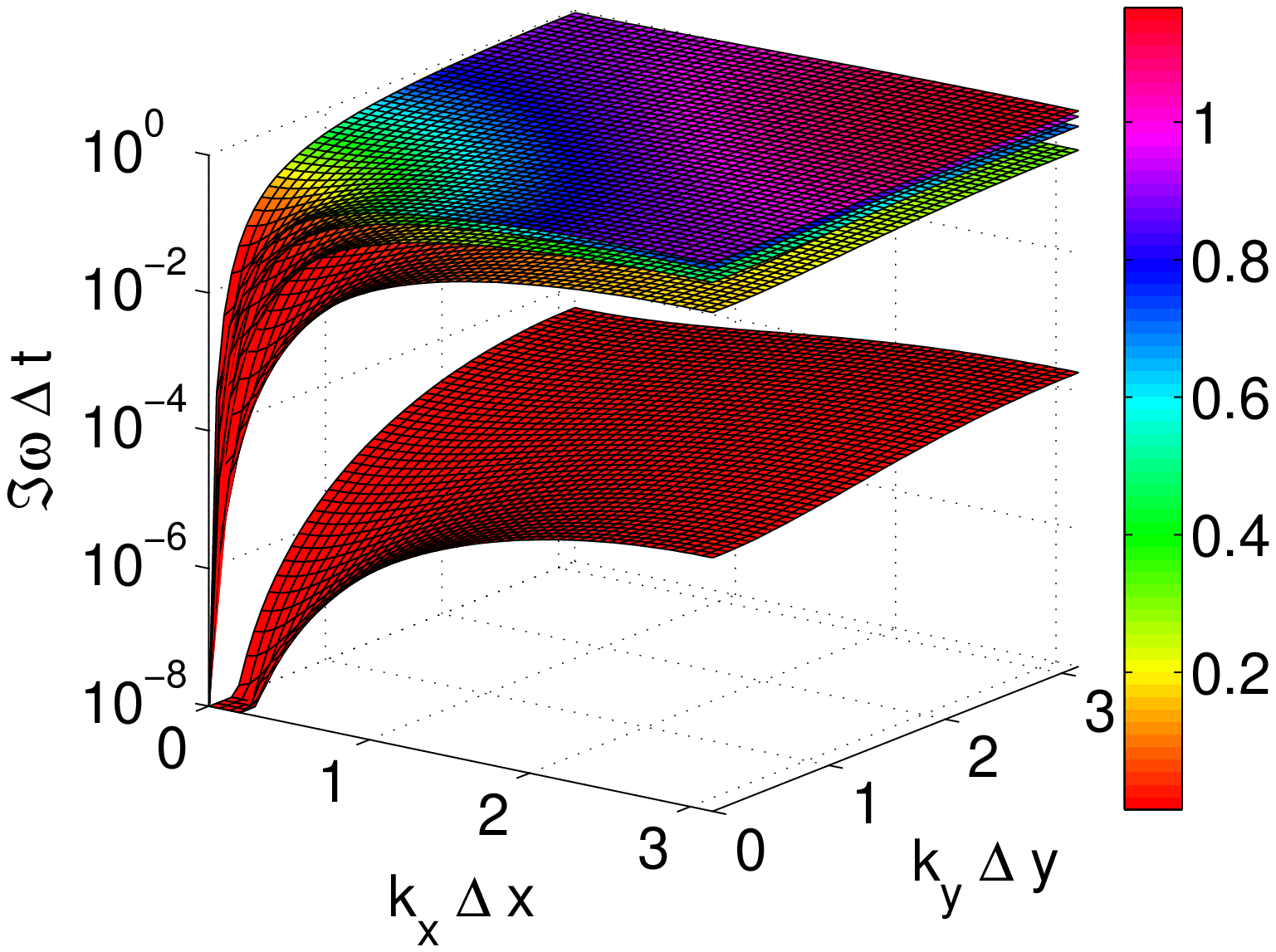}
\caption{Phase velocity (left) and damping rate $\Im \omega \Delta t$ (right) of the least damped root of Eq. (\ref{bp45}) as a function of
 $(k_x \Delta x, k_y \Delta y)$, for  different values of $c\Delta t/\Delta x = c\Delta t/\Delta y \in \{0.05, 0.66, 1.28, 1.9, 2.5\}$ (from top to bottom on the left and bottom to top on the right) and $\theta_f=1$. }
\label{fig301}
\end{figure}

\begin{figure}[htbp]
\includegraphics[width=7cm]{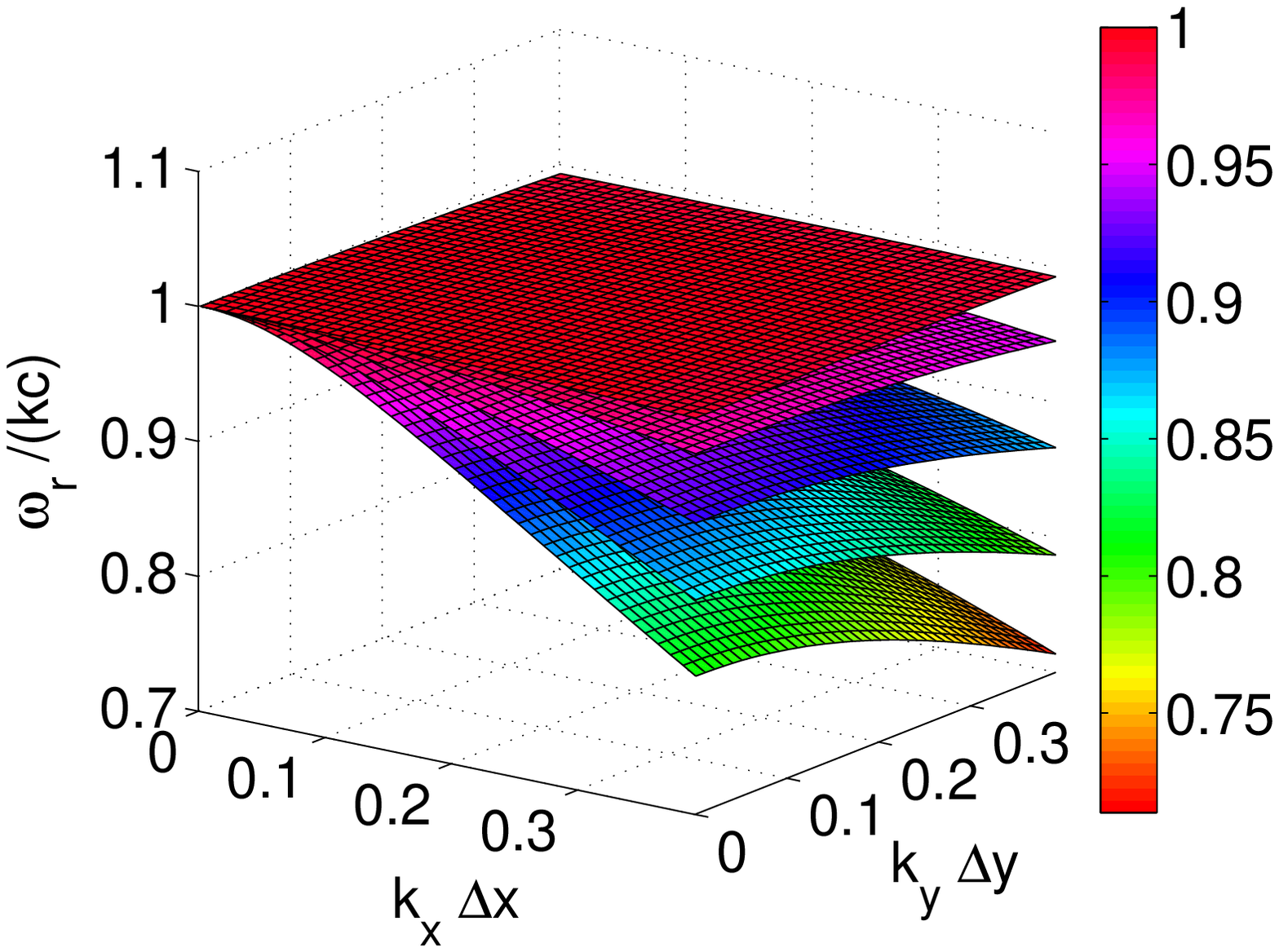}
\includegraphics[width=7cm]{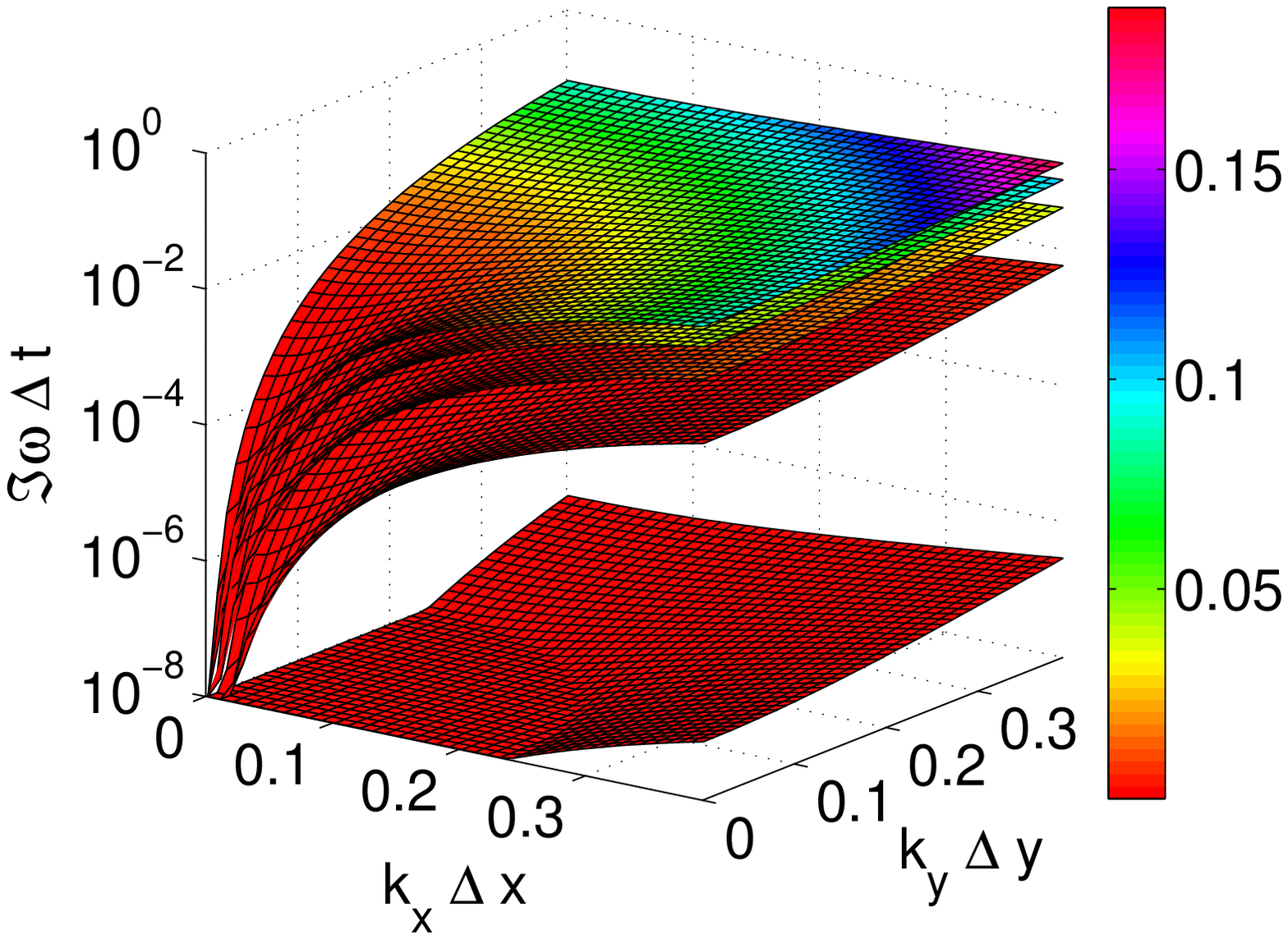}
\caption{Same as Fig. \ref{fig301} but with a narrower $(k_x\Delta x, k_y \Delta y)$ range.}
\label{fig302}
\end{figure}

\begin{figure}[htbp]
\includegraphics[width=7cm]{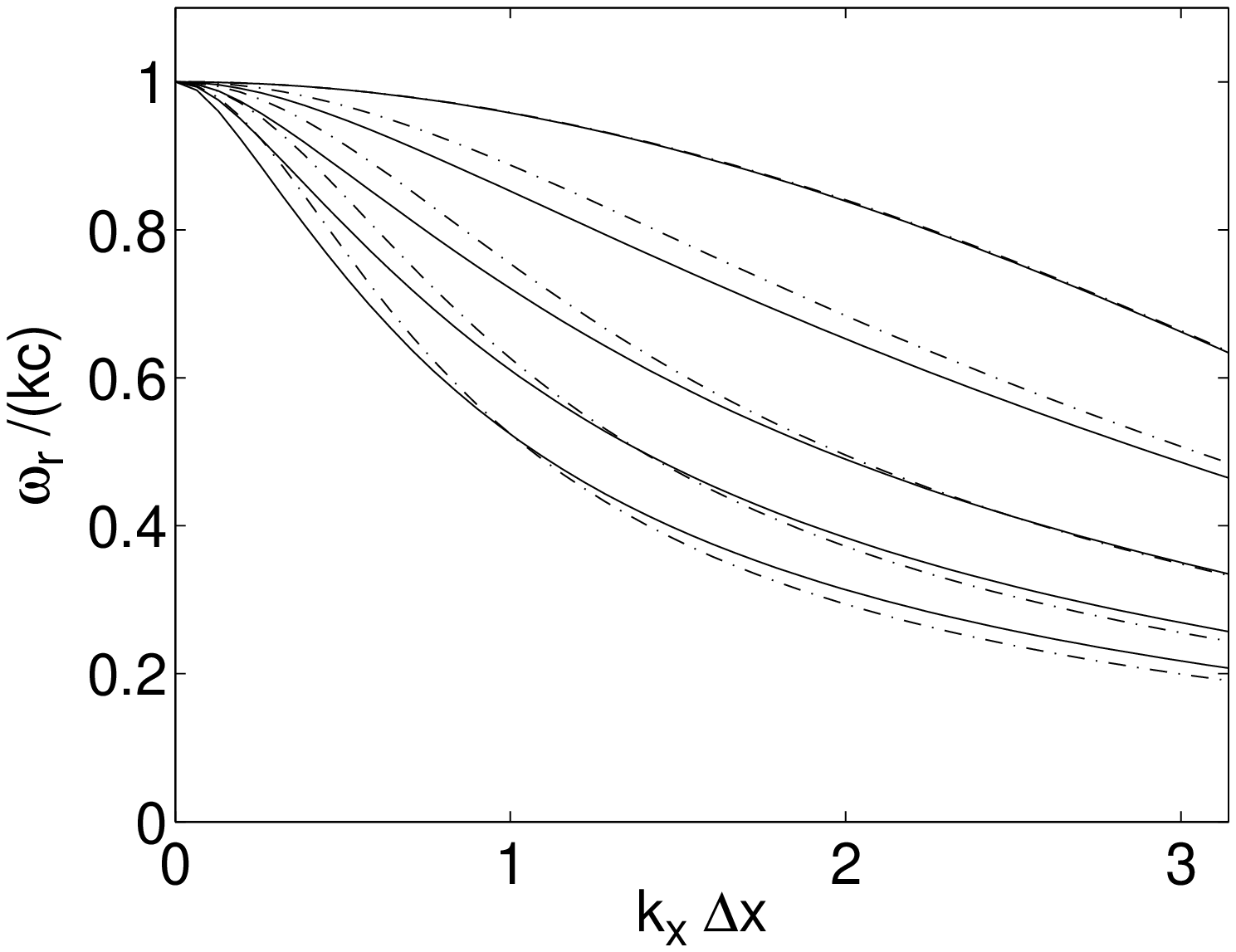}
\includegraphics[width=7cm]{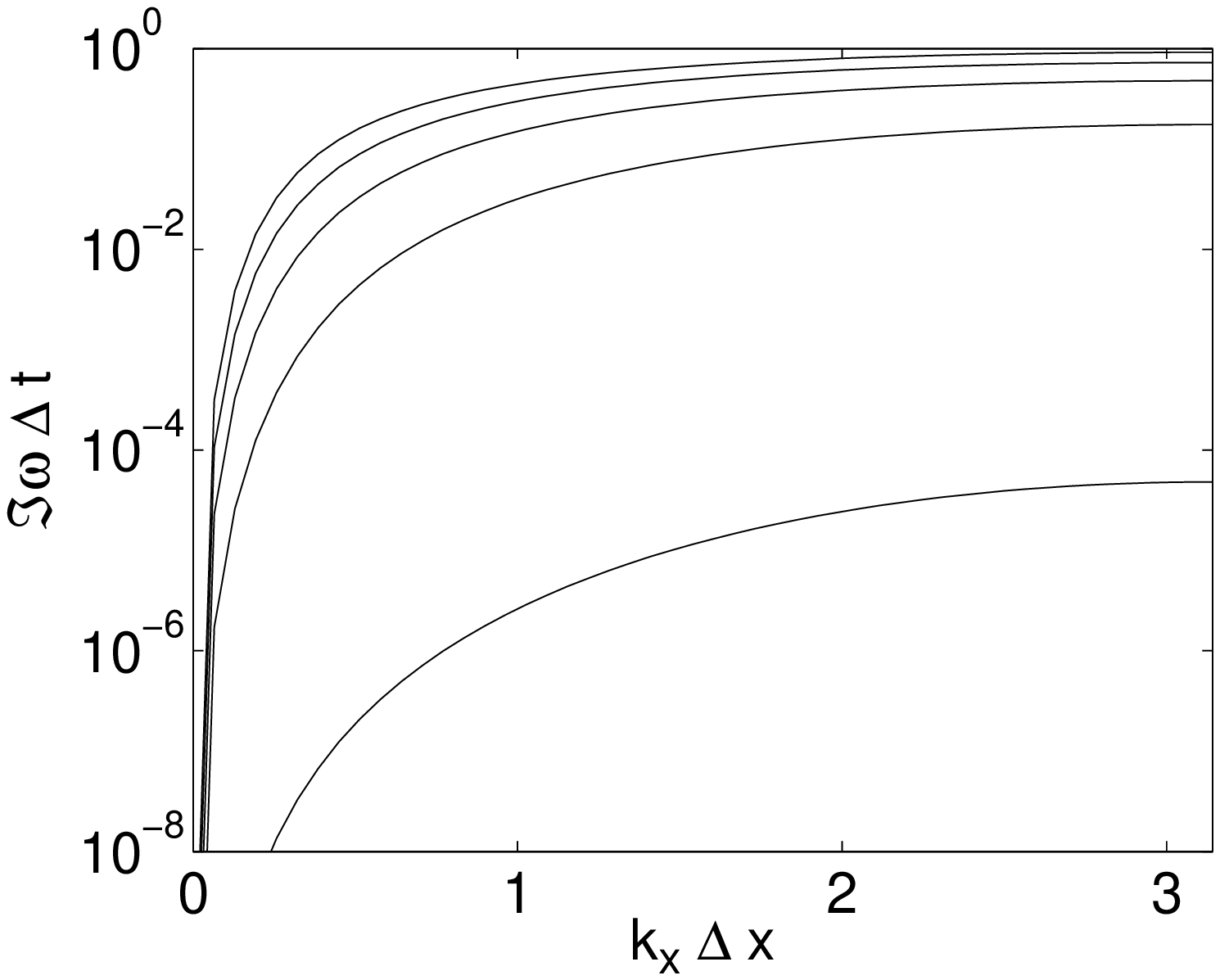}
\caption{Phase velocity (left) and damping rate $\Im \omega \Delta t$ (right) of the least damped root of Eq. (\ref{bp45}) as a function of
 $(k_x \Delta x)$, for  different values of $c\Delta t/\Delta x = c\Delta t/\Delta y \in \{0.05, 0.66, 1.28, 1.9, 2.5\}$ (from top to bottom on the left and bottom to top on the right) and $\theta_f=1$.
 Phase velocity without damping ($\theta_f=0$) is represented by dotted-dashed line.}
\label{fig303}
\end{figure}

\begin{figure}[htbp]
\includegraphics[width=7cm]{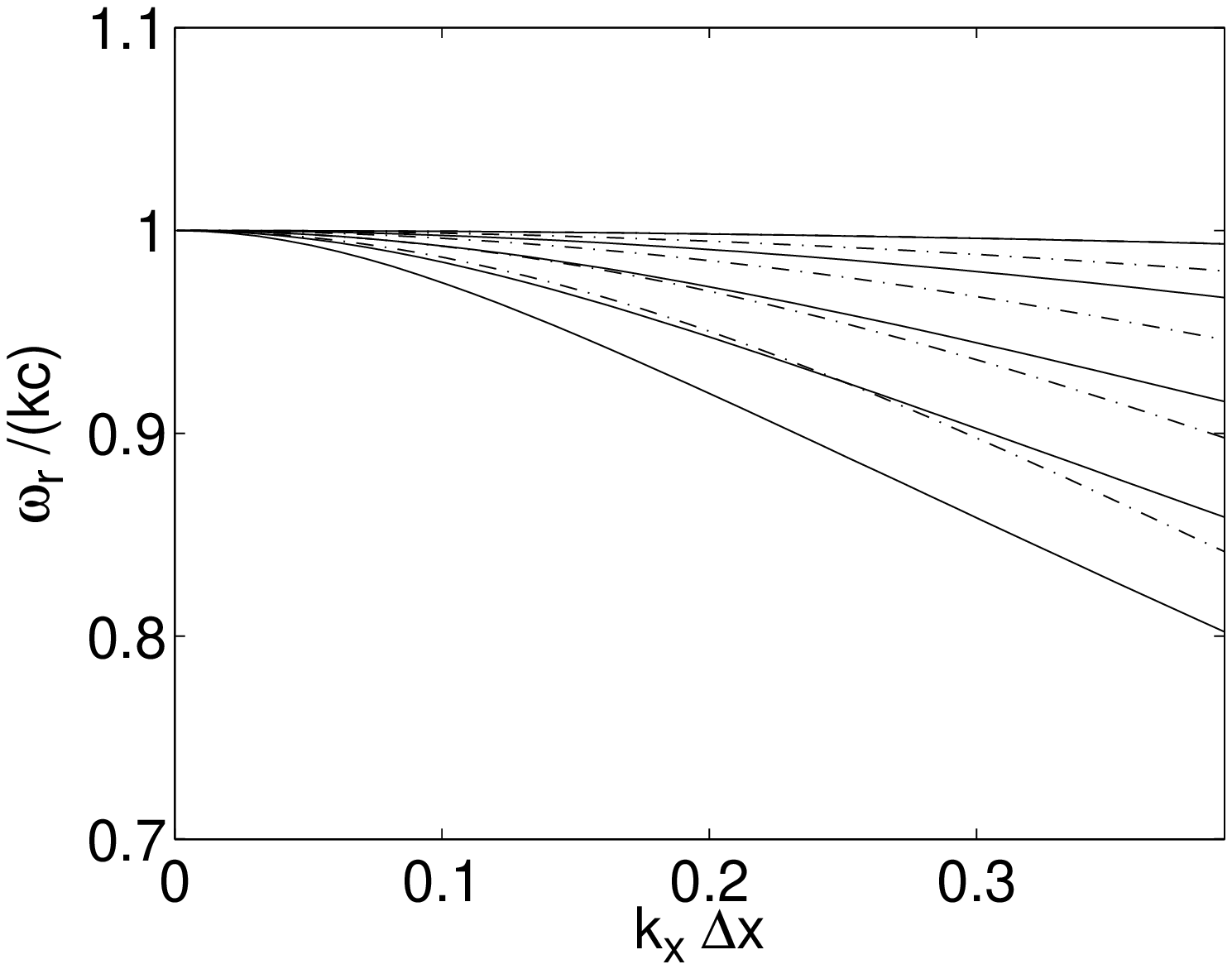}
\includegraphics[width=7cm]{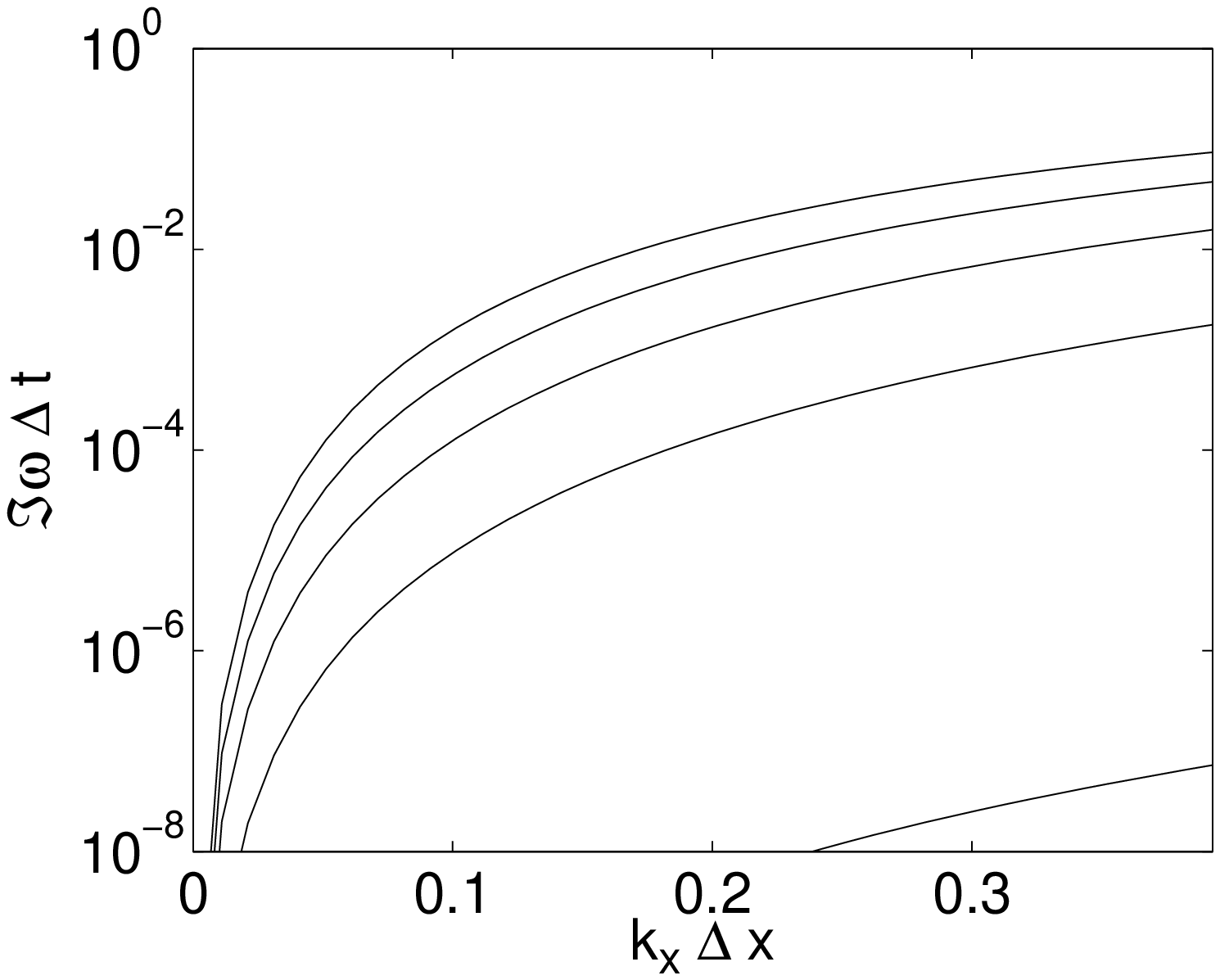}
\caption{Same as Fig. \ref{fig303} but with a narrower $(k_x\Delta x)$ range.}
\label{fig304}
\end{figure}

Let us now address the case of nonzero $\theta_f$. Figures \ref{fig301} and \ref{fig302} 
plot the normalized phase velocity $v_{\phi}/c$ (left) and damping rate $\Im \omega \Delta t$ (right) of the least damped root of 
Eq. (\ref{bp45}) as functions of $(k_x \Delta x, k_y \Delta y)$ for $\theta_f=1$. Cuts of these two quantities in the plane
$k_y=0$ are represented in Figures \ref{fig303} and \ref{fig304} respectively.
Again the phase velocity error grows for increasing $\Delta x$ and $\Delta t/\Delta x$.
A value $c\Delta t/\Delta x > 1$, therefore implies a reduced spatial step $k_x \Delta x
\lesssim 0.28$ ($c\Delta t/\Delta x = 1.27$) so as to keep phase velocity error below $5\%$. 
In this case the damping rate, which also increases with $\Delta x$ and $\Delta t/\Delta x$,
proves much too strong for applications relying on the propagation of an electromagnetic wave over several wavelengths. For example,
assuming $k_x\Delta x=0.2$ and $c\Delta t/\Delta x = 1$, a typical travel time of $200 \Delta t$ requires $\vert \Im \omega \vert \Delta t < 2.5\times 10^{-4}$ for a tolerable wave dissipation
($<5\%$). As seen in Fig. \ref{fig304}(right), this condition cannot be fulfilled when $\theta_f = 1$, which further demonstrates the need for an
adjustable-damping scheme for a proper modeling of laser-plasma interaction.

\subsection{Dispersion relation of electrostatic plasma waves}
\label{subsec:electostratic_waves}

We will now focus on the numerical relation dispersion of the electron plasma fluctuations in the case of a uniform,  nonrelativistic Maxwellian plasma
with a fixed neutralizing background. For this purpose, we shall adopt the formalism of Langdon \cite{langdon1979} that accounts for both finite space
and time steps, as well as allows for an arbitrary time-differencing scheme of the Lorentz equation. An infinite number of macroparticles is assumed,
yielding a continuous velocity distribution function (taken in the Maxwellian form). In this framework, as detailed in Appendix \ref{app:dispersion_relation},
the present adjustable-damping, direct implicit algorithm can be easily managed. The relation dispersion yielding the complex frequency $\omega$ as a function
of the wave number $k$ then reads
\begin{align}
\label{heat03}
&1 + \frac{(\Delta x/\lambda_D)^2}
{(k\Delta x)^2\left[ \frac{\sin\left( k\Delta x/2\right)}{k\Delta x/2}\right]^2}
\sum_{p=-\infty}^{+\infty}
\left[ \frac{\sin\left( k_p\Delta x/2\right)}{k_p\Delta x/2}\right]^{2m+2}
\frac{\sin(k_p \Delta x)}{k_p \Delta x}
\sum_{q=-\infty}^{+\infty} \left[ 1+ \xi_q \mathcal{Z}(\xi_q)\right] \notag \\
& + \frac{(\omega_p\Delta t)^2/2}{(k\Delta x)^2
\left[\frac{\sin\left(k\Delta x/2\right)}{k\Delta x/2}\right]^2}
\sum_{p=-\infty}^{+\infty} \left( k_p \Delta x \right)^2
\left[ \frac{\sin\left( k_p\Delta x/2\right)}{k_p\Delta x/2}\right]^{2m+2}
\frac{\sin(k_p \Delta x)}{k_p \Delta x} \mathcal{S}(\theta_f) = 0 \, ,
\end{align}
where $m$ is the order of the shape factor \cite{bird85}. $k_p=k-2\pi p/\Delta x$ and $\omega_q =\omega -2\pi q/\Delta t$
are the aliased wave number and frequency, respectively. $\mathcal{Z}$ denotes the plasma dispersion function \cite{frie61} whose argument is
$\xi_q =\omega_q /\sqrt{2}k_p v_t$ (where $v_t$ is the electron thermal velocity). Moreover, we have defined the function $\mathcal{S}$ as
\begin{equation}
\mathcal{S}(\theta_f) = \sum_{s=0}^{+\infty}
\frac{e^{i(\omega/\omega_p) s (\omega_p\Delta t)}}{(2/\theta_f)^s} e^{-\frac{1}{2}(\lambda_D/\Delta x)^2
s^2(k\Delta x)^2(\omega_p\Delta t)^2} \, ,
\end{equation}
with the value $\mathcal{S}(0)=1$. We have numerically solved Eq. (\ref{heat03}) using the nonlinear solver STRSCNE developed in Ref. \cite{bellavia04}
and the algorithm of Ref. \cite{weideman94} to compute the $\mathcal{Z}$ function. We will restrict the following analysis to systems characterized by
a crude resolution of the Debye length ($\Delta x/\lambda_D >1$), as is commonplace in simulations of large-scale, high-density plasmas.

\begin{figure}[h]
\includegraphics[width=4.5cm]{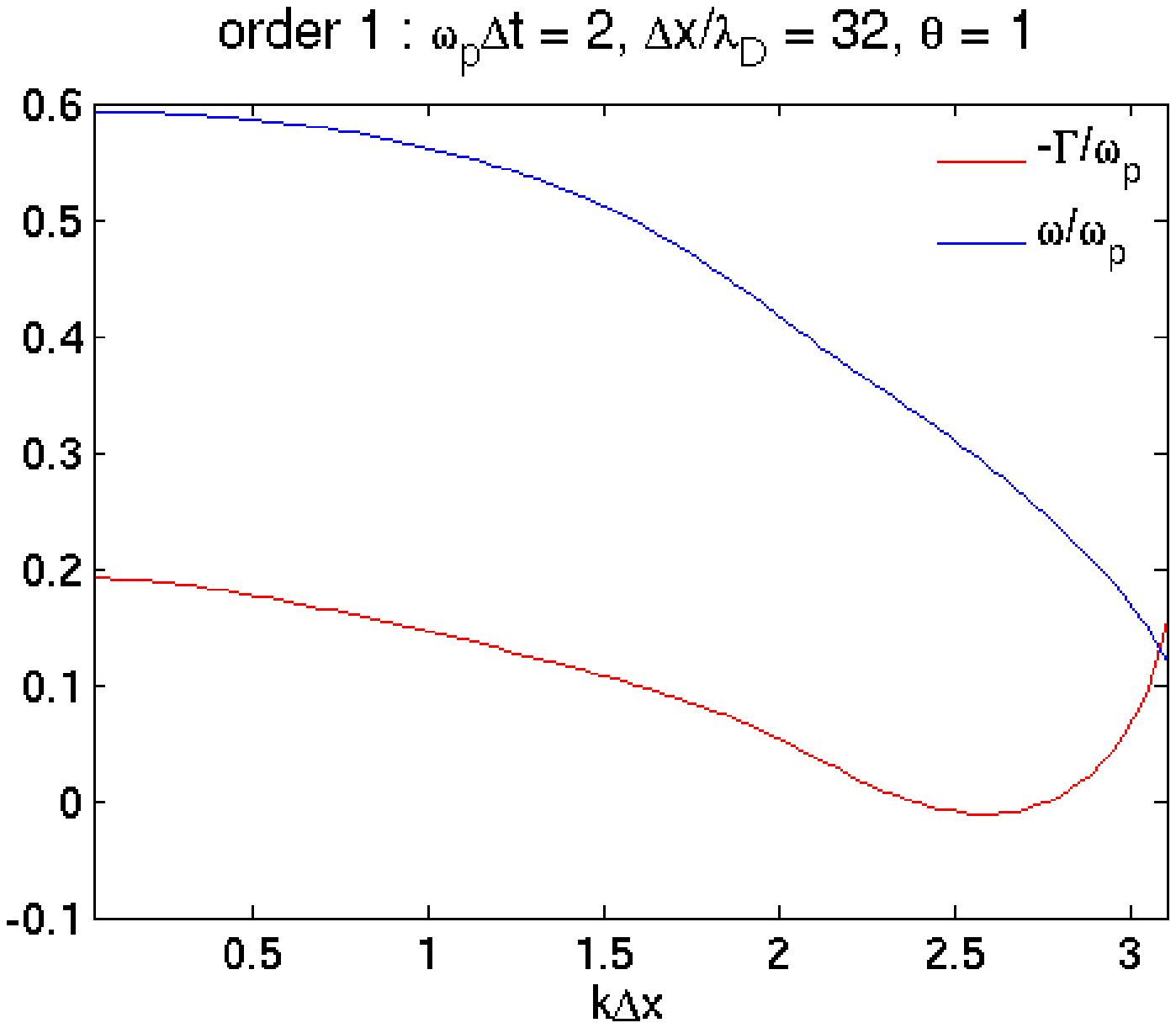}
\includegraphics[width=4.5cm]{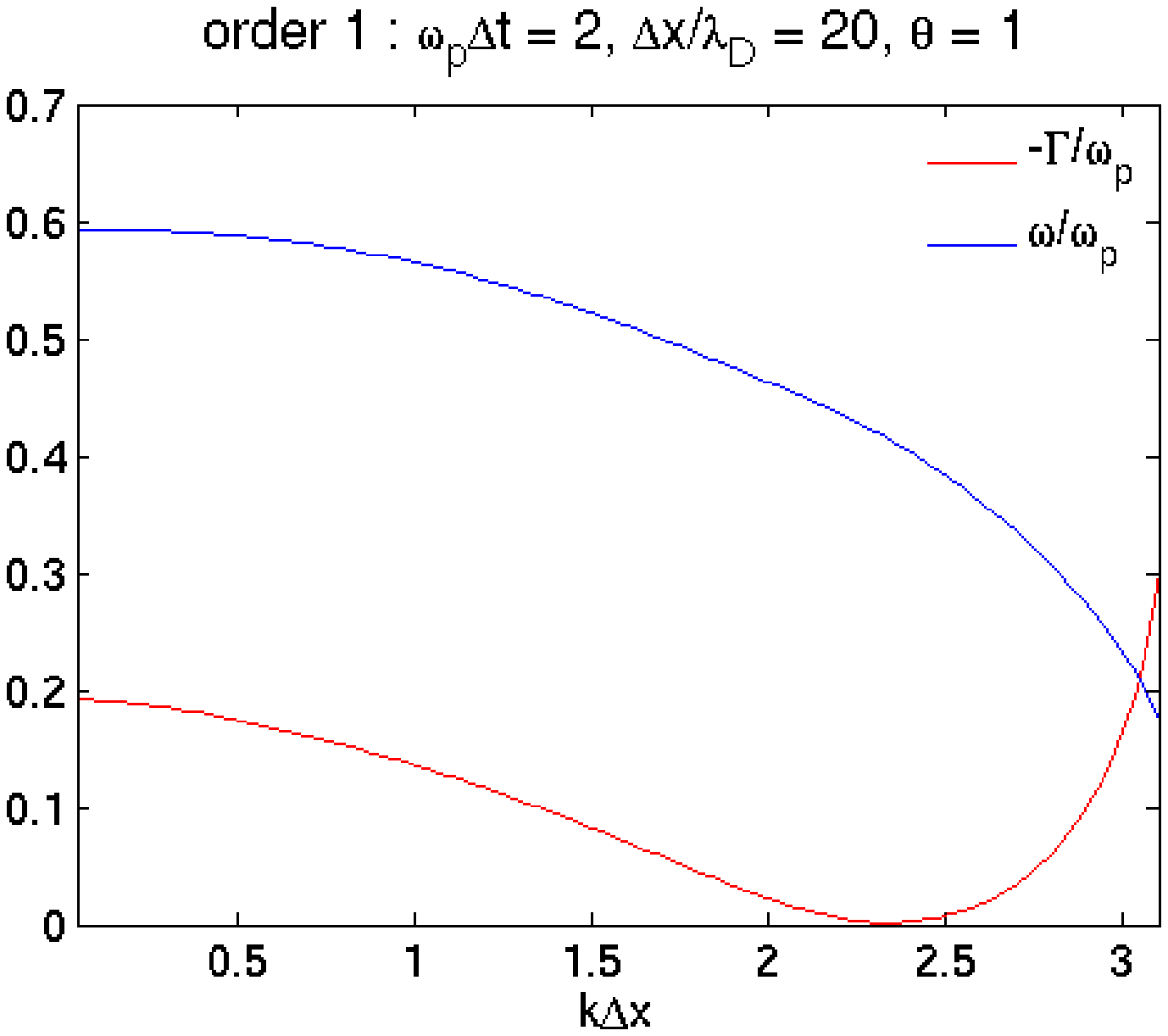}
\includegraphics[width=4.5cm]{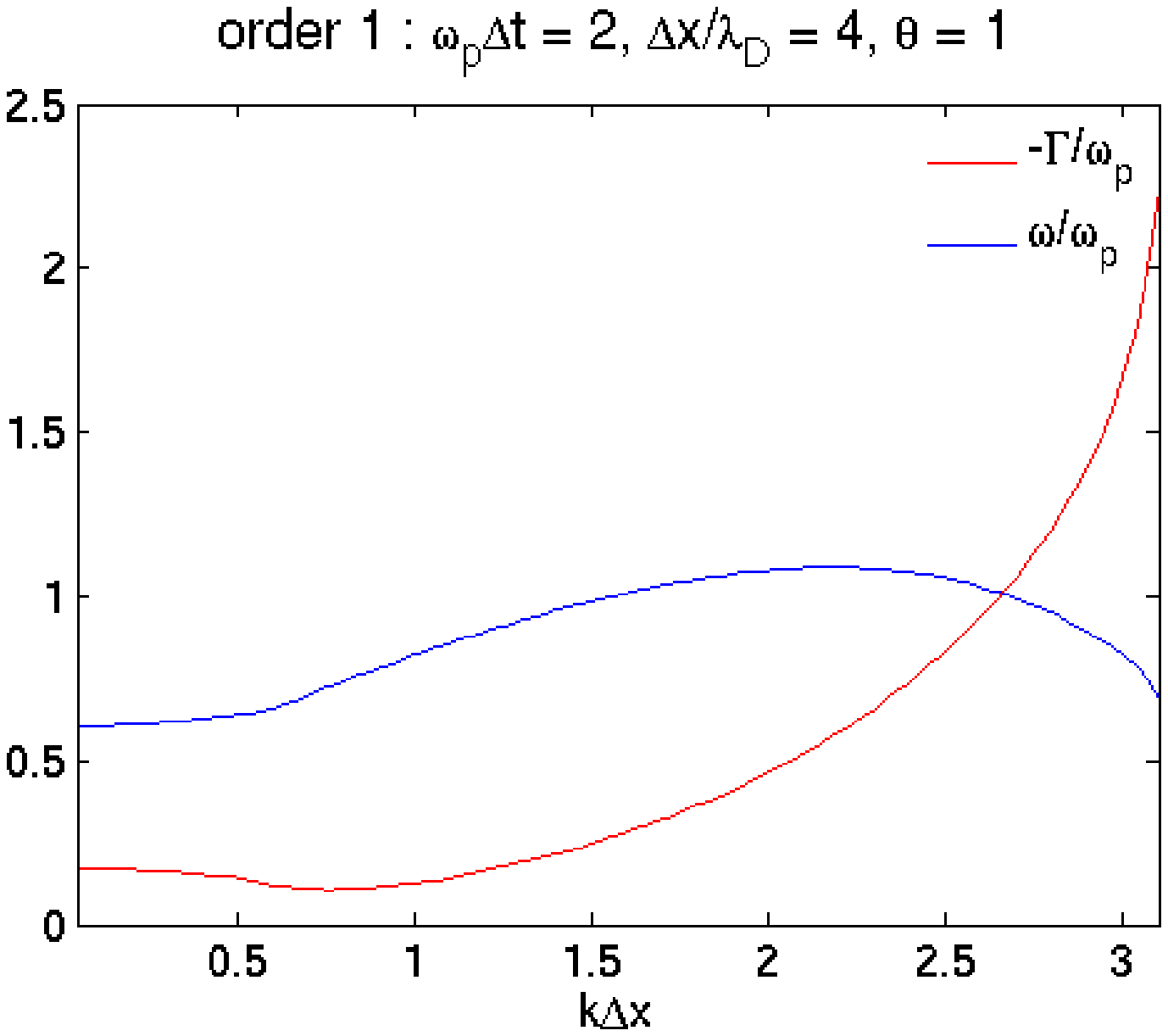}
\caption{Real frequency (blue) and growth rate (red) vs $k\Delta x$ of the dominant mode solving Eq. (\ref{heat03}) with $\omega_p \Delta t = 2$, $\theta_f=1$ and a linear
weight factor ($n=1$): $\Delta x/\lambda_D=32$ (left), 20 (center) and 4 (right).}
\label{heat001}
\end{figure}

Figure \ref{heat001} displays the $k$-dependence of the complex frequency of the fastest growing (or least damped) mode solution of Eq. (\ref{heat03}) for
$\theta_f =1$, $\omega_p \Delta t = 2$ and various values of $\Delta x/\lambda_D$. For $\Delta x/\lambda_D =32$ (\emph{i.e.}, $v_t \Delta t/\Delta x = 0.06$),
most of the $k$-spectrum is damped except for a bounded unstable region located near
$k\Delta x \sim 2.6$ with a maximum growth rate $\Im \omega/\omega_p \sim 0.011$.
This corresponds to the well-known
finite-grid instability \cite{bird85} commonly afflicting PIC simulations with $\Delta x/\lambda_D \gg 1$, and responsible for nonphysical field energy growth and plasma heating.
This instability originates from the interplay of the aliased wave numbers in Eq. (\ref{heat03}).
Note also the nonphysical $k$-dependence of the real frequency obtained at large $\omega_p\Delta t$ : $\Re \omega$ is
significantly below $\omega_p$ at $k=0$ and further drops with increasing $k\Delta x$.
As seen in Fig. \ref{heat001}, decreasing $\Delta x/\lambda_D$ eventually leads to a complete stabilization of the system along with a displacement of the dominant mode towards low $k$ values. For $\Delta x/\lambda_D= 4$ (\emph{i.e.}, $v_t \Delta t/\Delta x = 0.5$),
the least damped mode is thus located at $k\Delta x = 0.76$ with $\Im \omega/\omega_p \sim -0.1$. This evolution points to a transition between spatial step-dominated
and time-step-dominated regimes.

The dependence of the characteristics of the dominant mode on the ratio $\Delta x/\lambda_D \gg 1$ and the weight factor order is summarized in Table \ref{tabheat01}
for $\theta_f =1$ and $\omega_p \Delta t =2$. The benefit of a high-order interpolation scheme is clearly evidenced: the system turns out to be entirely stabilized
up to $\Delta x/\lambda_D = 32$ with a quadratic weight factor, and $\Delta x/\lambda_D = 64$ with a cubic weight factor. In addition, the wavenumber of the increasingly
damped dominant mode is shifted downward.

\begin{table}
\begin{center}
\begin{tabular}{lcccc}
\hline
\hline
 $\Delta x/\lambda_D$ & 14.3 & 22.6 & 32 & 64  \\
\hline
 linear & -0.024 & $3.3\times10^{-3}$  & 0.011 & 0.01 \\
  & (2.11) & (2.42) & (2.58) & (2.85)  \\
 quadratic & -0.04 & -0.015  & $-3.7\times10^{-3}$ & $2.8\times10^{-3}$ \\
  & (1.96) & (2.30) & (2.48) & (2.70) \\
 cubic & -0.039 & -0.018 & $-8.6\times10^{-3}$ & $-2\times 10^{-4}$ \\
  & (1.84) & (2.14) & (2.36) & (2.67) \\
\hline
\hline
\end{tabular}
\caption{Imaginary frequency $\Im \omega/\omega_p$ (wavenumber $k\Delta x$) of the dominant mode as a function of the ratio $\Delta x/\lambda_D$ and the
weight factor order for $\omega_p\Delta t=2$ and $\theta_f=1$.}\label{tabheat01}
\end{center}
\end{table}

A connection between the present calculations and previously published simulation results \cite{brac82,cohenlangdon1989} is provided by Tables \ref{tabheat02} and
\ref{tabheat03}, which display the dependence of the dominant mode on the ratio $v_t \Delta t/\Delta x = \omega_p \Delta t/(\Delta x/\lambda_D)$, as well as on the
damping parameter (the time step being fixed to  $\omega_p \Delta t=2$). An extensive set of implicit electrostatic PIC simulations using the $D_1$ scheme
(\emph{i.e.}, $\theta_f=1$) and linear interpolation has indeed revealed that satisfactory energy conservation can be achieved in the range \cite{brac82,cohenlangdon1989}
\begin{equation}
\label{heat06}
0.1 \lesssim v_t\frac{\Delta t}{\Delta x} \lesssim 1
\end{equation}
Even though the present stability analysis alone is not expected to account for the complex issue of numerical self-heating \cite{bird85,ueda94}, the results of Table \ref{tabheat02}
are found in reasonable agreement with the lower bound of the above heuristic range, as they indicate a complete stabilization of the system for $v_t \Delta t/\Delta x \gtrsim 0.1$ in case of a linear weigth factor and $\theta_f = 1$. For lower $\theta_f$ values, stabilization is reached for increased $v_t \Delta t/\Delta x$. Moreover, Table \ref{tabheat03} shows that the use of a quadratic weight factor permits to suppress the finite-grid instability at reduced $v_t \Delta t/\Delta x$ ($\gtrsim 0.06$ for $\theta_f = 1$). Similarly to Fig. \ref{heat001},
a clear transition from the high-$k$ spatial regime to the low-$k$ temporal regime is evidenced when raising $v_t \Delta t/\Delta x$.  As expected, a high-order ($m > 1$) weight factor, which enables to filter out high spatial frequencies, proves beneficial only in the high-$k$,  grid-instability regime (for $v_t \Delta t /\Delta x \lesssim 0.25$). Note that we have not
considered values $v_t \Delta t/\Delta x > 1$ since, in the present case, this would imply $\Delta x/\lambda_D< 2$, a parameter range of little practical interest for the aforementioned applications.

\begin{table}
\begin{center}
\begin{tabular}{lcccc}
\hline
\hline
 $ \theta_f$ &  0 & 0.1 & 0.5 & 1  \\
 $ v_t\Delta t/\Delta x$ &  &  &    \\
\hline
 0.05 & 0.0166 & 0.016 & 0.0150 & 0.012 \\
  & (2.64) & (2.64) & (2.67) & (2.67)  \\
 0.0625 & 0.0192 & 0.0187 & 0.0161 & 0.011 \\
  & (2.51) & (2.51) & (2.54) & (2.58) \\
 0.1 & 0.0204 & 0.0185  & 0.01 & $-1.8\times10^{-3}$ \\
  & (2.18) & (2.18) & (2.27) & (2.33) \\
 0.25 & $8\times10^{-4}$ & $-7.4\times10^{-3}$  & -0.04 & -0.08 \\
  & (1.05) & (1.11) & (1.28) & (1.46) \\
 0.5 & 0 & -0.01  & -0.0508 & -0.105 \\
  & (0.39) & (0.54) & (0.63) & (0.76) \\
 1 & 0 & -0.0102  & -0.0532 & -0.112 \\
  & (0.14) & (0.27) & (0.33) & (0.39) \\
\hline
\hline
\end{tabular}
\caption{Imaginary frequency $\Im \omega/\omega_p$ (wave number $k\Delta x$) of the dominant mode as a function of the ratio $v_t \Delta t/\Delta x$
 and the damping parameter $\theta_f$ for $\omega_p\Delta t=2$ and a linear weight factor.}\label{tabheat02}
\end{center}
\end{table}

\begin{table}[htbp]
\begin{center}
\begin{tabular}{lcccc}
\hline
\hline
 $\theta_f$ &  0 & 0.1 & 0.5 & 1  \\
 $v_t\Delta t/\Delta x$ &  &  &    \\
\hline
 0.05 & $5.3\times10^{-3}$ & $5\times10^{-3}$ & $3.5\times10^{-3}$ & $10^{-4}$ \\
  & (2.54) & (2.54) & (2.58) & (2.61)  \\
 0.0625 & $5.4\times10^{-3}$ &  $4.8\times10^{-3}$ & $1.8\times10^{-3}$ & $-3.7\times10^{-3}$ \\
  & (2.39) & (2.39) & (2.45) & (2.48)  \\
 0.1 & $3.2\times10^{-3}$ & $1.1\times10^{-3}$  & $-8\times10^{-3}$ & -0.0207 \\
  & (1.99) & (2.02) & (2.14) & (2.24) \\
 0.25 & 0 & $-8.1\times10^{-3}$  & -0.039 & -0.078 \\
  & (0.81) & (1.05) & (1.22) & (1.4) \\
 0.5 & 0 & $-9.7\times10^{-3}$  & -0.05 & -0.103 \\
  & (0.33) & (0.54) & (0.64) & (0.76) \\
 1 & 0 & -0.01  & -0.053 & -0.11 \\
  & (0.14) & (0.27) & (0.33) & (0.39) \\
\hline
\hline
\end{tabular}
\caption{Imaginary frequency $\Im \omega/\omega_p$ (wave number $k\Delta x$) of the dominant mode as a function of the ratio $v_t \Delta t/\Delta x$
 and the damping parameter $\theta_f$ for $\omega_p\Delta t=2$ and a quadratic ($n=2$) weight factor.}
 \label{tabheat03}
\end{center}
\end{table}

Further insight into the stability properties of the adjustable-damping scheme is given by fixing the ratio $v_t \Delta t/\Delta x = 0.09$ and varying accordingly
the space and time steps.  Equivalently, within the laser-plasma context which we propose to address, this can be achieved by fixing the parameters
$\omega_0 \Delta x/c$ and $\omega_0 \Delta t$ (where $\omega_0$ is the incident laser frequency) and varying the plasma density. The resulting data is displayed in
Table \ref{heattab04} in the ranges $1.26 \le \omega_p t \le 8.94$ and $14.3 \le \Delta x/\lambda_D \le 101.1$. One can see that  a linear shape factor proves
rather inappropriate  for most of the parameter range considered. By contrast, complete stabilization is achieved for $n \ge 2$ weight factors. It is worth noting that,
in terms of laser-plasma parameters, the rightmost column of Table \ref{heattab04}  corresponds to a $2000 n_c$, 1 keV plasma (where $n_c$ is the critical density
at the laser frequency $\omega_0$) discretized with $\omega_0 \Delta t= 0.2$ and $\omega_0\Delta x/c = 0.1$. In addition to accessing such extreme plasma conditions,
employing a cubic weight factor may give the opportunity to reduce the damping parameter $\theta_f$.

\begin{table}
\begin{center}
\begin{tabular}{lcccccccc}
\hline
\hline
$\omega_p \Delta t $ & 1.26 & 2 & 2.83 & 3.46 & 4 & 5.66 & 6.32 & 8.94 \\
$\Delta x/\lambda_D $ & 14.3 & 22.6 & 32 & 39.1 & 45.2 & 64 & 71.5 & 101 \\
\hline
linear & -0.0036 & 0.0034 & 0.0048 & 0.0047 & 0.0044 &
0.0036 & 0.0033 & 0.0024 \\
 & (2.09) & (2.41) & (2.59) & (2.67) & (2.74) & (2.85) & (2.87) & (2.96) \\
quadratic & -0.021 & -0.015 & -0.01 & -0.0078 & -0.0066 &
-0.0044 & -0.0039 & -0.0026 \\
 & (1.95) & (2.3) & (2.5) & (2.62) & (2.68) & (2.82) & (2.85) & (2.92) \\
cubic &  -0.022 & -0.019 & -0.015 & -0.013 & -0.011 &
-0.0079 & -0.0071 & -0.0051 \\
 & (1.83) & (2.16) & (2.36) & (2.48) & (2.56) & (2.7) & (2.76) & (2.85) \\
 \hline
 \hline
\end{tabular}
\caption{Imaginary frequency $\Im \omega/\omega_p$ (wave number $k\Delta x$) of the dominant mode as a function of the space and time steps and
the weight factor order, for a fixed ratio  $v_t \Delta t/\Delta x = 0.09$ and $\theta_f=1$.}
\label{heattab04}
\end{center}
\end{table}

\section{Numerical applications}
\label{sec:applications}

\subsection{Wave propagation in vacuum}
\label{subsec:wave_propagation}

\begin{figure}[htbp]
\begin{center}
\includegraphics[width=0.45\textwidth]{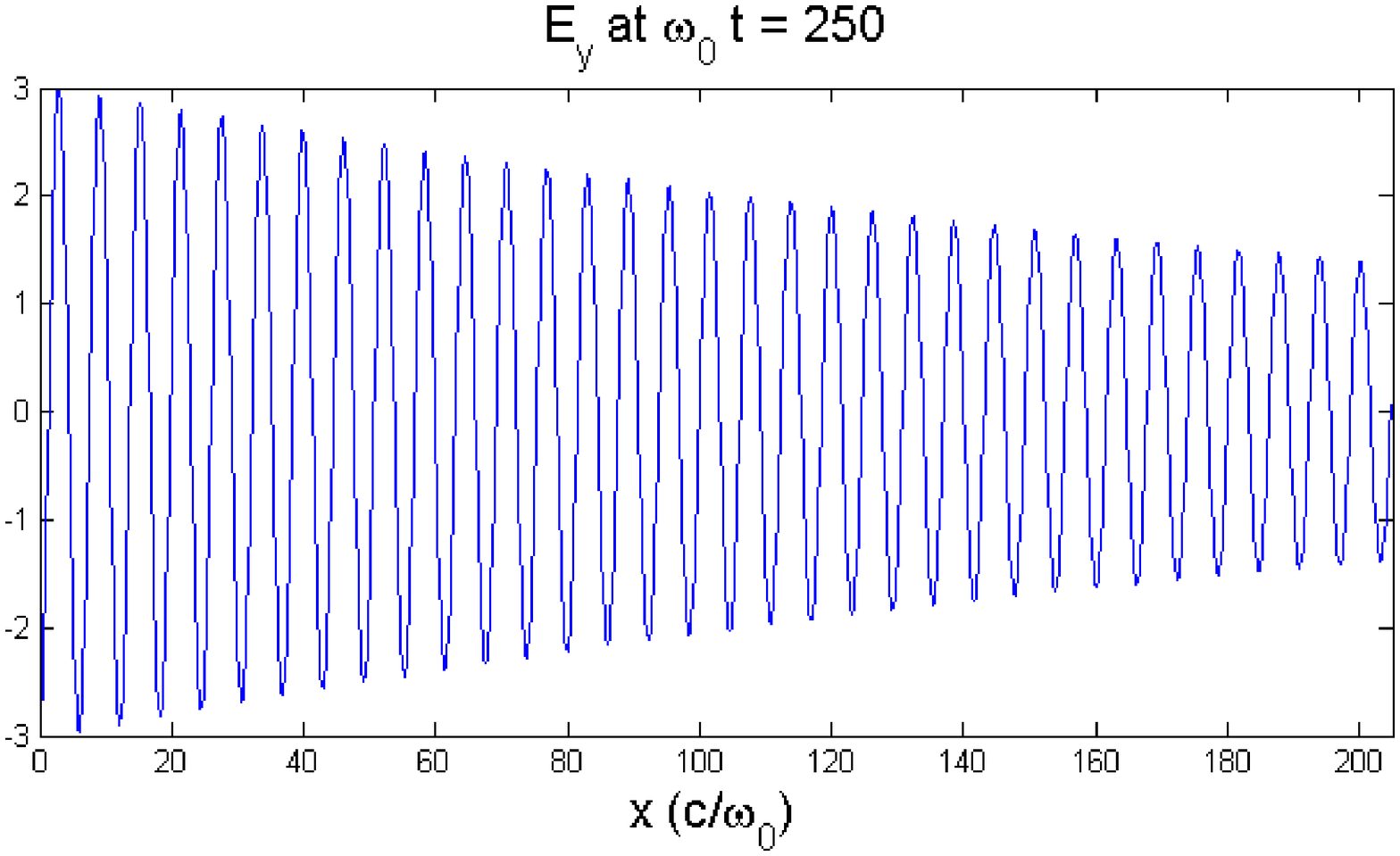}
\includegraphics[width=0.45\textwidth]{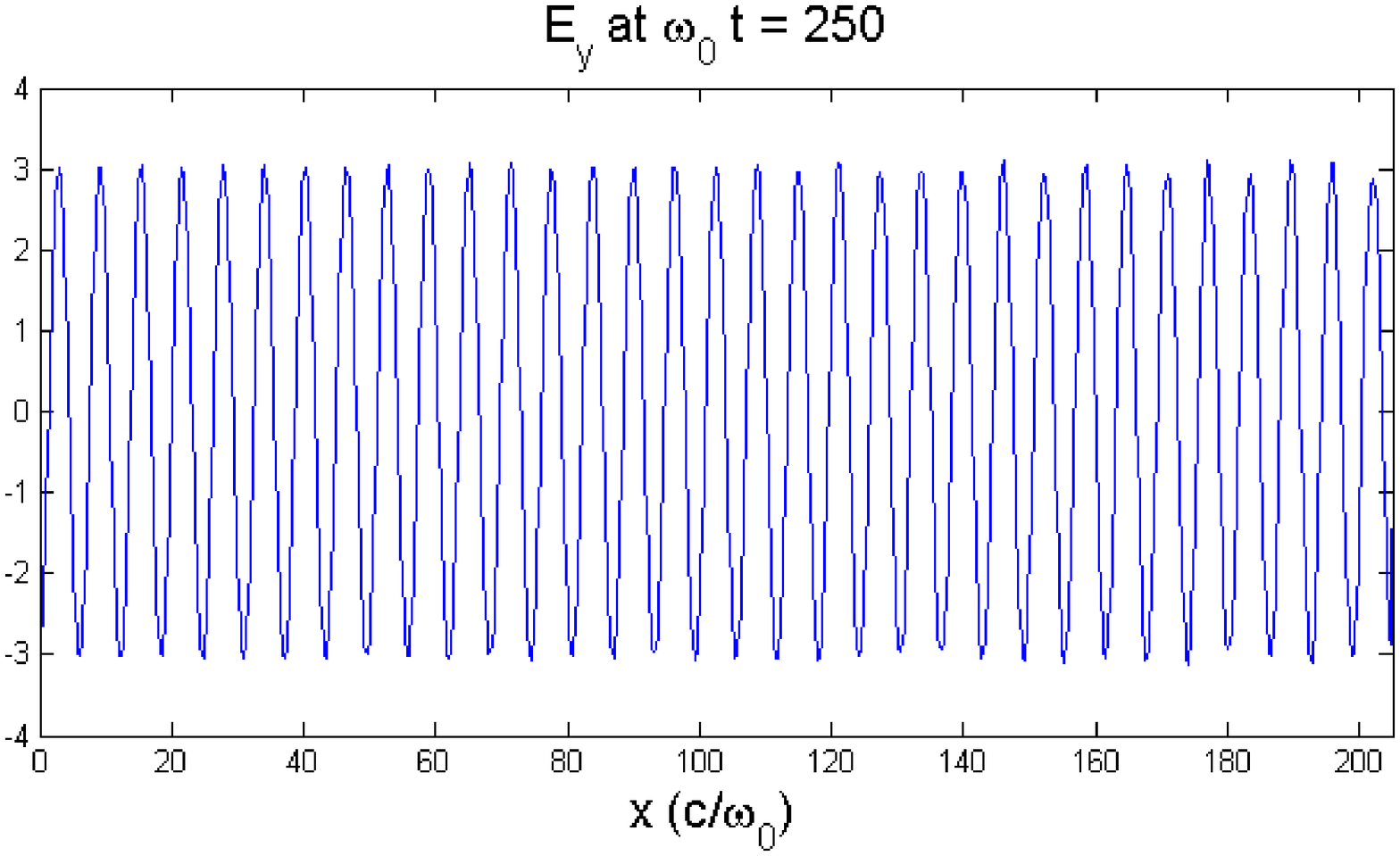} \\
\includegraphics[width=0.45\textwidth]{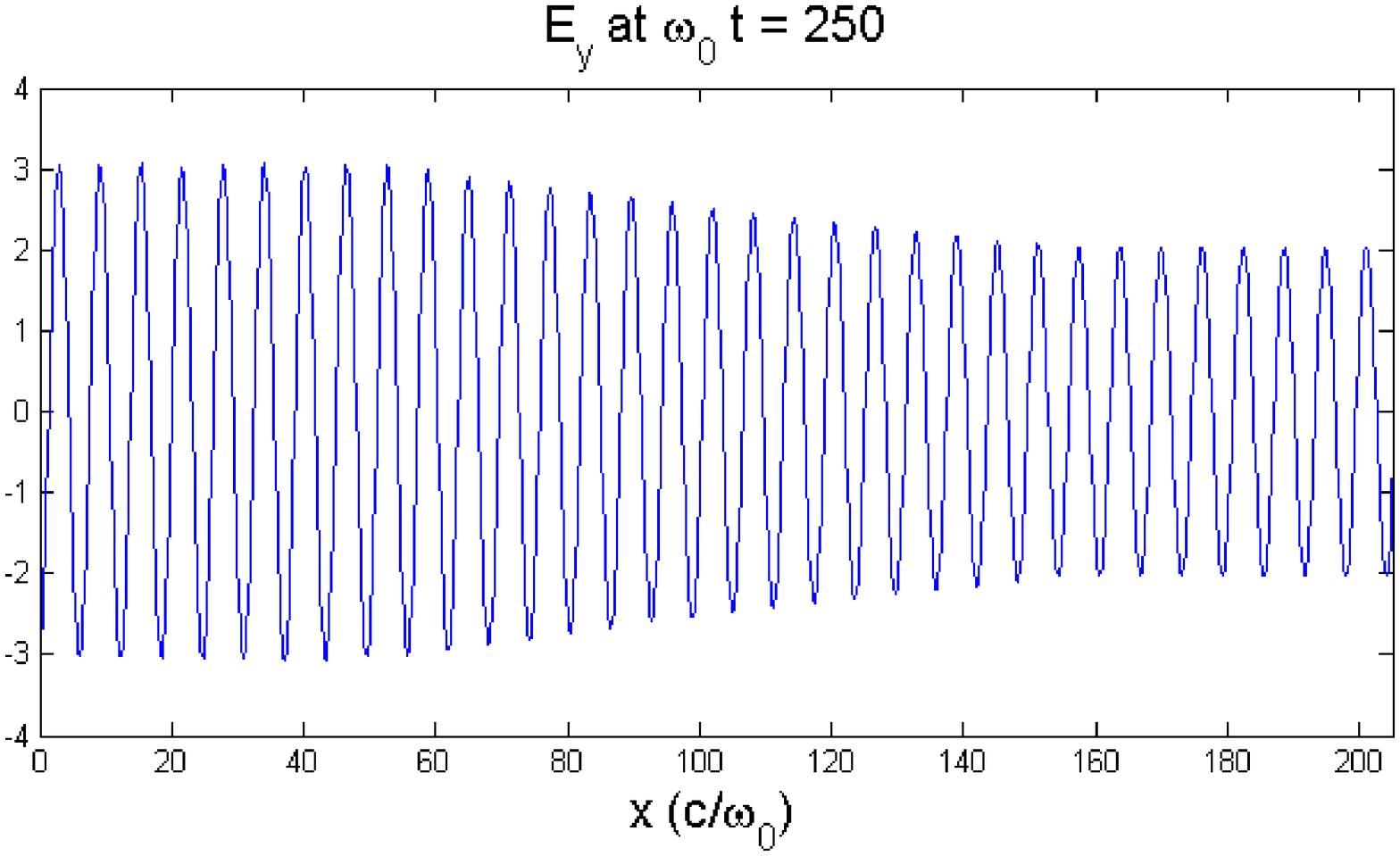}
\caption{Propagation of a plane wave with $\theta_f=1$ (top, left), $\theta_f = 0$ (top, right), and a spatially varying $\theta_f$ profile according to Eq. (\ref{waveq01}) (bottom).}
\label{wave01}
\end{center}
\end{figure}

Here, we illustrate the capability of the adjustable damping, implicit scheme implemented in the code ELIXIRS to manage the propagation of electromagnetic
waves in vacuum. Let us consider a plane wave, with normalized vector potential $a_0=3$ and frequency $\omega_0$, entering the left-hand side of a
$1024 \Delta x \times 4 \Delta y$ box, with $\Delta x=0.2 c/\omega_0$, $\Delta y=0.8 c/\omega_0$ and $\Delta t=0.2 \omega_0^{-1}$.  The wave is injected
and absorbed using the procedure detailed in \ref{subsec:boundary_conditions}. Figure \ref{wave01}(left) shows the expected monotonous damping of the
incident wave induced when a spatially uniform damping parameter $\theta_f=1$ is applied. After propagating across the simulation box, the wave amplitude
is measured to be $46\%$ of the initial value, which is close to the theoretical value ($49\%$). The opposite,  dissipation-free case corresponding to $\theta_f=0$ is
displayed in Fig. \ref {wave01}(right). Finally, with the problem of laser plasma interaction in mind, we address the case of a spatially varying $\theta_f$
profile in the form
\begin{equation}
\label{waveq01}
\left\{
\begin{array}{ll}
\theta_f=0, & 0< \omega_0 x/c < 51.2 \\
\theta_f=1, & 51.2< \omega_0 x/c< 153.6 \\
\theta_f=0, & 153.6< \omega_0 x/c< 204.8
\end{array}
\right.
\end{equation}
Figure \ref {wave01}(center) shows that the discontinuity in $\theta_f$ does not cause significant spurious effects. This sought-for property is of major interest for
modeling laser-plasma  interaction as it allows the laser wave to travel unperturbed in vacuum over several wavelengths before reaching the overcritical target,
whose numerical stability calls for finite numerical damping. For the sake of completeness, we have checked that the weak ($\sim 0.1\%$ in the present case) reflection
arising at the discontinuity  surface is consistent with Fresnel's formula $R = (N(1)- N(0))^2/ (N(1)+N(0))^2$, where $N(\theta_f) = c/v_\phi (\theta_f)$ is the numerical
refraction index derived in Sec. \ref{subsec:electromagnetic_waves}.

\subsection{Plasma expansion into vacuum: benchmarking against explicit simulations}

As a first test of the implicit Vlasov-Maxwell solver, we simulate the dynamics of a plasma slab freely expanding into vacuum. The results of the implicit code ELIXIR are
confronted to refined, explicit simulations performed with the code CALDER \cite{calder2003}. We consider a $0.6c/\omega_p$ plasma slab composed of hot (10 keV) electrons
and cold ions. In the implicit case, the simulation box is $103 \Delta x \times 4\Delta y$ large, with $\Delta x = 2c/\omega_p$ and $\Delta y = 0.4c/\omega_p$ (yielding the
ratios $\Delta x/\lambda_D = 14$ and $v_t\Delta t/\Delta x = 0.14$), whereas the explicit simulation handles a $1024 \Delta x \times 8 \Delta y$ box, with
$\Delta x = \Delta y = 0.2c/\omega_p$. A linear weight factor is used in all cases.

\begin{figure}[htbp]
\includegraphics[width=0.45\textwidth]{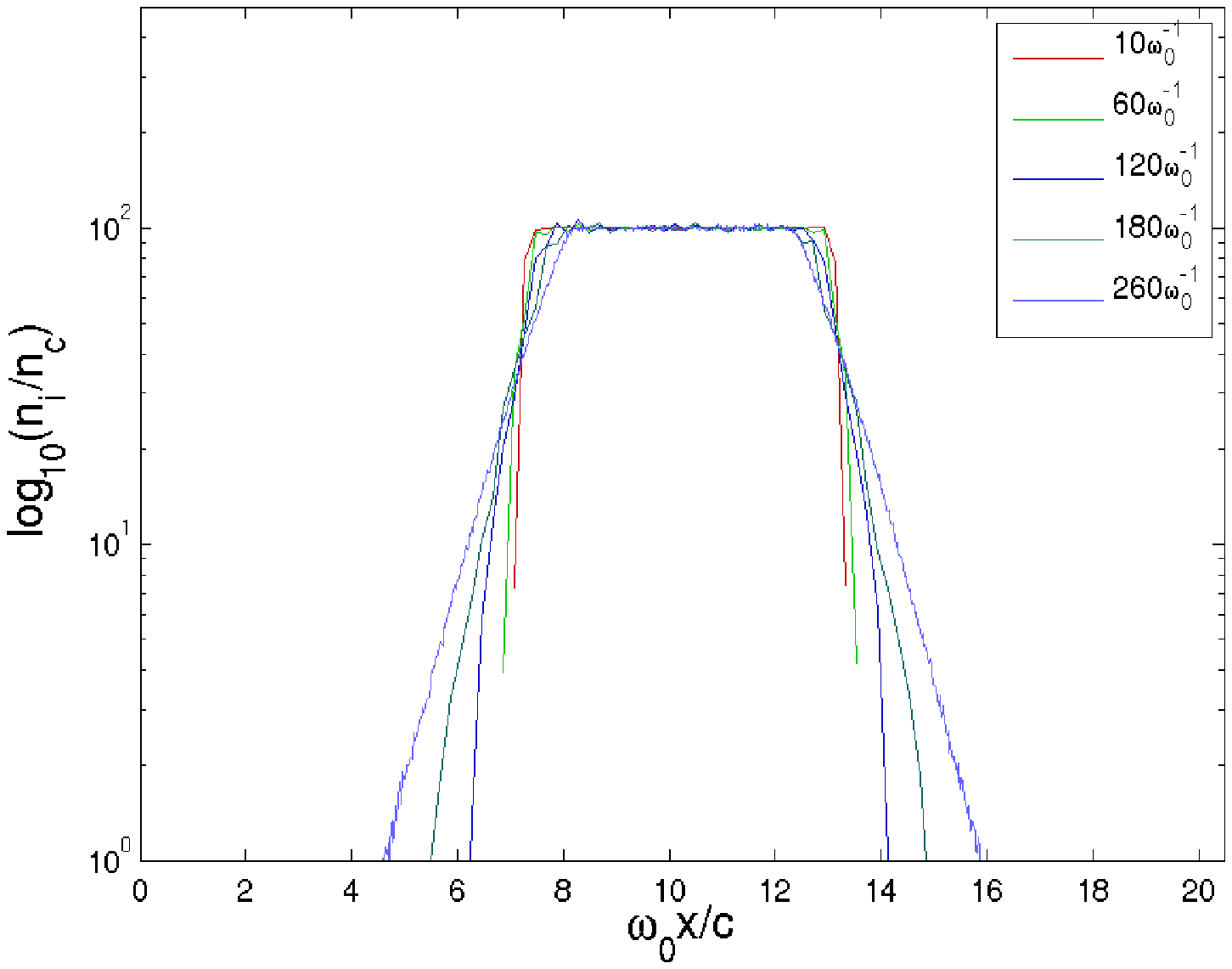}
\includegraphics[width=0.45\textwidth]{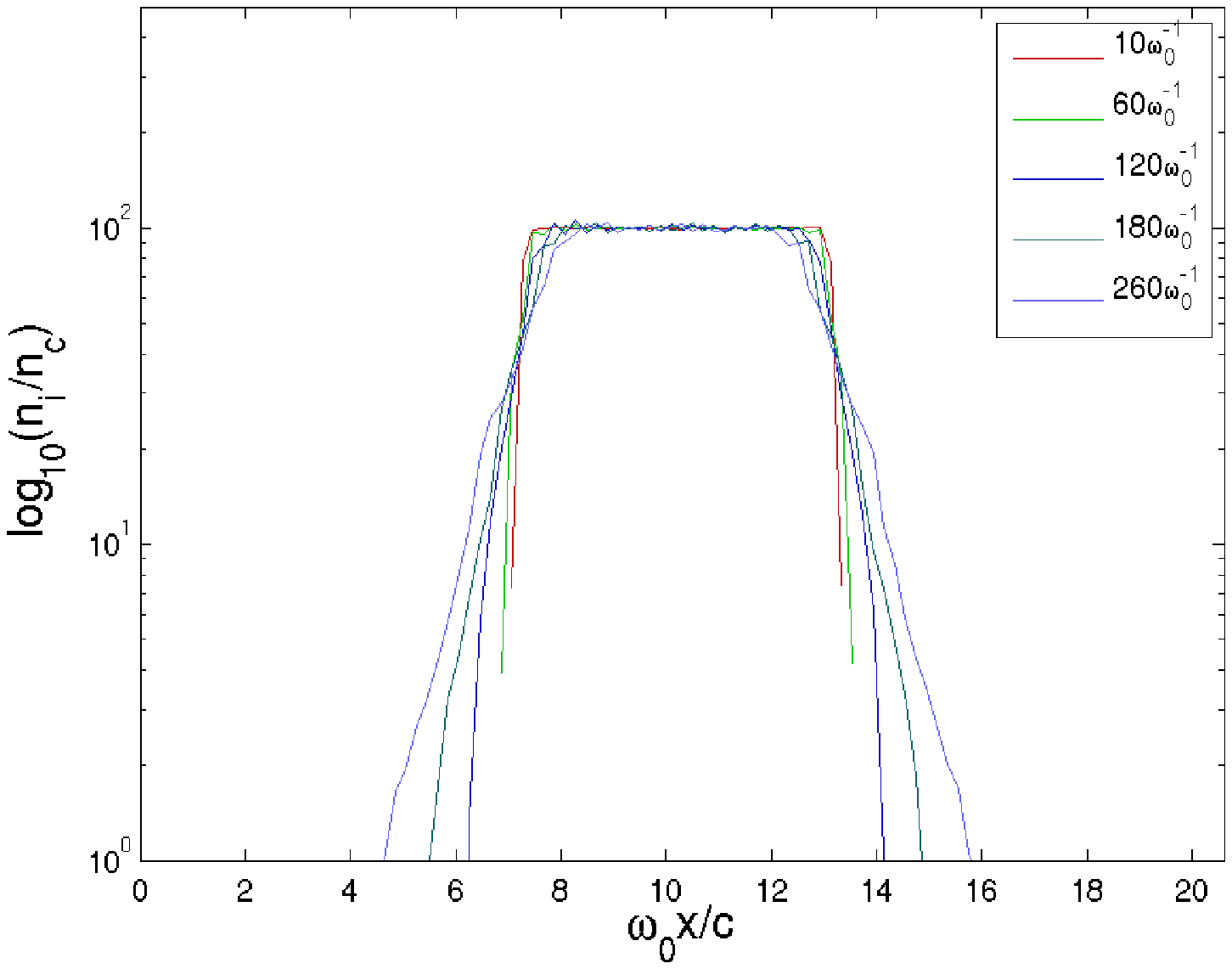}
\caption{Time evolution of the ion density profile: explicit (left) and implicit (right) simulations with $\Delta x= 0.2 c/\omega_p$, $\Delta t= 0.1 \omega_p^{-1}$,
$N_p = 6\times10^5$ and $\Delta x= 2 c/\omega_p$, $\Delta t= 2 \omega_p^{-1}$, $N_p = 6\times10^4$, respectively. The implicit damping parameter
is $\theta_f=1$.}
\label{expand01}
\end{figure}

\begin{figure}[htbp]
\includegraphics[width=0.45\textwidth]{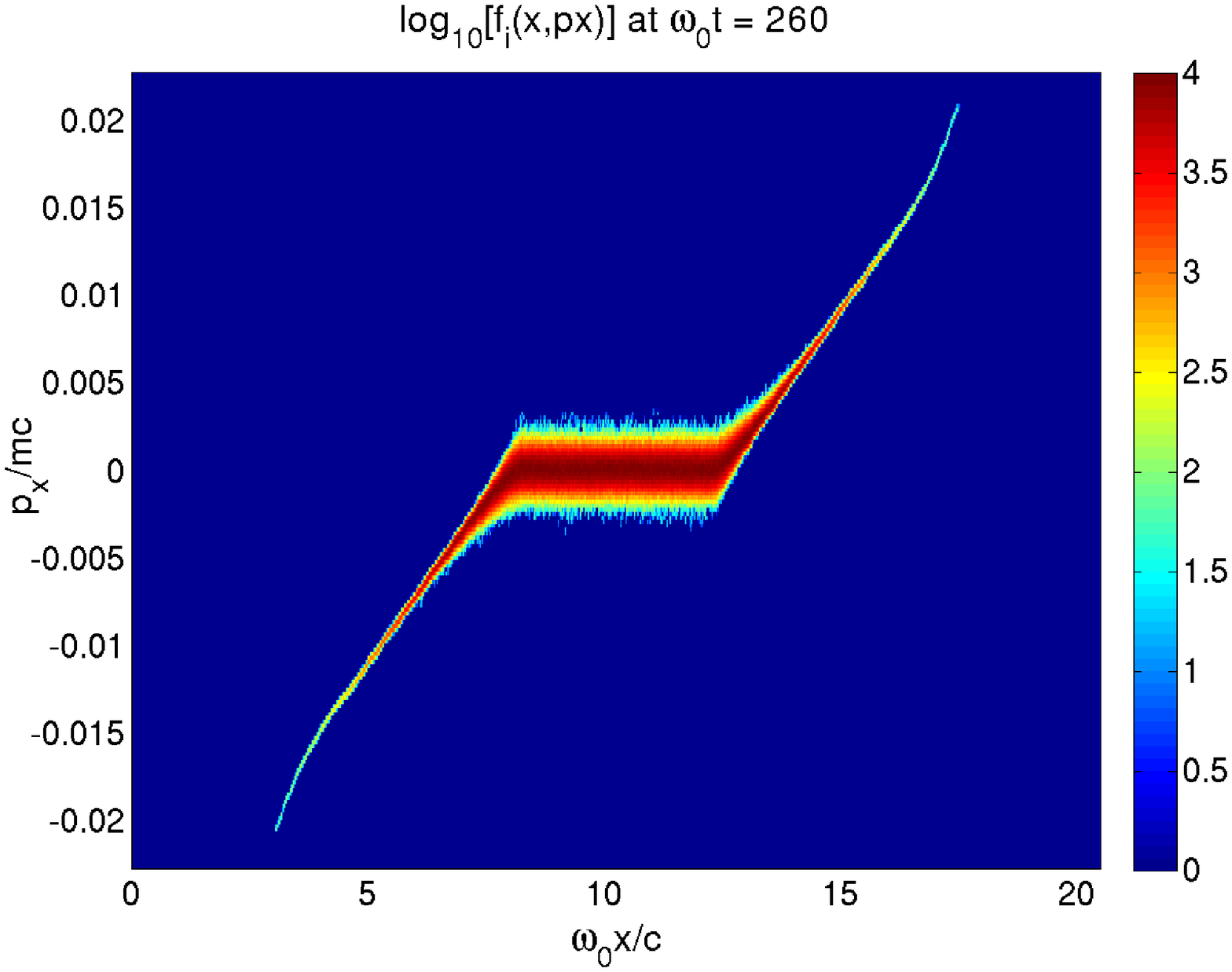}
\includegraphics[width=0.45\textwidth]{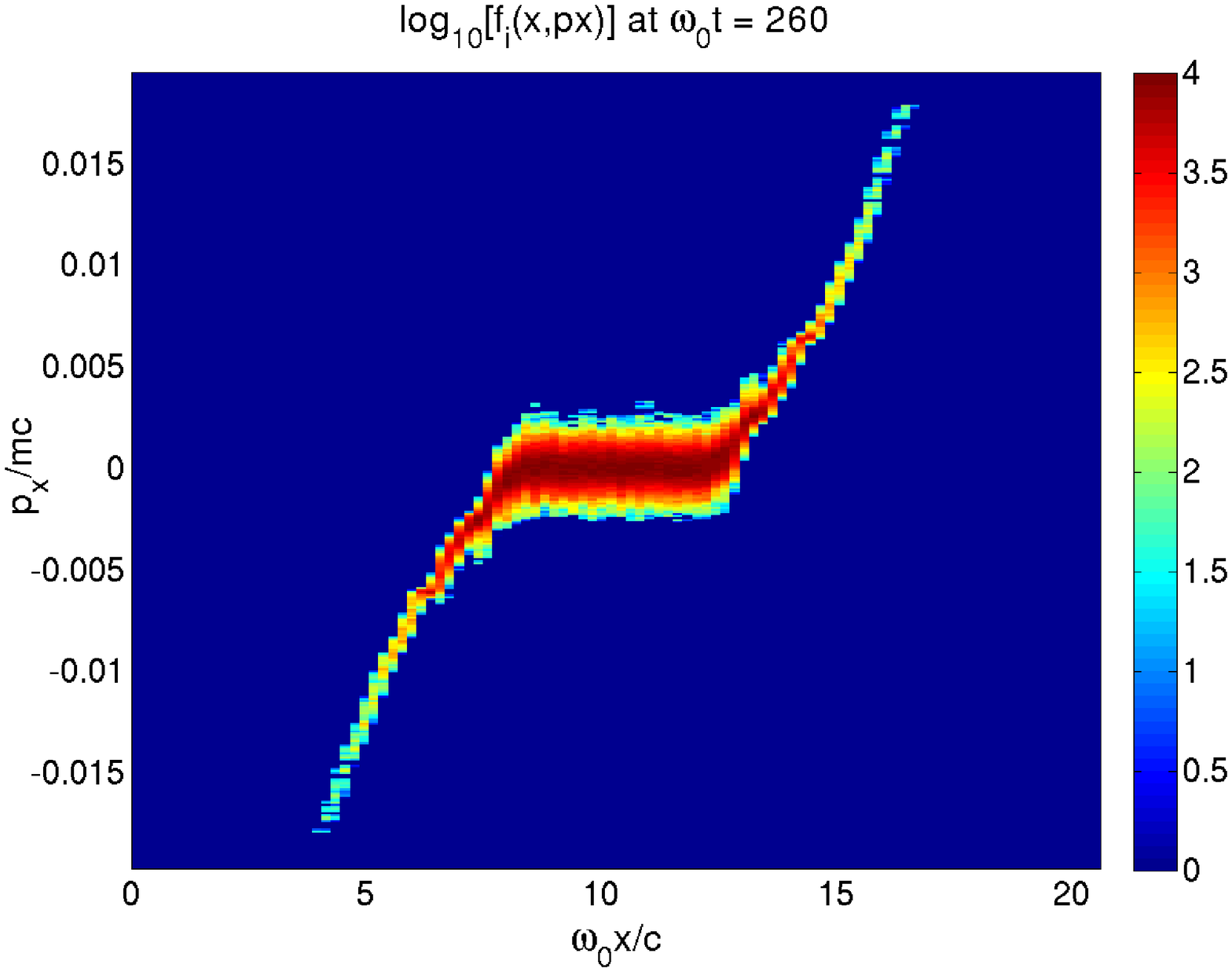}
\caption{Ion phase space at $t=2600 \omega_p^{-1}$: explicit (left) and implicit (right) simulations with $\Delta x= 0.2 c/\omega_p$, $\Delta t= 0.1 \omega_p^{-1}$,
$N_p = 6\times10^5$ and $\Delta x= 2 c/\omega_p$, $\Delta t= 2 \omega_p^{-1}$, $N_p = 6\times10^4$, respectively. The implicit damping parameter
is $\theta_f=1$.}
\label{expand03}
\end{figure}

\begin{figure}[htbp]
\includegraphics[width=0.4\textwidth]{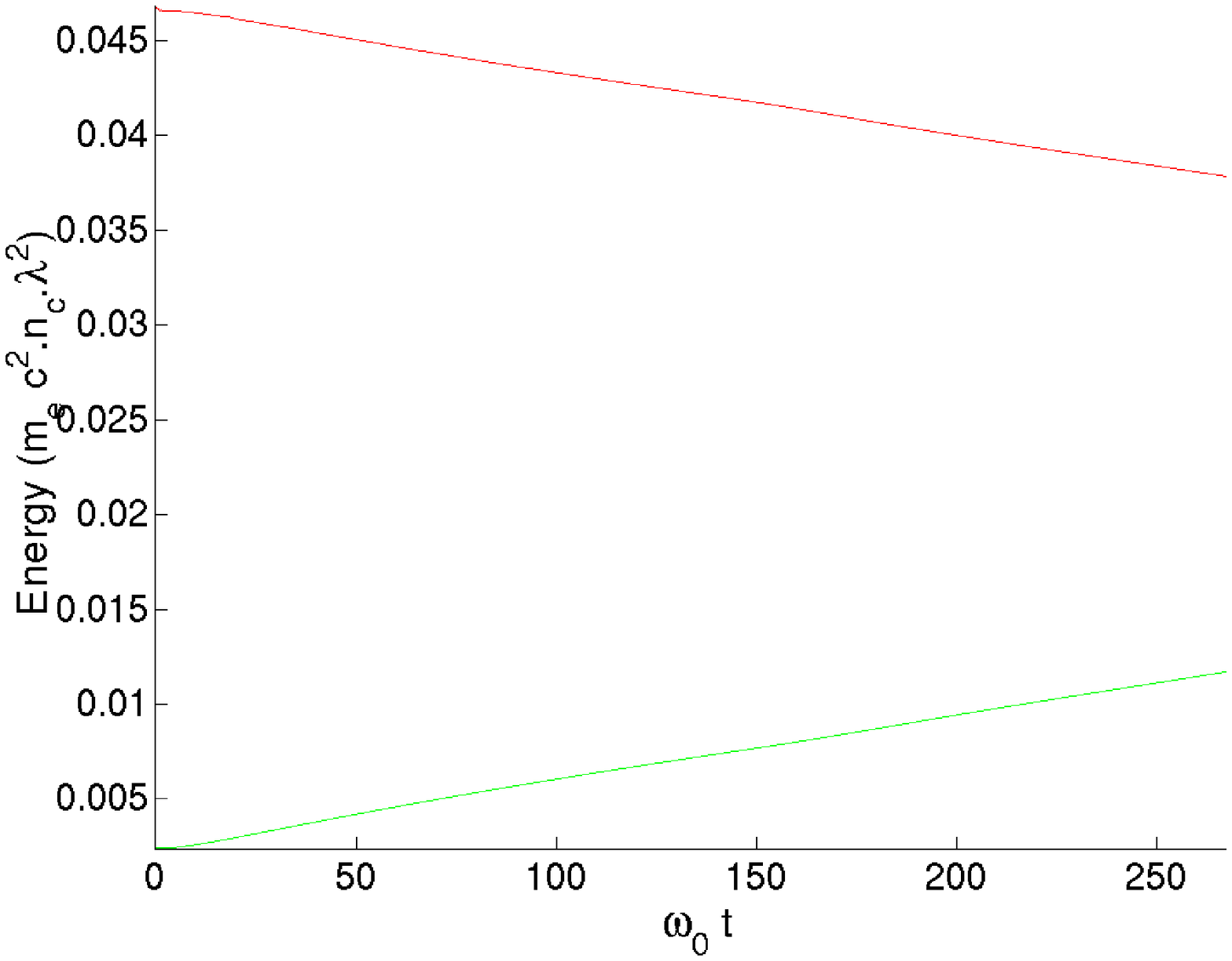}
\includegraphics[width=0.4\textwidth]{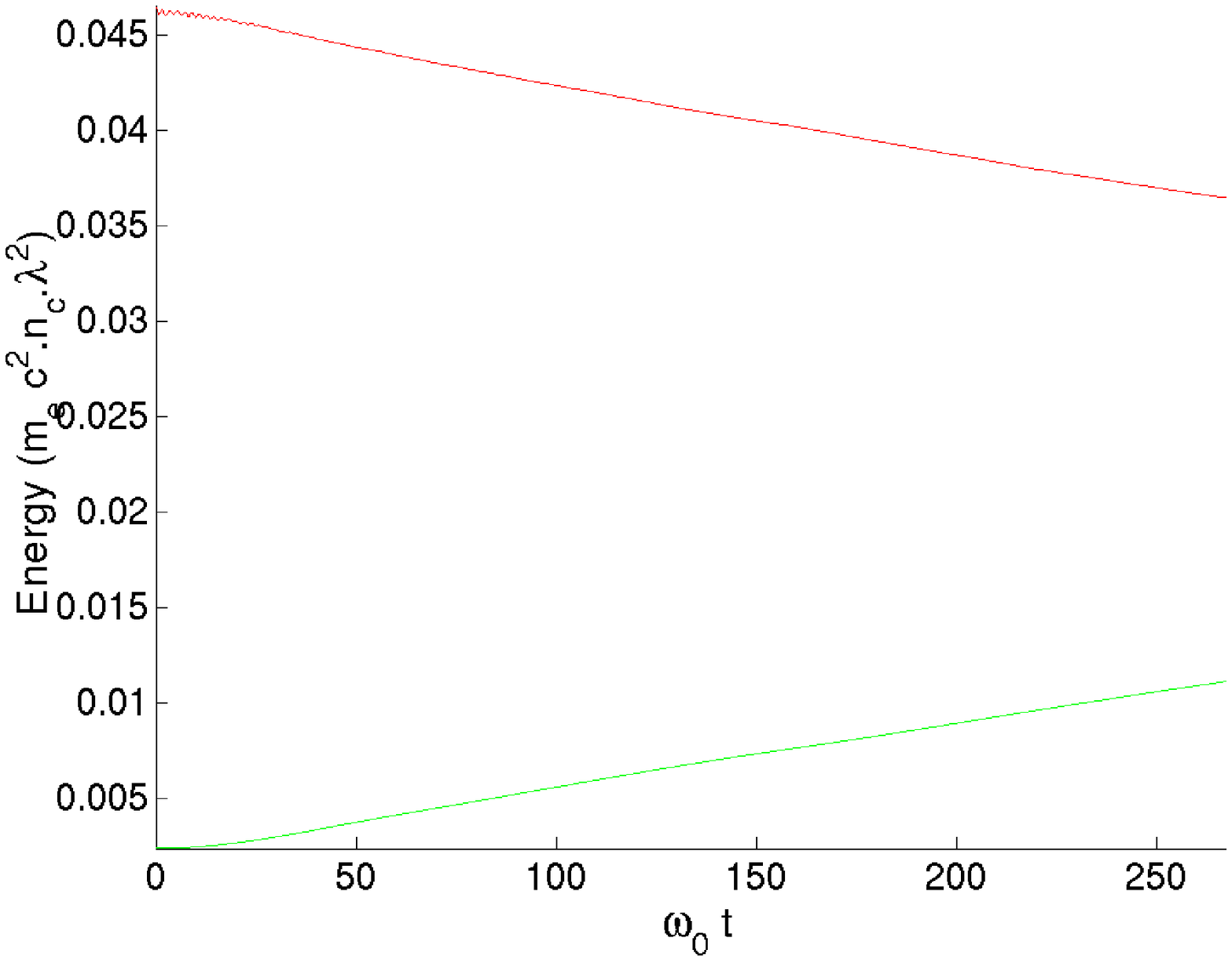}
\caption{Time evolution of the electron (red) and ion (green) kinetic energies: explicit (left) and implicit (right) simulations with $\Delta x= 0.2 c/\omega_p$,
$\Delta t= 0.1 \omega_p^{-1}$, $N_p = 6\times10^5$ and $\Delta x= 2 c/\omega_p$, $\Delta t= 2 \omega_p^{-1}$, $N_p = 6\times10^4$, respectively.
The implicit damping parameter
is $\theta_f=1$.}
\label{expand05}
\end{figure}

Figures \ref{expand01}, \ref{expand03} and \ref{expand05} plot the time evolution of the ion density profile, the ion phase space and the time evolution of the plasma kinetic
energies, as simulated by the implicit and explicit codes. The implicit damping parameter is chosen to be $\theta_f=1$, whereas the total number of macroparticles $N_p$ is
$6\times 10^4$ and $6\times 10^5$ in the implicit and explicit cases, respectively. Overall, albeit roughly resolved and strongly damped (as expected from Table \ref{tabheat01}),
the implicit scheme manages to satisfactorily capture the finely resolved, explicit results. Yet, the wave damping gives rise to artificial electron cooling, which results
into a weakened ion acceleration as seen in Figs. \ref{expand03} and \ref{expand05}. More quantitatively, the total energy drops by $\sim 3\%$, yielding a maximum ion energy
of $\sim 160$ keV, as compared to $\sim 220$ keV in the explicit case. For the sake of completeness, we have carried out additional calculations so as to assess the influence
of the damping parameter and the number of macroparticules. For each simulation, we have measured the energy variation and the peak ion energy. The data thus obtained
is summarized in Tables \ref{tabexp01} and \ref{tabexp02}. The implicit scheme behaves reasonably well up to $\theta_f = 0.15$ with an energy variation $<10\%$, comparable
or better than its explicit counterpart for an equal number of macroparticles. Increasing the latter from $6\times10^4$ to $6\times10^5$ approximately halves the energy
variation but hardly changes the peak ion energy.  The transition from numerical electron cooling and heating occurs between $\theta_f = 1$ and $\theta_f = 0.5$. Finally,
the undamped ($\theta_f$=0) case is subject to a much stronger, if still limited, electron heating, which translates into a twofold overestimate of the peak ion energy.

\begin{table}[h]
\begin{center}
\begin{tabular}{lcc}
\hline
\hline
   & $\Delta E/E_0$  & Ion peak energy (keV) \\
\hline
Explicit                  &  +9.3 \% & 232  \\
Implicit ($\theta_f=1$)   &  -2.8\%  & 162   \\
Implicit ($\theta_f=0.5$) &  +3.1\%  & 208  \\
Implicit ($\theta_f=0.15$)&  +9\%    & 273  \\
Implicit ($\theta_f=0$)   &  +19.7\% & 451  \\
\hline
\hline
\end{tabular}
\caption{Total energy variation and ion peak kinetic energy (keV) at $2600 \omega_p^{-1}$ with $N_p=6\times 10^4$.}
\label{tabexp01}
\end{center}
\end{table}

\begin{table}[h]
\begin{center}
\begin{tabular}{lcc}
\hline
\hline
   & $\Delta E/E_0$  & Ion peak energy (keV)  \\
\hline
Explicit                  & +1  \%  & 221 \\
Implicit ($\theta_f=1$)   & -1.4\%  & 162 \\
Implicit ($\theta_f=0.5$) & +1.5\%  & 198 \\
Implicit ($\theta_f=0.15$)& +4.5\%  & 256 \\
Implicit ($\theta_f=0$)   & +12.4\% & 418 \\
\hline
\hline
\end{tabular}
\caption{Total energy variation and ion peak kinetic energy (keV) at $2600 \omega_p^{-1}$ with $N_p=6\times 10^5$.}
\label{tabexp02}
\end{center}
\end{table}

\subsection{A parametric study of plasma self-heating and cooling}
\label{subsec:parametric_study}

We have carried out a series of simulations of the free evolution of an electron-ion plasma to gauge the potential discrepancy between the idealized linear analysis
of Sec. \ref{subsec:electostratic_waves} and the actual predictor-corrector numerical scheme. Evidently, the objective is to gain further insight into the energy
conservation properties of the latter and the predictive capability of the former. These calculations draw upon and extend the work of Ref. \cite{cohenlangdon1989} to
the electromagnetic regime. The system consists of a bounded electron-ion plasma with $T_e=T_i=1$ keV and $m_i/m_e = 900$, extending over half a
$300 \Delta x \times 4\Delta y$ simulation box. We have scanned the $(\Delta x/\lambda_D, \omega_p \Delta t)$ parameter space in the range  $[5, 60]\times [1,5]$.
In practice, after introducing $\omega_0$, the frequency of a fictitious electromagnetic wave, and $n_c$, the corresponding critical density, we have set
$\Delta x = 0.2 c/\omega_0$ and varied the ratio $n_e/n_c$ and the time step so that $\Delta x/\lambda_D \in \{5, 10, 20, 30, 60\}$ and $\omega_p \Delta t \in \{1,2,5\}$.
The damping parameter is $\theta_f = 1$. The total simulation time is kept fixed at $1000 \omega_0^{-1}$. For each simulation, we have calculated the relative variation of
the total kinetic energy per time step $(\Delta K/K_0)/N$ (where $\Delta K$ is the kinetic variation, $K_0$ the initial kinetic energy and $N$ the number of time steps).
To be complete, we have also performed electrostatic calculations, whereby the electric field is directly computed through the Poisson equation (\ref{eqcharge03}).

\begin{table}[h]
\begin{center}
\begin{tabular}{lccccc}
\hline
\hline
 $\Delta x/\lambda_D$&  5 & 10 & 20 & 30 & 60  \\
 $\omega_p\Delta t$&  &  & &    \\
\hline
 1 & $3.2\times10^{-5}$ & $3.2\times10^{-4}$ & $1.1\times10^{-3}$  & $2.1\times10^{-3}$  &
 $4.7\times10^{-3}$  \\
 2 & $-9.2\times10^{-5}$  & $1.5\times10^{-4}$  & $7.9\times10^{-4}$ & $1.5\times10^{-3}$  &
 $4.5\times10^{-3}$  \\
 5 & $0$  & $-1.6\times10^{-4}$ &  $1.2\times10^{-4}$  & $4.7\times10^{-4}$ &
 $1.7\times10^{-3}$ \\
 \hline
 \hline
\end{tabular}
\caption{Relative variation of the total kinetic energy per time step $(\Delta K/K_0)/N$: electrostatic case and linear weight factor. }
\label{tabheat04}
\end{center}
\end{table}

\begin{table}[h]
\begin{center}
\begin{tabular}{lccccc}
\hline
\hline
 $\Delta x/\lambda_D$&  5 & 10 & 20 & 30 & 60  \\
 $\omega_p\Delta t$&  &  & &    \\
\hline
 1 & $2.8\times10^{-5}$ & $3.2\times10^{-4}$ & $9.9\times10^{-4}$  & $1.7\times10^{-3}$  &
 $2.8\times10^{-3}$  \\
 2 & $-1.1\times10^{-4}$  & $1.3\times10^{-4}$  & $6.9\times10^{-4}$ & $1.2\times10^{-3}$  &
 $2.6\times10^{-3}$  \\
 5 & $-5.8\times10^{-5}$  & $-2.4\times10^{-4}$ &  $2.9\times10^{-5}$  & $3\times10^{-4}$ &
 $9.5\times10^{-4}$ \\
\hline
\hline
\end{tabular}
\caption{Relative variation of the total kinetic energy per time step $(\Delta K/K_0)/N$: electromagnetic case and linear weight factor.}
\label{tabheat05}
\end{center}
\end{table}

\begin{table}[h]
\begin{center}
\begin{tabular}{lccccc}
\hline
\hline
 $\Delta x/\lambda_D$&  5 & 10 & 20 & 30 & 60  \\
 $\omega_p\Delta t$&  &  & &    \\
 \hline
 1 & $-3\times10^{-5}$ & $4\times10^{-5}$ & $2.3\times10^{-4}$  & $4.6\times10^{-4}$  &
 $1.1\times10^{-3}$  \\
 2 & $-1.1\times10^{-4}$  & $-3.5\times10^{-5}$  & $1.4\times10^{-4}$ & $3.2\times10^{-4}$  &
 $8.3\times10^{-4}$  \\
 5 & $-1.3\times10^{-4}$  & $-2.2\times10^{-4}$ &  $-10^{-4}$  & $0$ &
 $2.4\times10^{-4}$ \\
 \hline
 \hline
\end{tabular}
\caption{Relative variation of the total kinetic energy per time step $(\Delta K/K_0)/N$: electromagnetic case and quadratic weight factor.}
\label{tabheat06}
\end{center}
\end{table}

\begin{figure}[htbp]
\includegraphics[width=\textwidth]{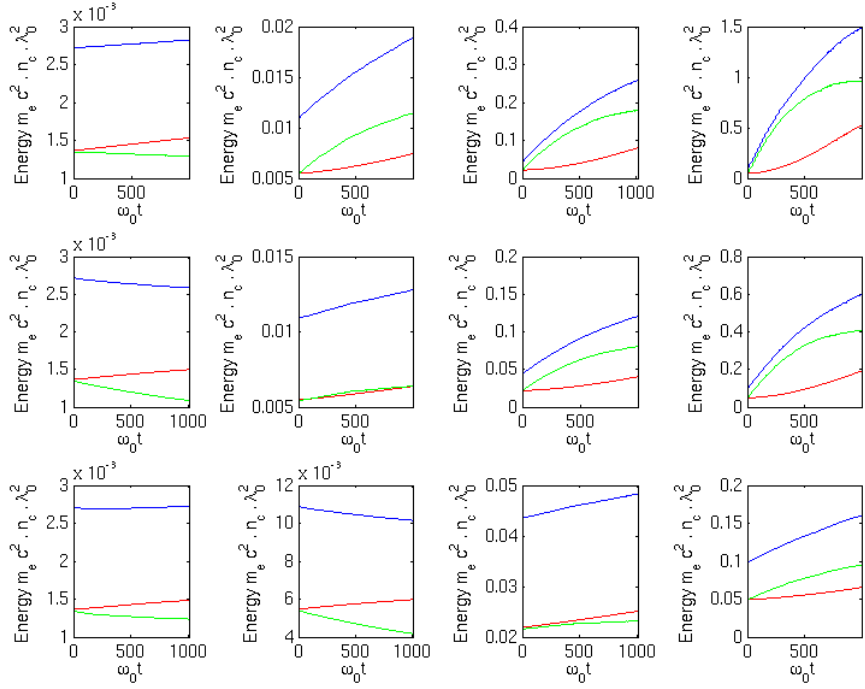}
\caption{Time evolution of the total (blue), ion (red) and electron (green) energies: electrostatic case with linear weight factor.  $\Delta x/\lambda_D=(5,10,20,30)$
from left to right and $\omega_p \Delta t = (1,2,5)$ from top to bottom. }
\label{figheat06}
\end{figure}

\begin{figure}[htbp]
\includegraphics[width=\textwidth]{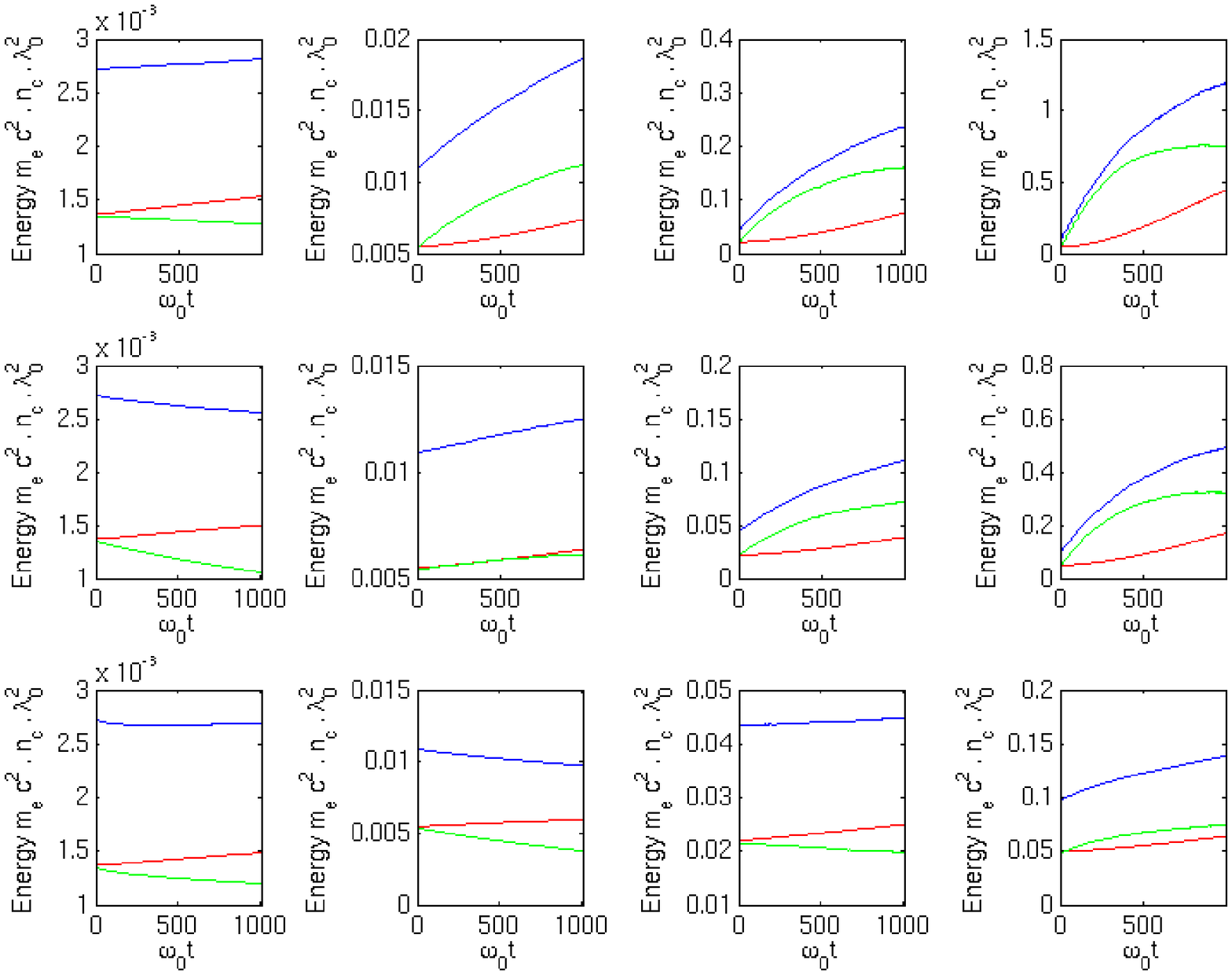}
\caption{Time evolution of the total (blue), ion (red) and electron (green) energies: electromagnetic case with linear weight factor. $\Delta x/\lambda_D=(5,10,20,30)$
from left to right and $\omega_p \Delta t = (1,2,5)$ from top to bottom.}
\label{figheat07}
\end{figure}

\begin{figure}[htbp]
\includegraphics[width=\textwidth]{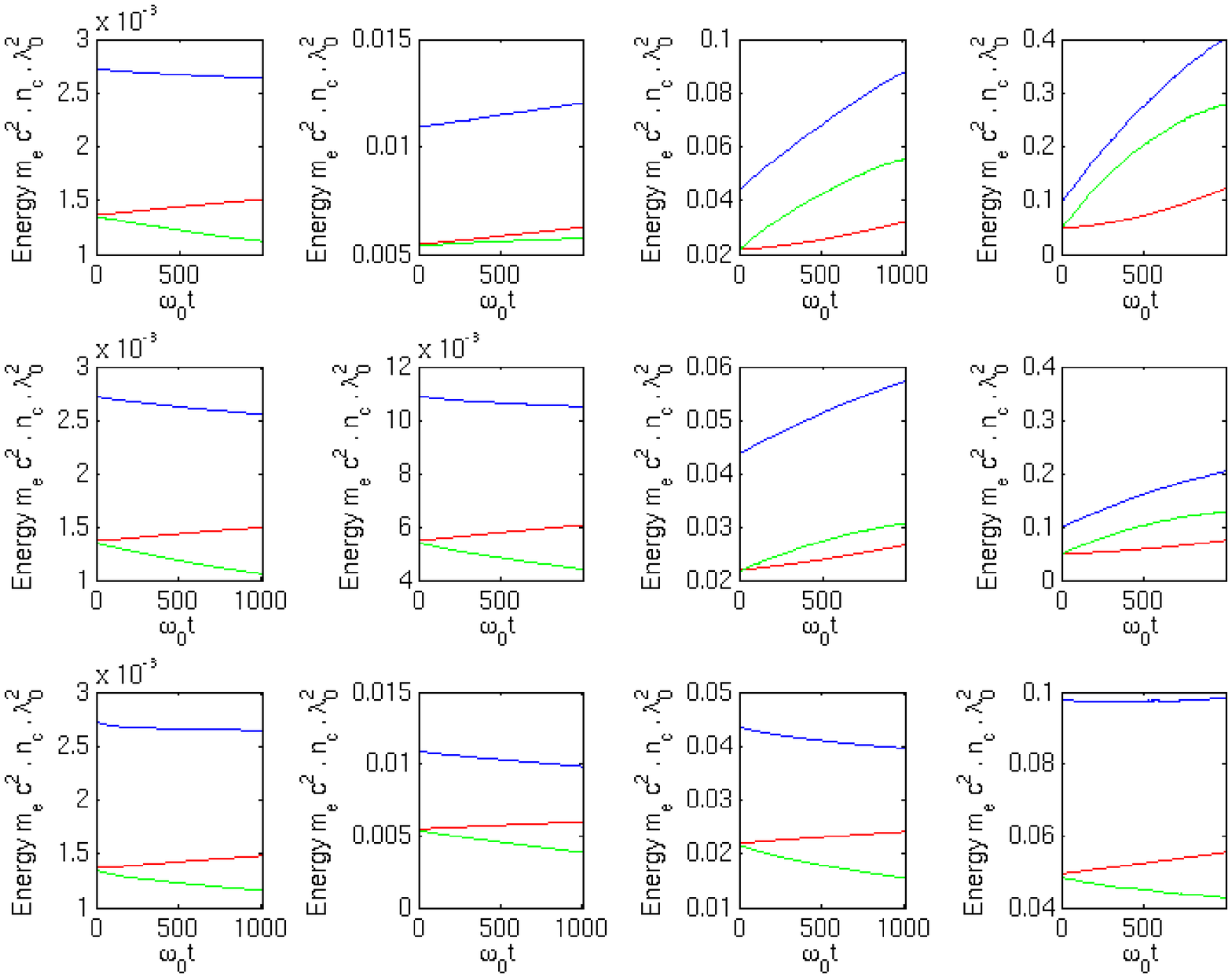}
\caption{Time evolution of the total (blue), ion (red) and electron (green) energies: electromagnetic case with quadratic weight factor.}
\label{figheat08}
\end{figure}

The results are summarized in Tables \ref{tabheat04}-\ref{tabheat06}. The associated plots of the kinetic energies are shown in Figs. \ref{figheat06}- \ref{figheat08}:
each column corresponds to a specific value of $\Delta x/\lambda_D$ and each line to a specific value of $\omega_p\Delta t$. Note that we have excluded in these plots
the case $\Delta x/\lambda_D=60$ as it always gives rise to significant numerical heating. We have checked that the plasma  kinetic energy makes up for most of the system
energy. Overall, the electrostatic results prove close to the electromagnetic ones. Satisfactory energy conservation ($\lesssim 10^{-4}$) is obtained for
$v_t\Delta t/\Delta x \gtrsim 0.2$ and  $v_t\Delta t/\Delta x \gtrsim 0.1$ in the linear and quadratic interpolation cases, respectively. These lower bound values
are in fairly good agreement, albeit slightly higher, with the linear results of Sec. \ref{subsec:electostratic_waves}. Larger $v_t\Delta t/\Delta x$ ratios eventually
lead to plasma cooling,

\subsection{High intensity laser interaction with an overdense plasma slab}
\label{subsec:laser_interaction}

\subsubsection{Quasi-one-dimensional simulation}
\label{subsec:1D_simulation}

Let us now address the problem of the interaction of a relativistic-intensity laser pulse with an overcritical plasma, which is the prime motivation behind this work.

As a first illustration, we consider the case of a quasi-1D laser-plasma system. The irradiated target consists of a $60 c/\omega_0$-long, 1 keV, $200n_c$ plasma slab
preceded by a $18c/\omega_0$-long density ramp rising linearly from 0 to $200n_c$ . The incident electromagnetic plane wave has a $120 \omega_0^{-1}$ constant-intensity
profile with a $22 \omega_0^{-1}$ rise time and a normalized amplitude $a_0 = eE_0/m_e c\omega_0=3$. The implicit simulation employs a $2048 \Delta x \times 4 \Delta y$ grid,
with $\Delta x = \Delta y = 0.1c/\omega_0$ and $\Delta t =0.14 \omega_0^{-1}$, yielding, in terms of plasma parameters, $\Delta x/\lambda_D = 32$ and $\omega_p \Delta t  = 2$
($v_t \Delta t /\Delta x = 0.06$). The damping parameter in the electromagnetic solver, as well as in the particle pusher,
is set to zero in the vacuum region and the moderately dense plasma region up to $n_e = 60n_c$,
and to unity in the denser plasma region. Guided by the results of
Sec. \ref{subsec:parametric_study}, we make use of a quadratic weight factor to reduce the numerical heating. The number of macroparticles per cell $N_p$ is varied from
100 to 1300. These calculations are compared with explicit simulations using the same parameters except for a decreased time step $\Delta t = 0.05\omega_0^{-1}$ so as to
fulfill the Courant stability condition.

\begin{table}[h]
\begin{center}
\begin{tabular}{lccc}
\hline
\hline
  & Explicit & Implicit ($\theta_f = 0$) & Implicit ($\theta_f = 1$ if $n_e>60n_c$) \\
\hline
$N_p= 1300$ &  $+14.4\%$  &  $+6\%$    & $-3\%$  \\
$N_p= 400$  &  $+15.3\%$  &  $+10.5\%$ & $-1\%$  \\
$N_p= 100$  &  $+22\%$    &  $+25.5\%$ & $+12.7\%$ \\
\hline
\hline
\end{tabular}
\caption{Quasi-1D laser-plasma interaction: energy variation in the explicit simulations with $\Delta t=0.05 \omega_0^{-1}$ and the implicit simulations with
$\Delta t = 0.14 \omega_0^{-1}$ and varying $\theta_f$. See text for other simulation parameters. }
\label{table02}
\end{center}
\end{table}

Table \ref{table02} compares the values of the total energy variation (calculated after complete reflection of the laser pulse) as obtained in the explicit and implicit cases.
Results from implicit simulations with zero damping are also displayed. Overall, except for $N_p =100$, for which case the three schemes behave similarly, the implicit simulations
are found to achieve better energy conservation than their explicit counterparts. The benefit of a strongly damped scheme
in the densest region of the plasma is mostly evidenced for $N_p=1300$ and 400. The not-so-good
performances of the explicit calculations prompted us to carry out an additional, more refined explicit simulation that can serve more properly as a reference calculation. This
simulation made use of a $4096 \Delta x \times 8 \Delta y$ grid with $\Delta x = \Delta y = 0.05 c/\omega_0$ and $\Delta t =0.03 \omega_0^{-1}$, as well as of a third-order
weight factor with $N_p= 650$. It yielded a total energy variation of $4\%$.

\begin{figure}[htbp]
\includegraphics[width=0.45\textwidth]{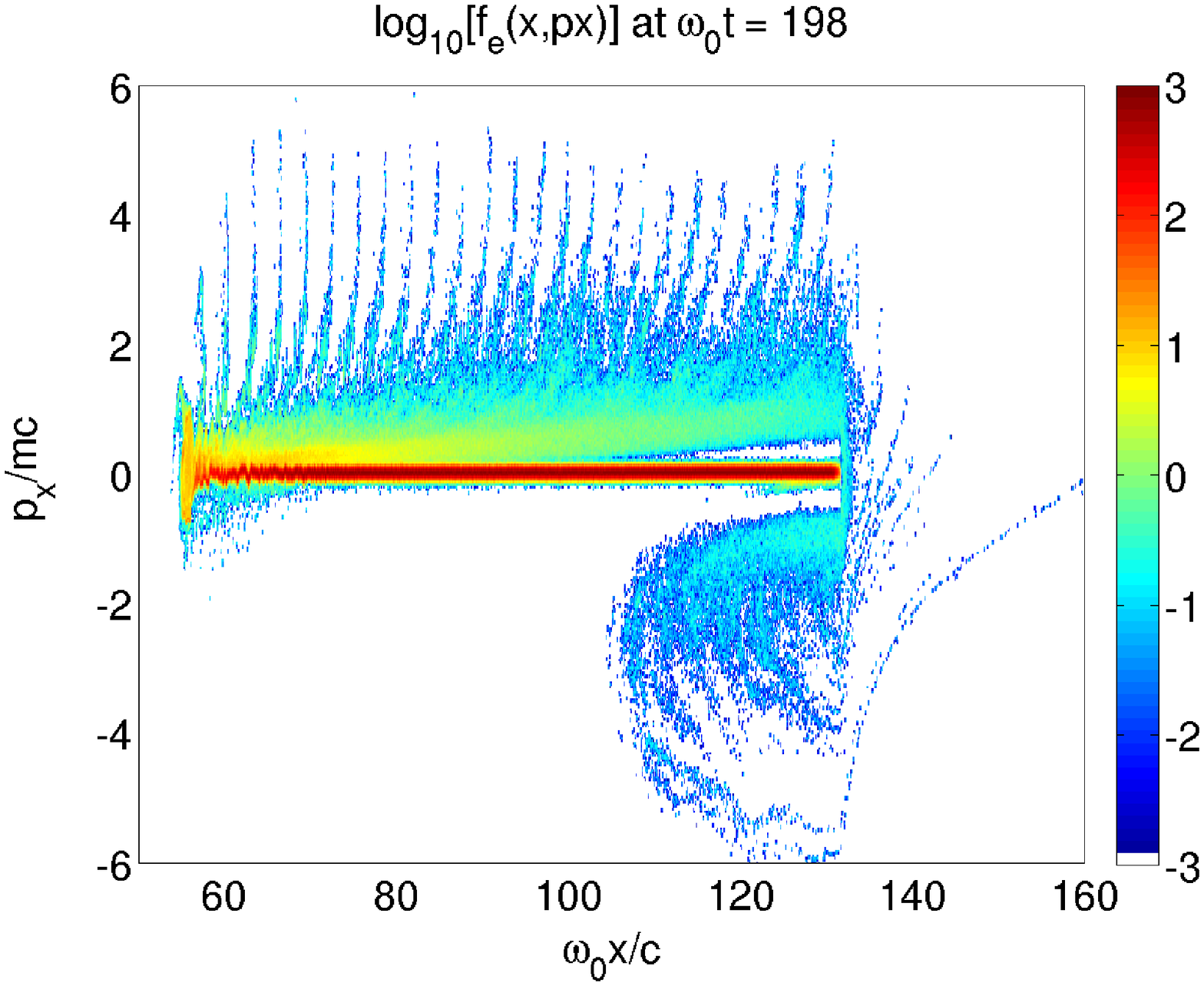}
\includegraphics[width=0.45\textwidth]{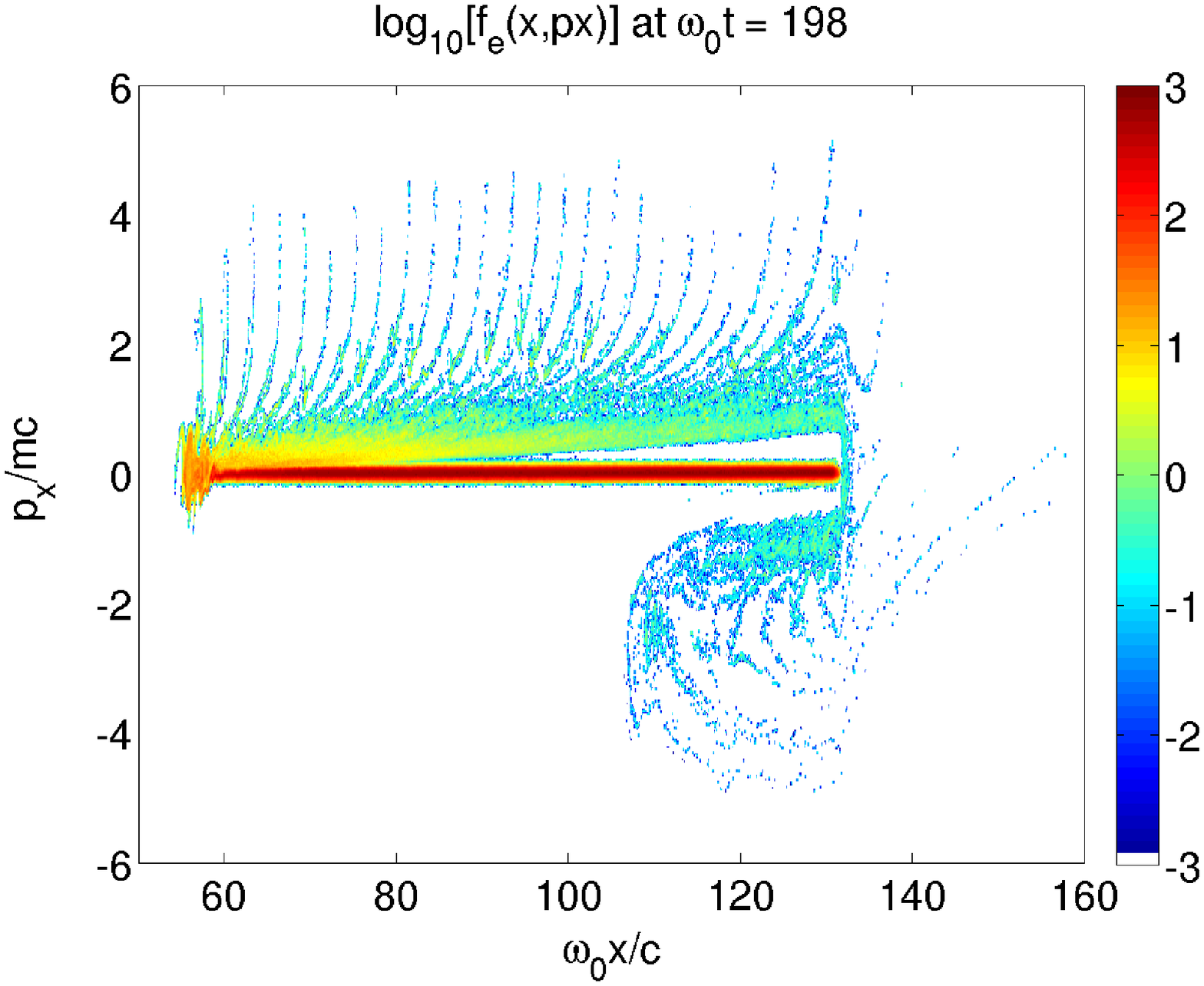}
\caption{Electron $(x,p_x)$ phase space at $t = 198 \omega_0^{-1}$: explicit simulation (left) and implicit simulation with $\theta_f=1$ (right). In both cases, $N_p =1300$.
See text for other simulation parameters.}
\label{ilp14}
\end{figure}

\begin{figure}[htbp]
\includegraphics[width=1\textwidth]{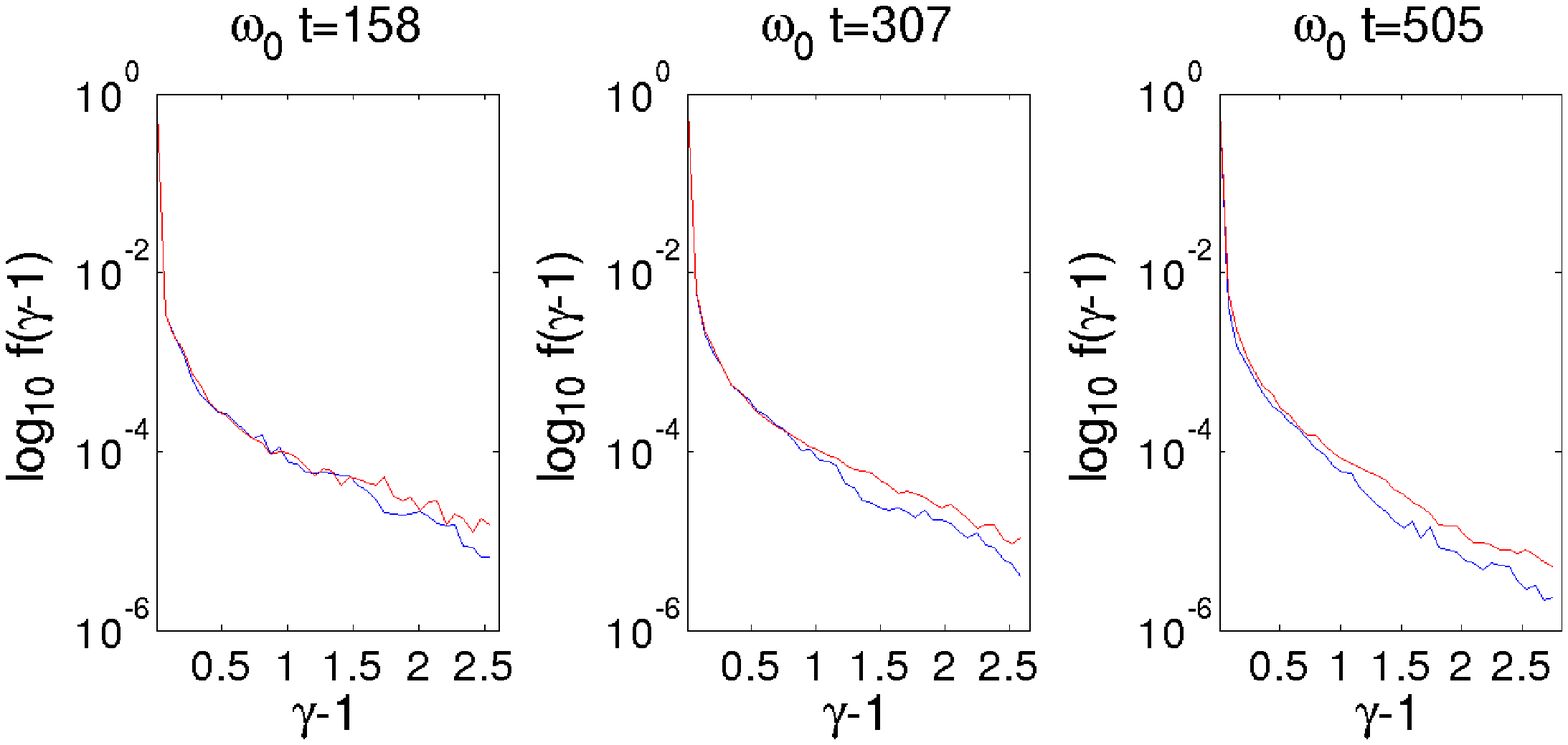}
\caption{Electron energy distribution at different times: explicit simulation (red) and implicit simulation (blue). Energy is normalized by $m_e c^2$.}
\label{ilp15}
\end{figure}

\begin{figure}[htbp]
\includegraphics[width=0.45\textwidth]{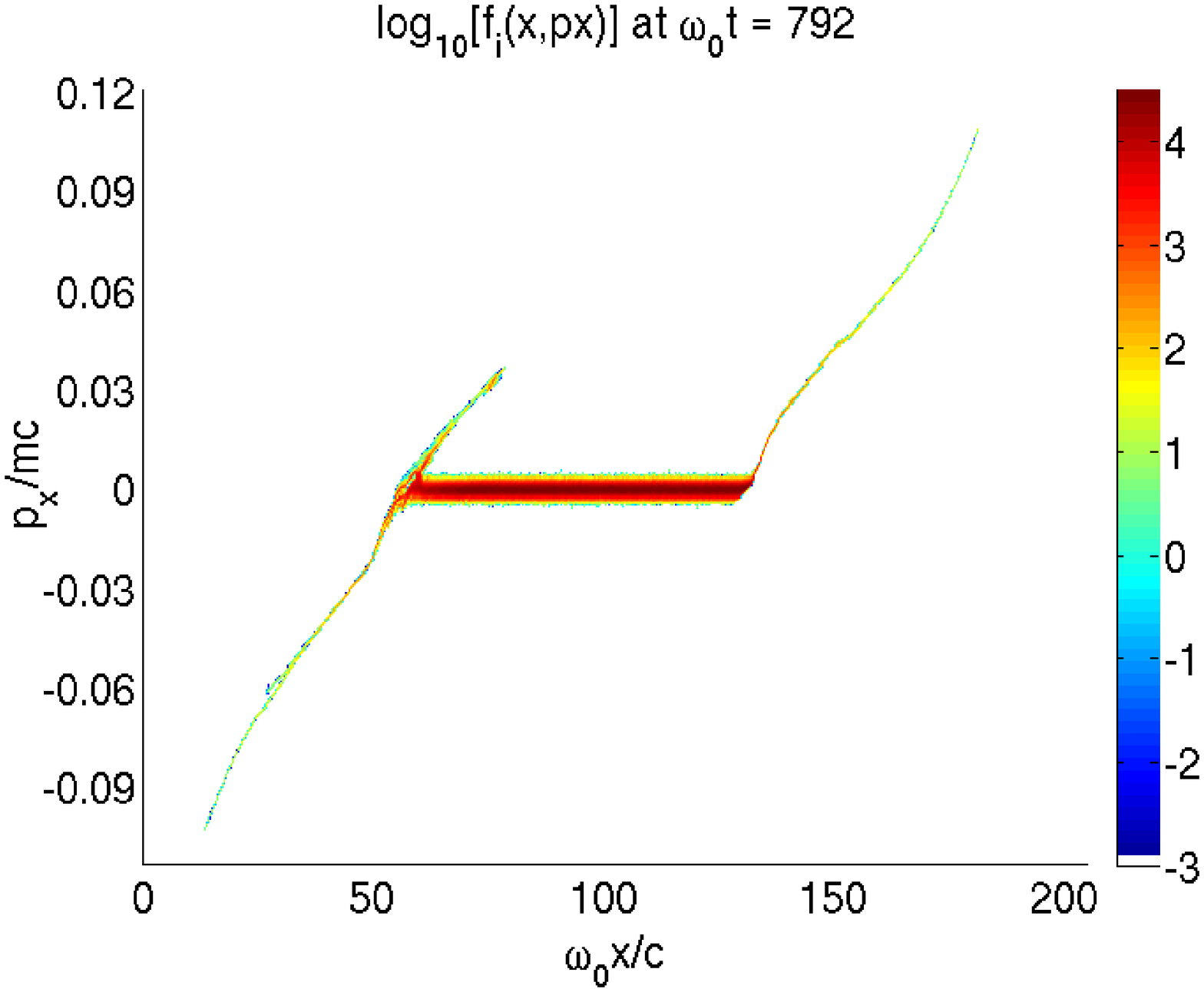}
\includegraphics[width=0.45\textwidth]{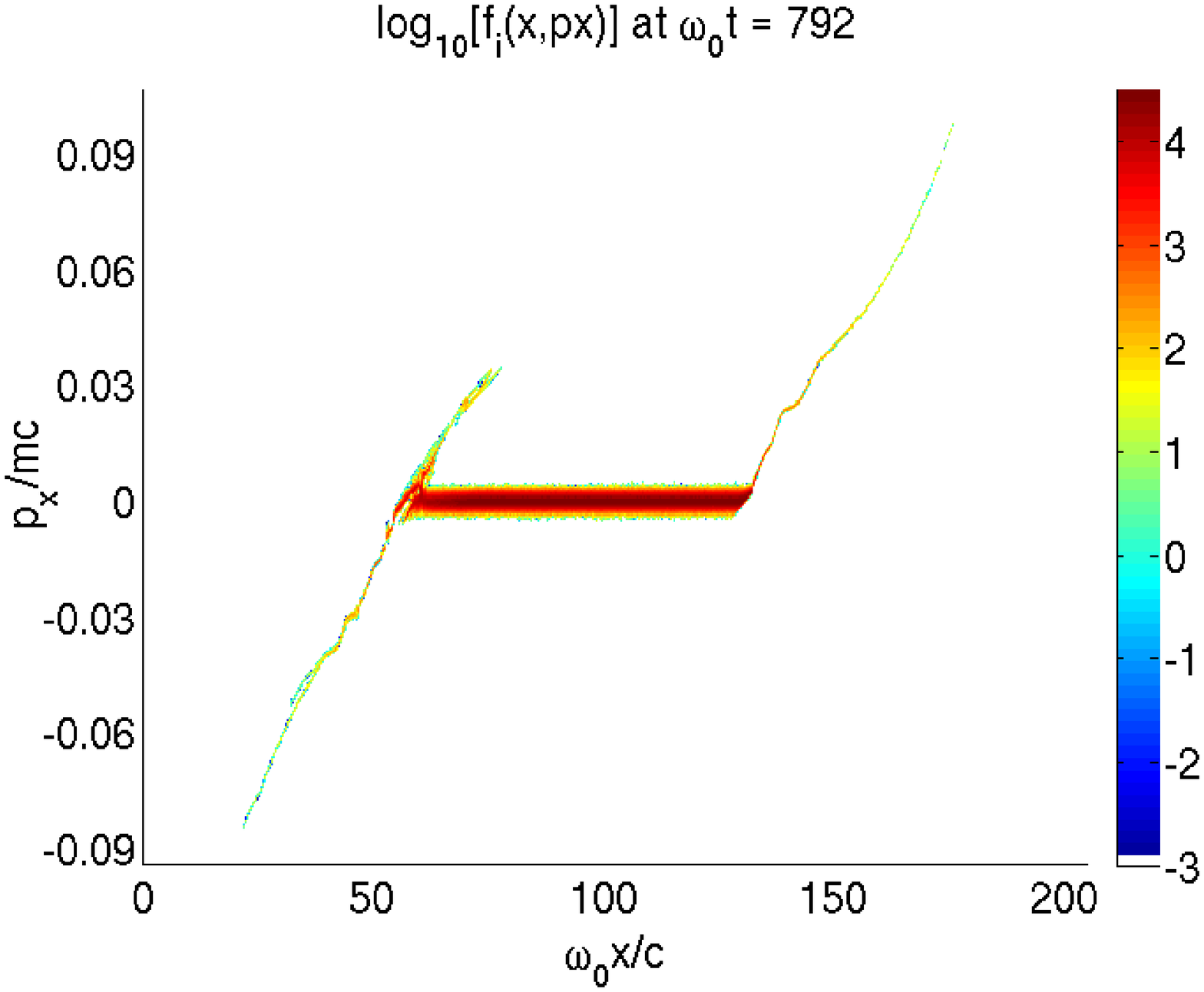}
\caption{Ion  $(x,p_x)$  phase space at $t=792 \omega_0^{-1}$: explicit simulation (left) and implicit simulation with $\theta_f=1$ (right). In both cases, $N_p =1300$.
See text for other simulation parameters.}
\label{ilp16}
\end{figure}

\begin{figure}[htbp]
\includegraphics[width=0.45\textwidth]{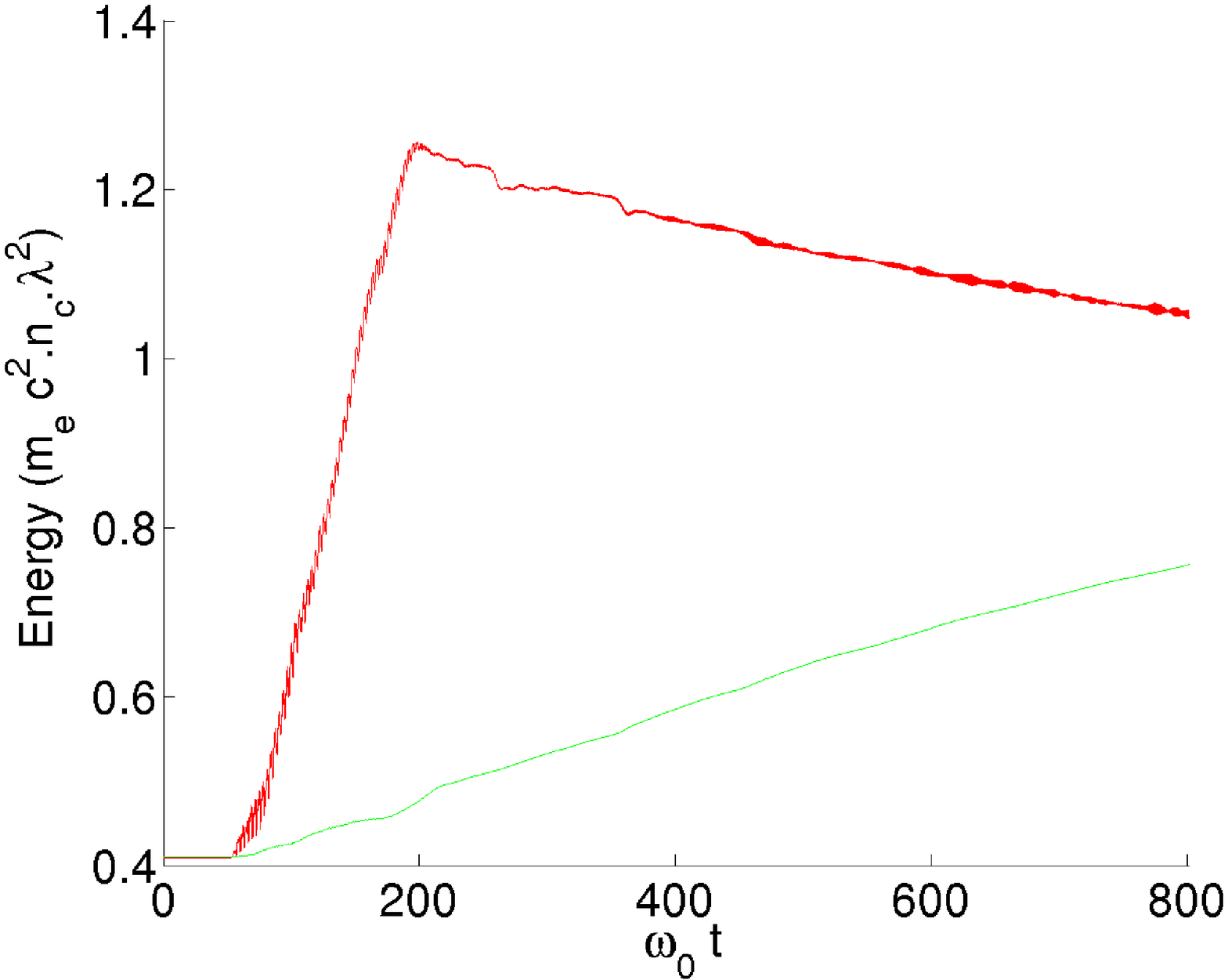}
\includegraphics[width=0.45\textwidth]{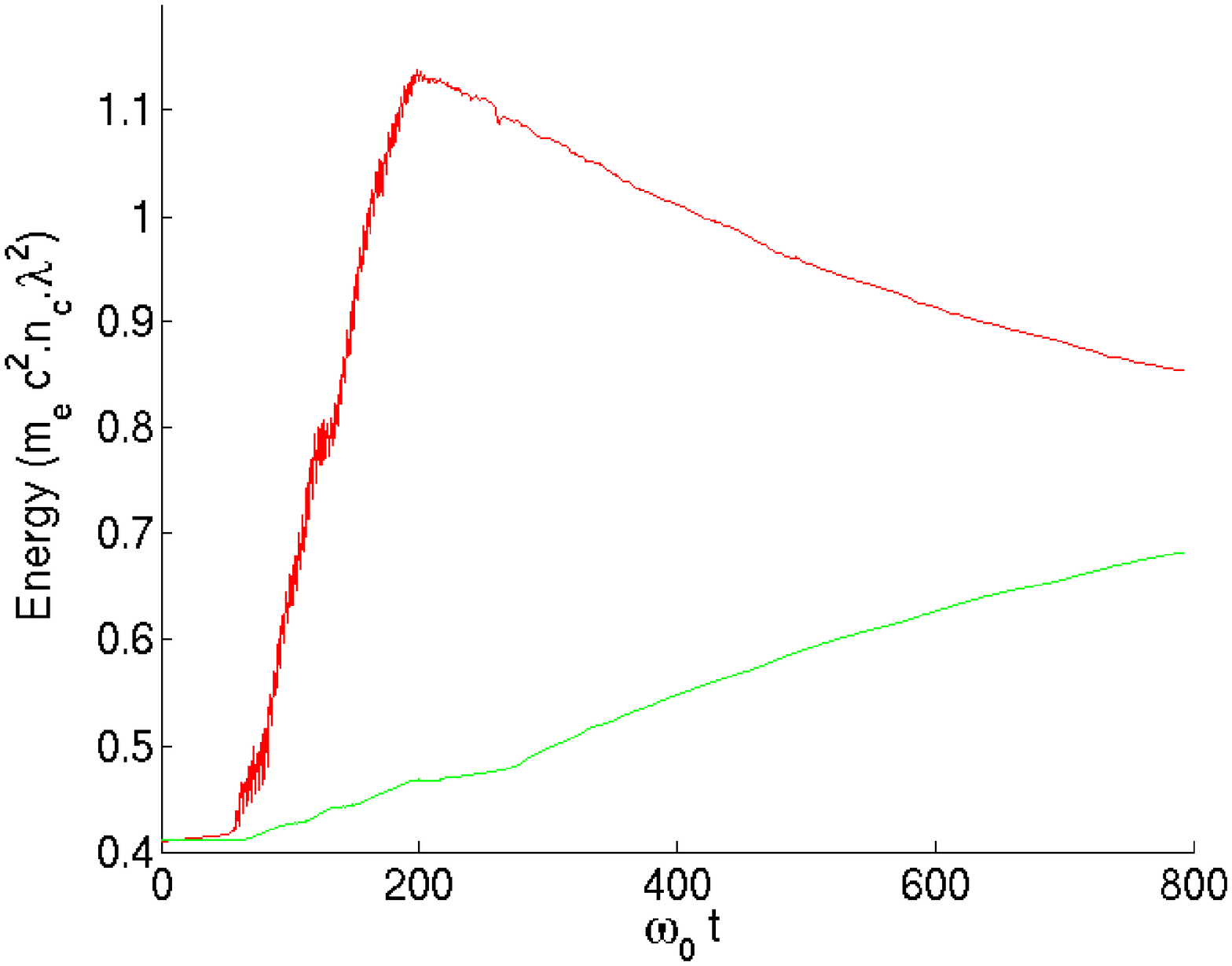}
\caption{Time evolution of the electron (red) and ion (green) kinetic energies: explicit simulation (left) and implicit simulation with $\theta_f=1$ (right).  In both cases, $N_p =1300$.
See text for other simulation parameters.}
\label{ilp10}
\end{figure}

The electron $(x,p_x)$ phase space (integrated in the $y$-direction) is displayed in Fig. \ref{ilp14} for both explicit and implicit schemes. Consistently with the well-known
ponderomotive heating mechanism arising at relativistic laser intensities, fast electrons are accelerated into the target as bunches separated by half the laser wavelength
\cite{adam06}. The explicit simulation predicts maximum electron momenta about 20\% higher than that predicted by the implicit simulation. Also, as a result of the damping
of longitudinal beam-plasma modes, the implicit simulation exhibits a longer-lived separation between the thermal electrons and the fast electrons as the latter propagate through
the target. In an actual solid-density configuration, though, the beam-plasma wave mixing observed in the explicit case should be suppressed by collisions as demonstrated in
Ref. \cite{kemp06}. Yet, these discrepancies do not translate into major differences in the electron energy distribution as shown at three successive times in Fig. \ref{ilp15}.
In particular, the slope of the high-energy tail of the spectra is satisfactorily reproduced. The reduced electron heating gives rise in turn
to a $\sim 15\%$ slower,
space-charge-driven ion acceleration into vacuum as depicted by the ion  $(x,p_x)$ phase spaces of Fig. \ref{ilp16}.

\subsection{Two-dimensional simulations}
\label{sec:2D_simulations}

We now consider a fully two-dimensional laser-plasma system. The electron-ion plasma slab has a peak density of $200n_c$, a temperature of 1 keV and a thickness of $6c/\omega_0$.
A $12 c/\omega_0$-long linear density ramp is added in front of the target. The simulation box consists of a $1024 \times 512$ grid with $\Delta x = \Delta y = 0.1 c/\omega_0$
($\Delta x/\lambda_D = 32$). The incoming laser pulse has unchanged parameters except for a $12c/\omega_0$ FWHM Gaussian transverse profile. Open and periodic boundary
conditions are applied for the electromagnetic fields along the $x$- and $y$-axis, respectively. Due to memory constraints, we use a rather small number of macroparticles $N_p = 40$.
So as to stabilize the system, in addition to using a quadratic weight factor, the time step is significantly increased as compared to the previous simulations: $\Delta t = 0.3 \omega_0^{-1}$,
which corresponds to $\omega_p \Delta t = 4.2$ and $v_t \Delta t/\Delta x = 0.13$.  Particles are subject to periodic boundary conditions in the $y$-direction, and reinjected with their
initial temperature in the $x$-direction.
The damping parameter in the electromagnetic solver, as well as in the particle pusher, is set to zero in the vacuum region and the moderately dense plasma
region up to $n_e=30n_c$.
Two maximum values of the spatially varying damping parameter have been tried in the denser plasma region: $\theta_f =0.1$ and 0.5.
The explicit simulation of reference
makes use of a third-order weight factor with the parameters  $\Delta x=\Delta y=0.08 c/\omega_0$, $\Delta t=0.05\omega_0^{-1}$ and $N_p = 160$. This parallel calculation takes 4.5h
on 64 \mbox{1.6 GHz} Itanium 2 processors. By contrast, the (sequential) implicit simulations take 27h
on a \mbox{2.66 GHz} Intel Xeon X5355 processor.

\begin{figure}[htbp]
\begin{center}
\includegraphics[width=0.3\textwidth]{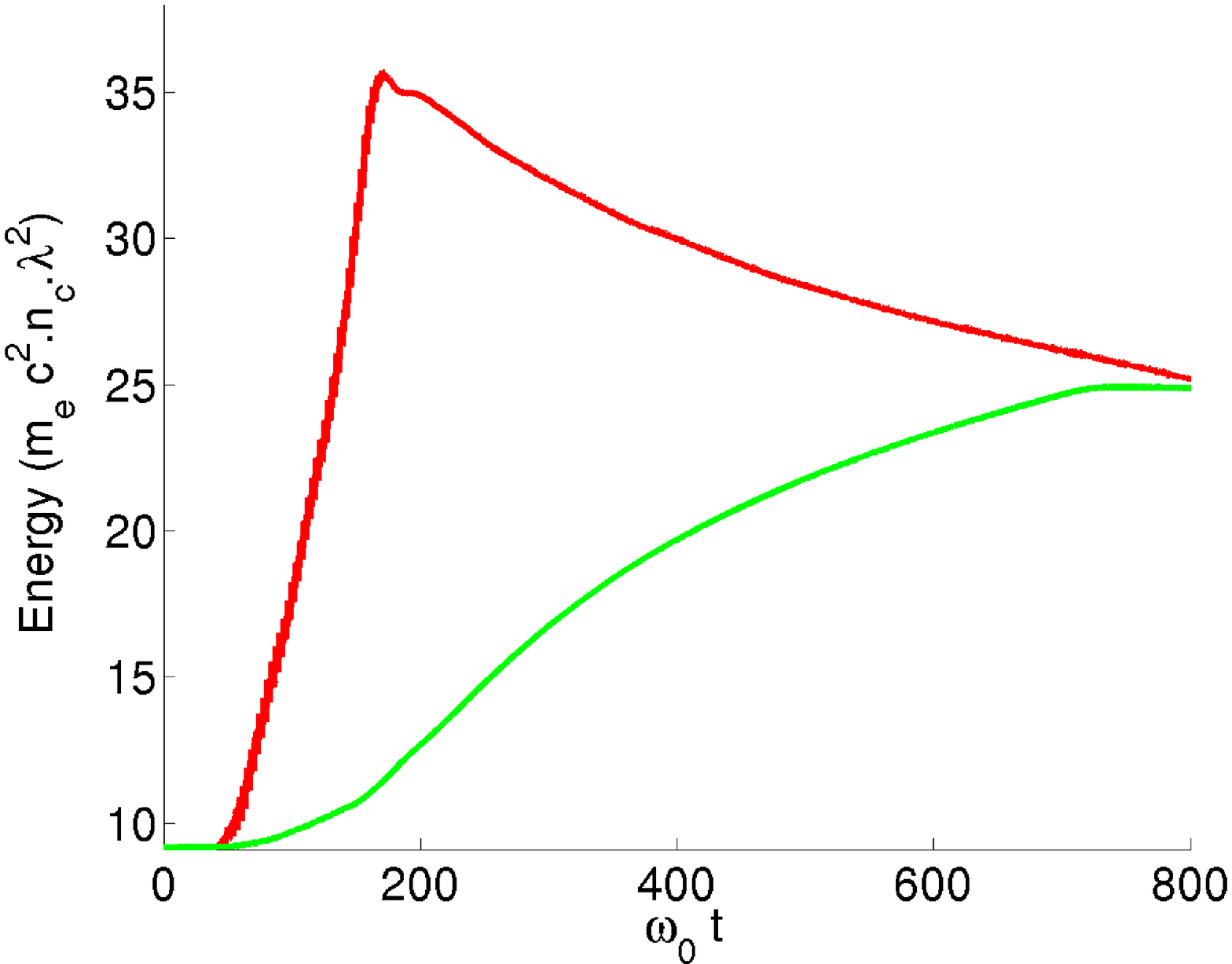}
\includegraphics[width=0.3\textwidth]{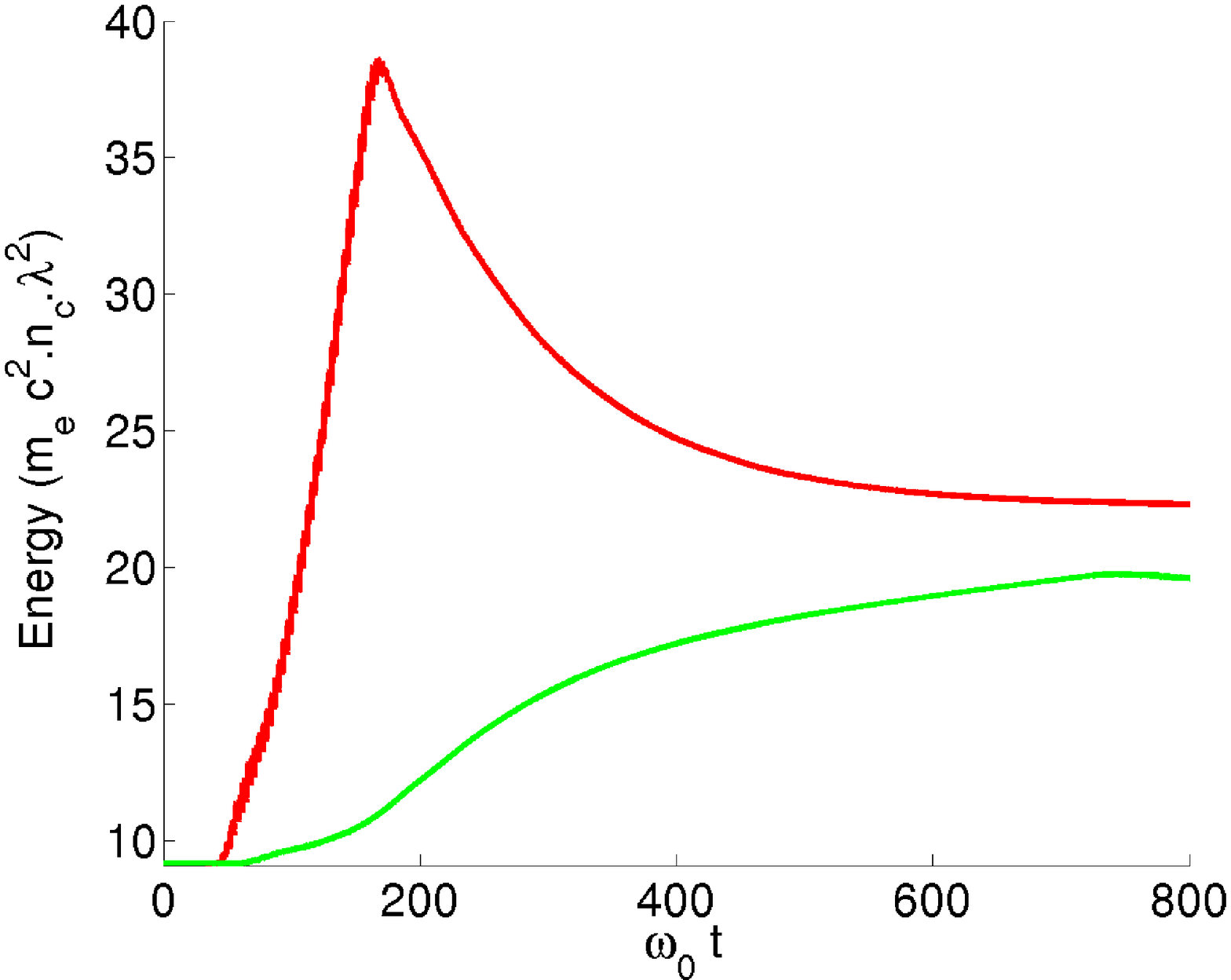}
\includegraphics[width=0.3\textwidth]{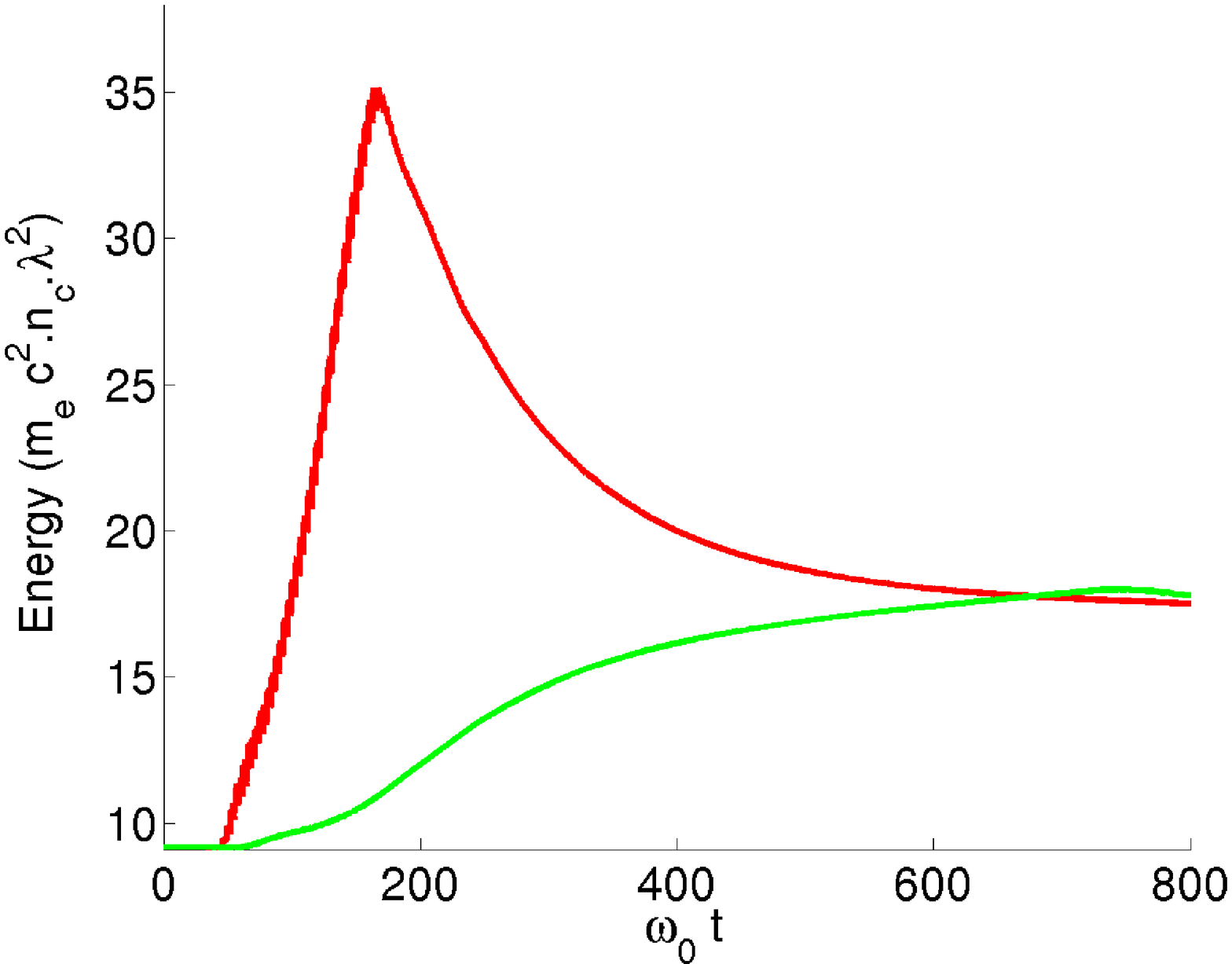}
\caption{Time evolution of the electron (red) and ion (green) kinetic energies: explicit simulation (left), implicit simulations with $\theta_f=0.1$ (center) and  $\theta_f=0.5$ (right).}
\label{ilp2d001}
\end{center}
\end{figure}

The time evolution of the particle kinetic energies is displayed in Fig. \ref{ilp2d001}. All simulations predict about the same peak electron energy. Yet, the damped implicit calculations
yield a faster decreasing electron energy. The total energy variation, evaluated over the time interval $215 < \omega_0 t < 715$ (that is, after complete reflection of the laser pulse
and before the fastest ions hit the box boundaries) is $-12\%$ and $-15\%$ for the $\theta_f=0.1$ and $\theta_f=0.5$ implicit cases, respectively, as compared to $+5\%$ in the explicit case.

\begin{figure}[htbp]
\begin{center}
\includegraphics[width=0.3\textwidth]{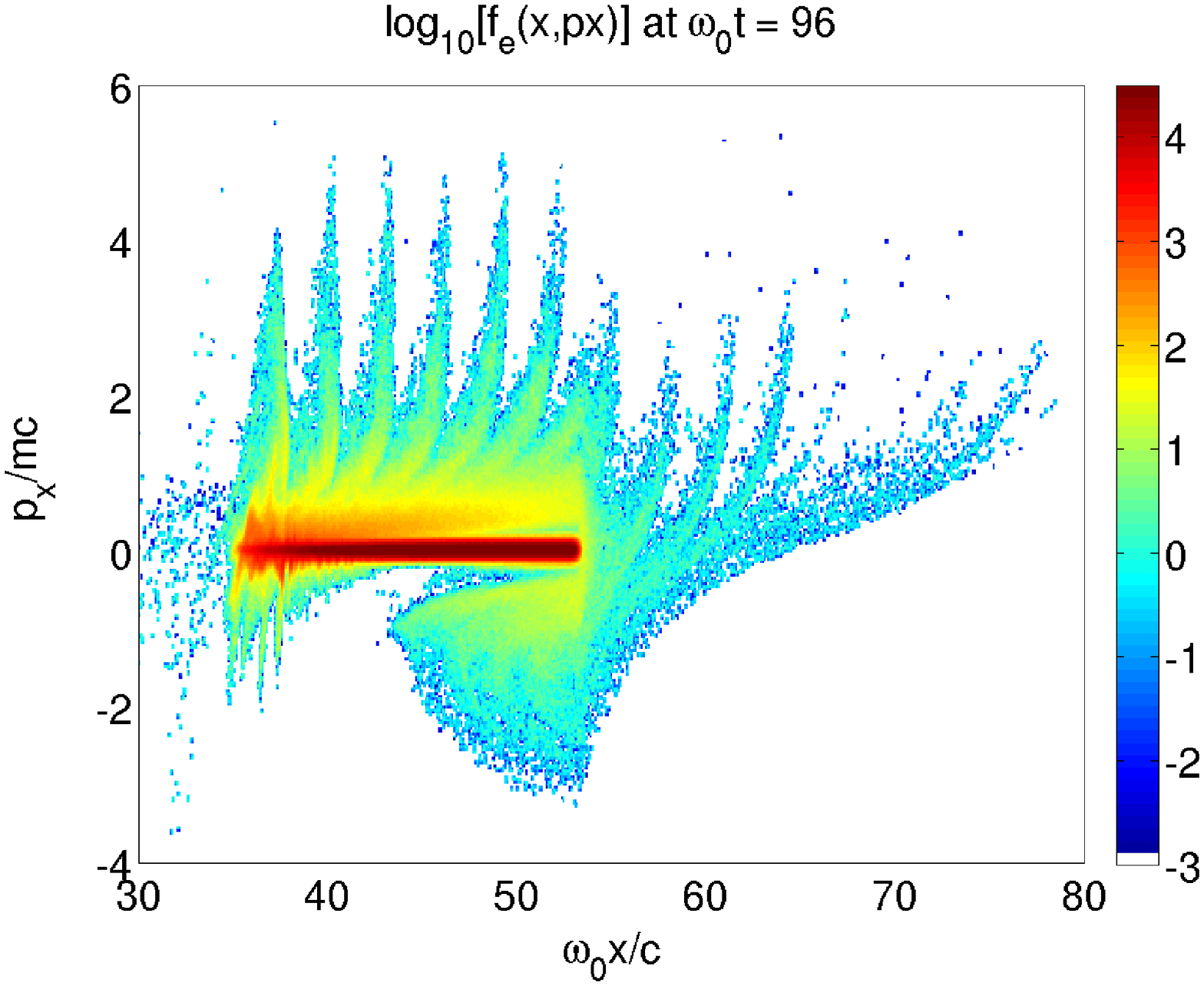}
\includegraphics[width=0.3\textwidth]{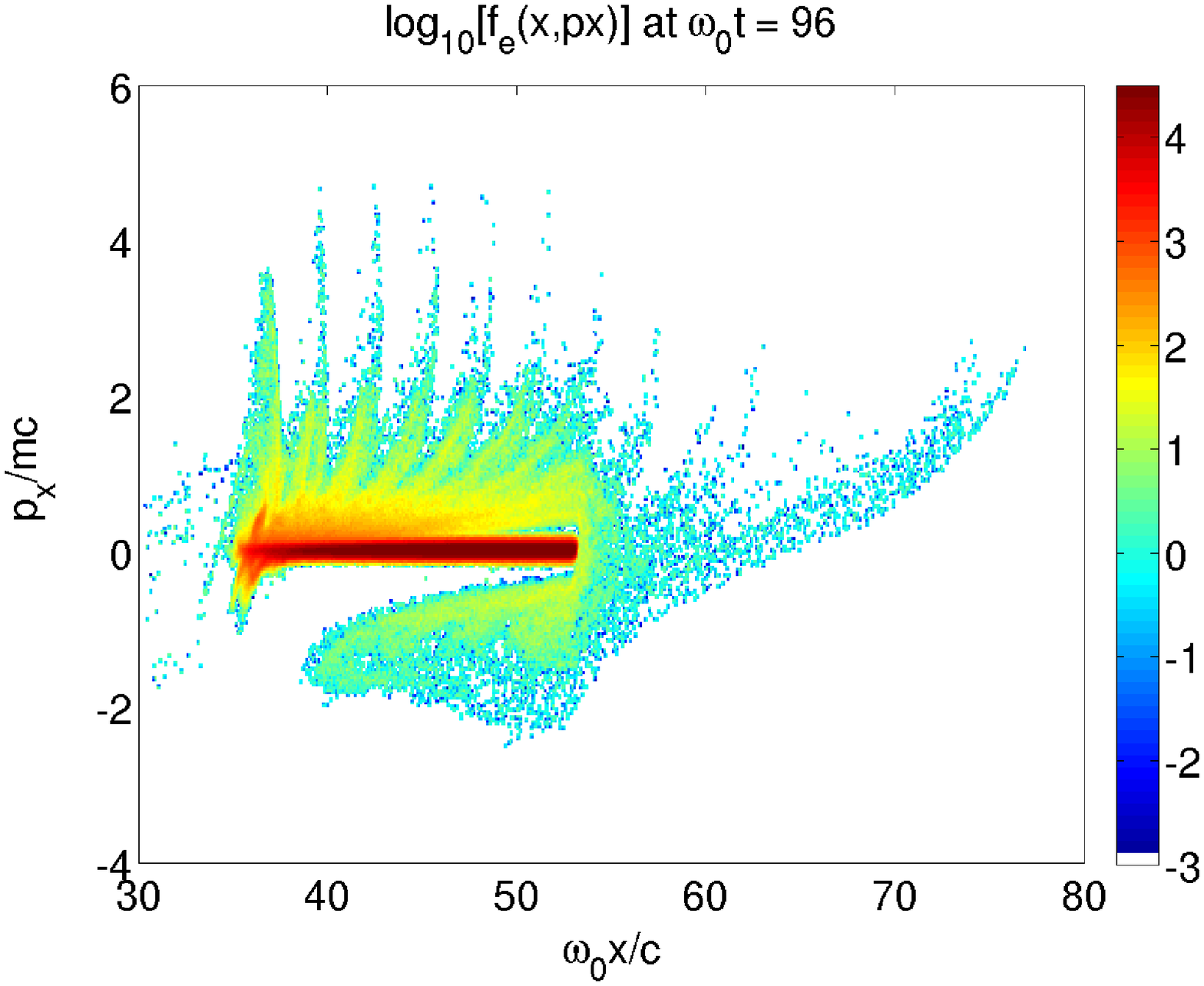}
\includegraphics[width=0.3\textwidth]{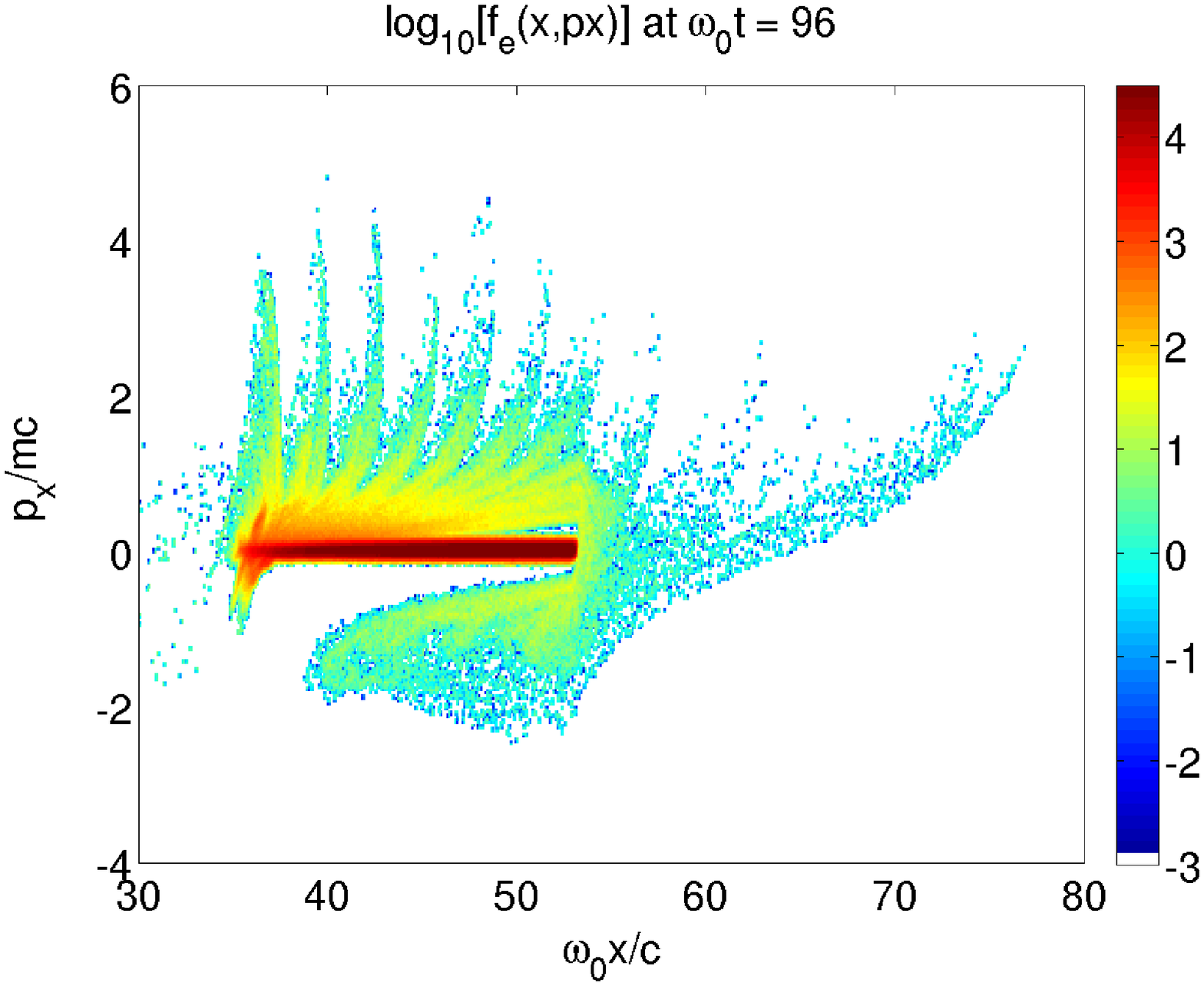}
\caption{Electron $(x,p_x)$ phase space at $t=96 \omega_0^{-1}$: explicit simulation (left) and implicit simulations with $\theta_f=0.1$ (center) and $\theta_f=0.5$ (right).}
\label{ilp2d01}
\end{center}
\end{figure}

\begin{figure}[htbp]
\begin{center}
\includegraphics[width=0.3\textwidth]{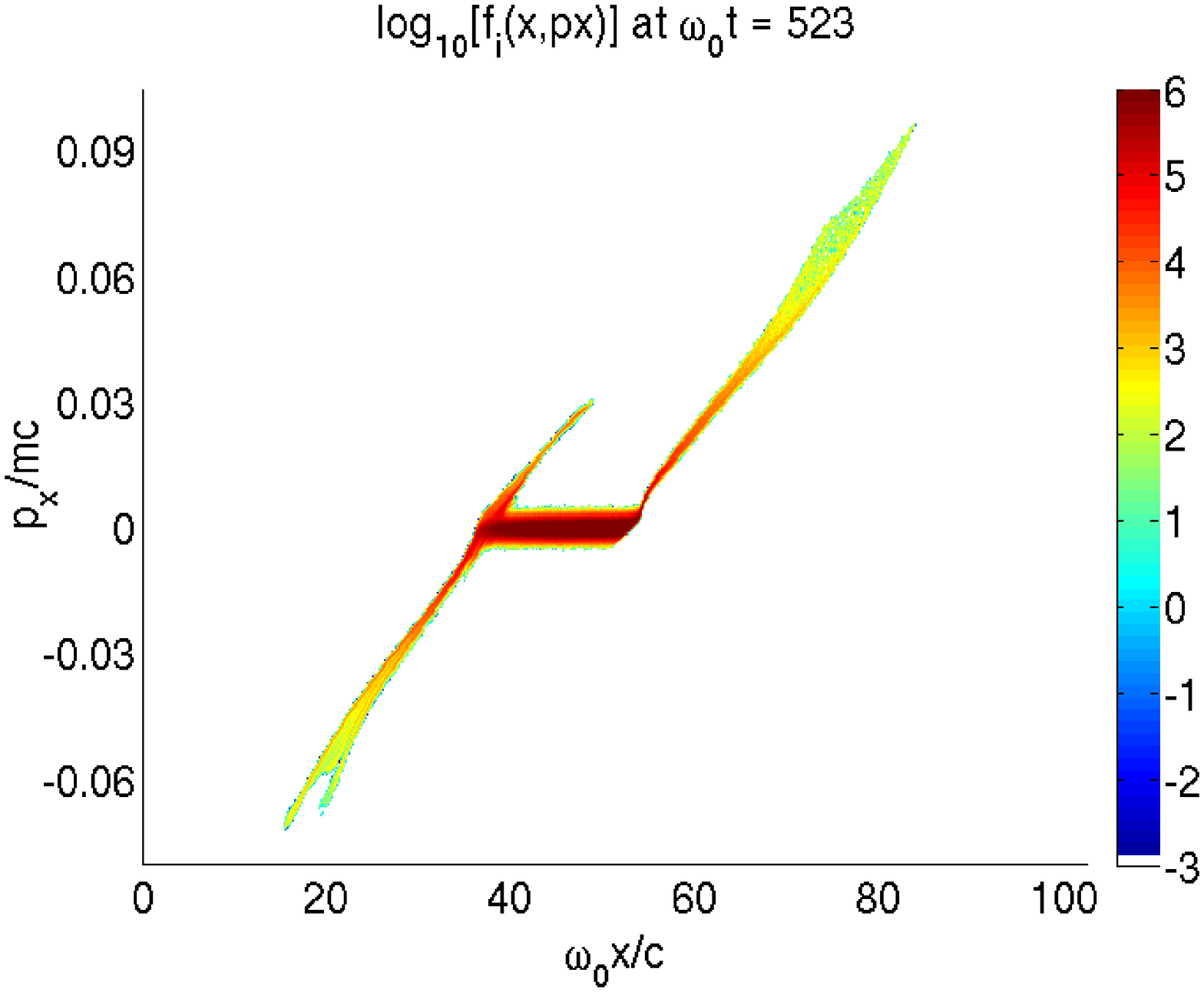}
\includegraphics[width=0.3\textwidth]{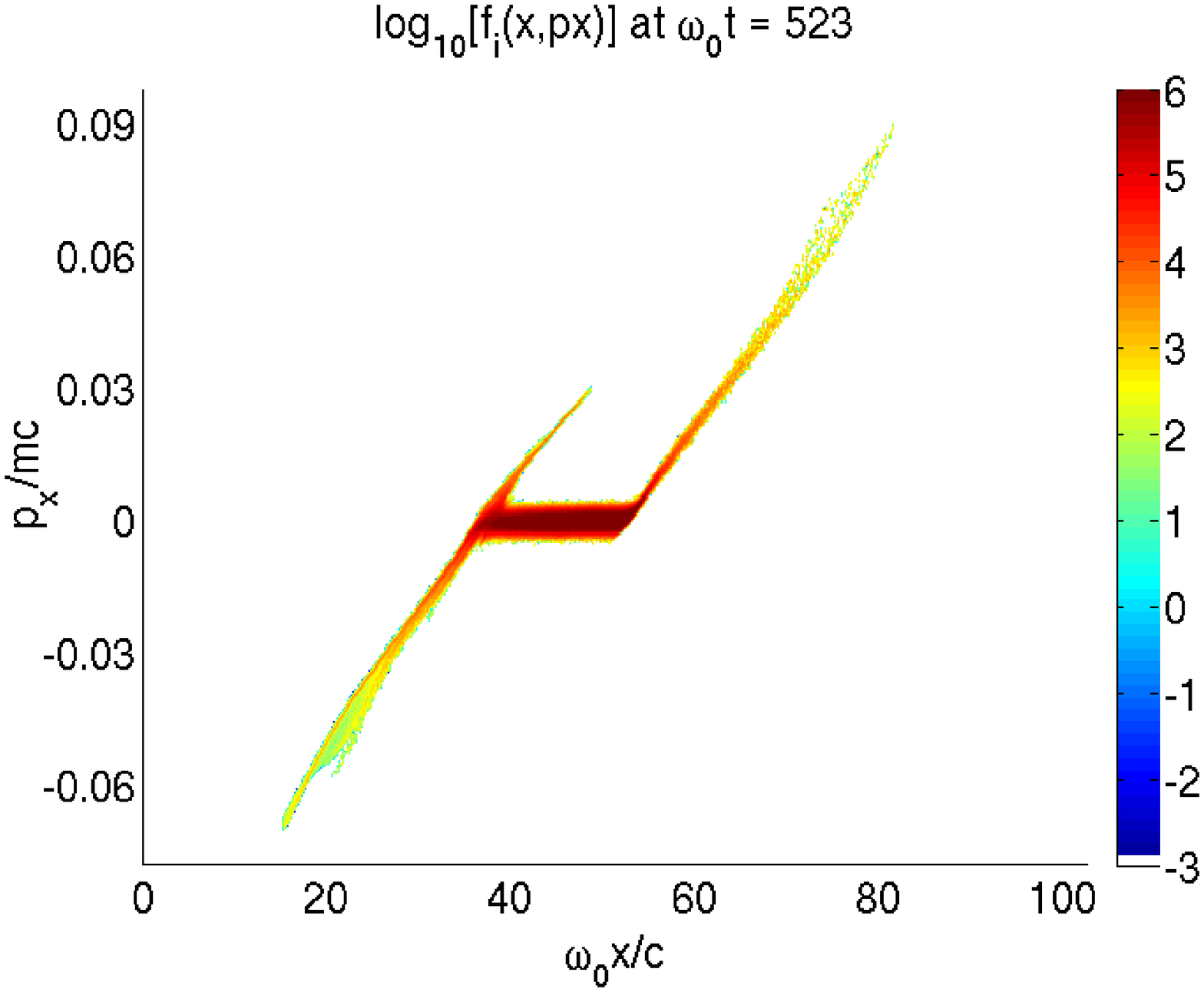}
\includegraphics[width=0.3\textwidth]{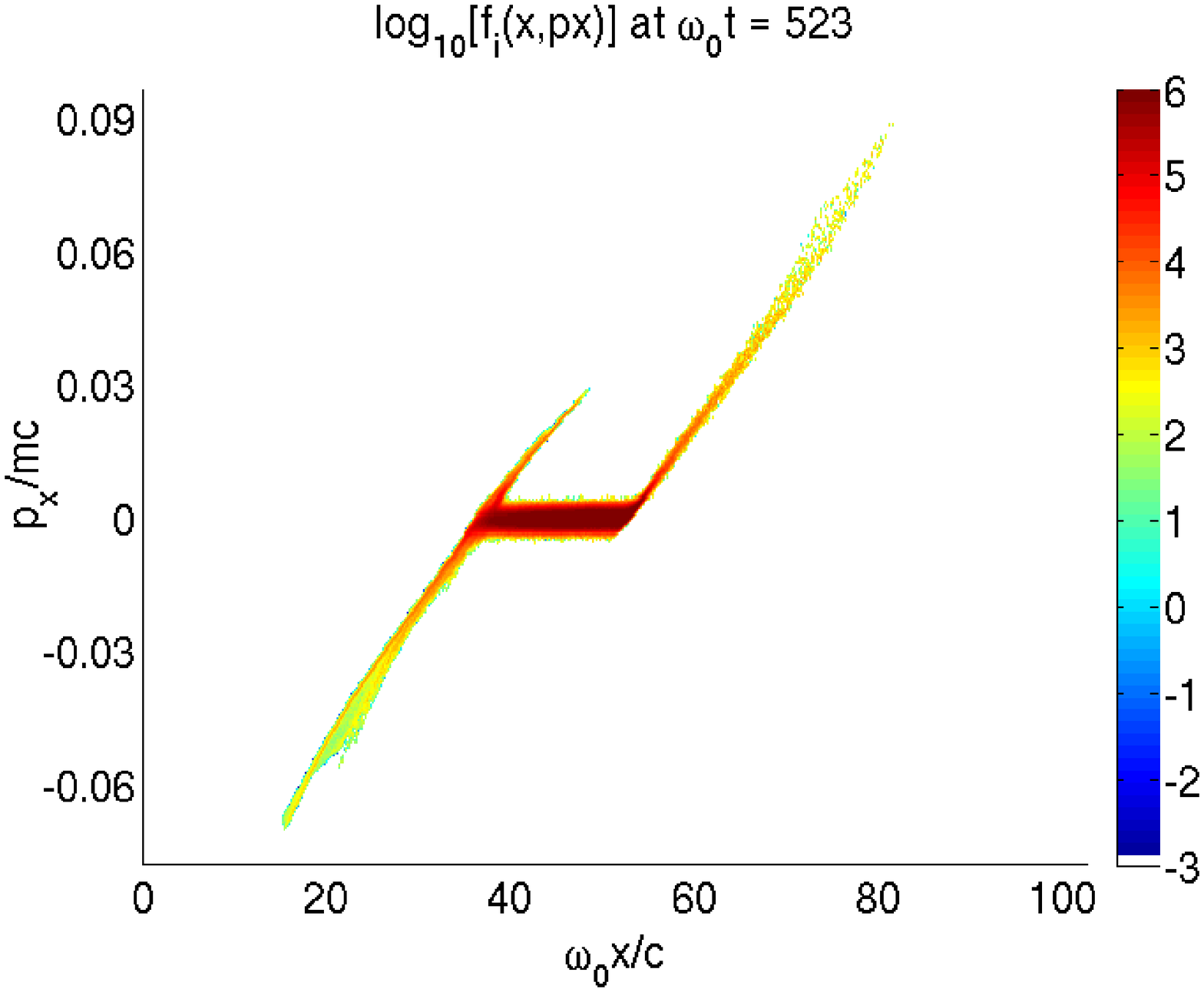}
\caption{Ion $(x,p_x)$ phase space at $t=523 \omega_0^{-1}$: explicit simulation (left) and implicit simulations with $\theta_f=0.1$ (center) and $\theta_f=0.5$ (right).}
\label{ilp2d02}
\end{center}
\end{figure}

Despite their crude time resolution and limited number of macroparticles, the implicit calculations manage to reproduce quite accurately the salient features of the fast electron and ion
generation. This is evidenced by the electron and ion $(x,p_x)$ phase spaces of Figs. \ref{ilp2d01} and \ref{ilp2d02}, as well as by the electron energy spectra of Fig. \ref{ilp2d03}.
As in the previous Section, if to a lesser extent due to the weaker numerical damping employed here, the implicit simulations somewhat underestimate the maximum electron energies. A 2-D picture
of the fast electron generation is provided by the map of the electron kinetic energy density shown in Fig. \ref{ilp2d04}. A reasonable agreement is observed between the three cases,
each calculation showing the characteristic $2\omega_0$-bunched propagation of the fast electrons and their breakout into vacuum.

\begin{figure}[htbp]
\begin{center}
\includegraphics[width=1\textwidth]{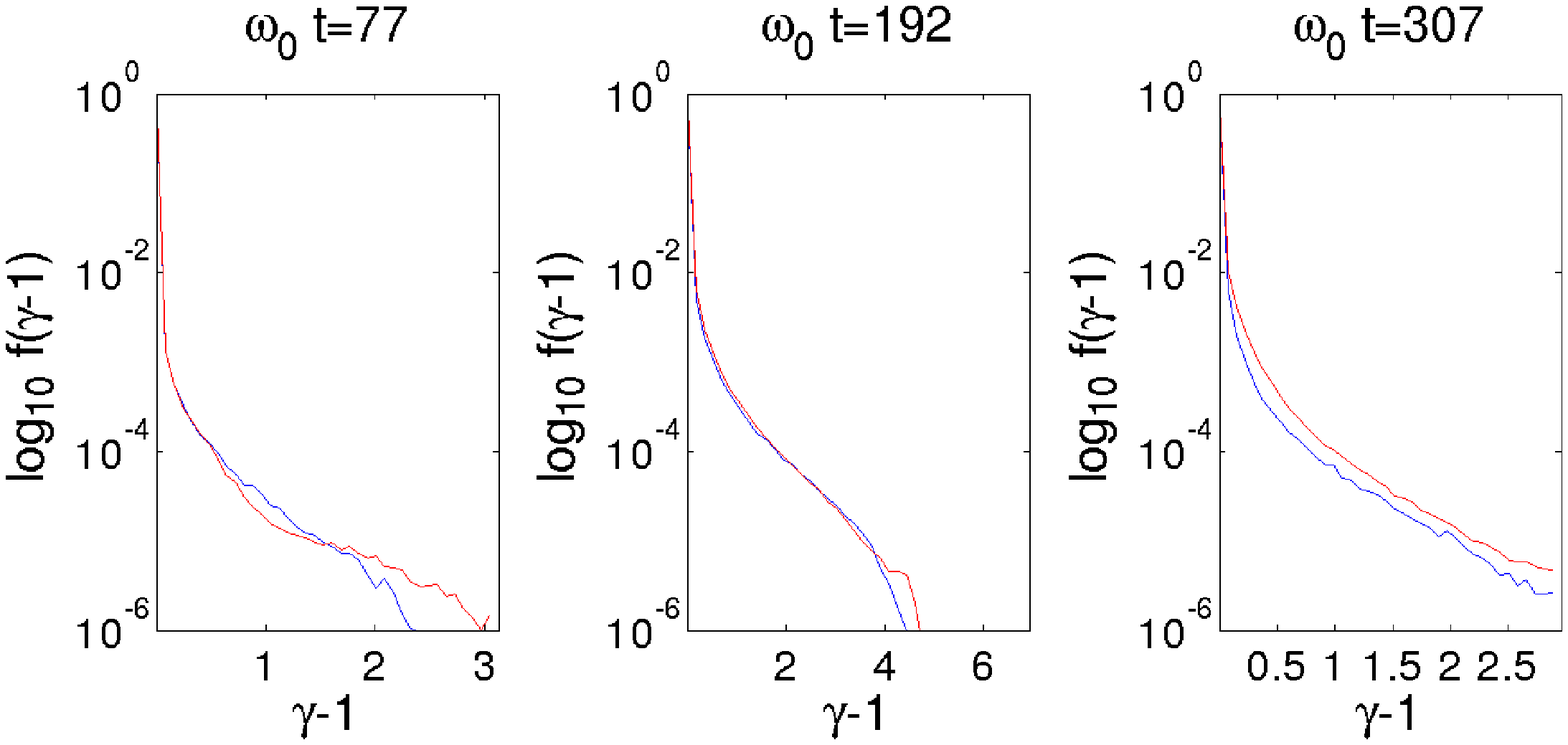}
\caption{Electron energy distribution at different times: explicit simulation (red) and implicit simulation with
$\theta_f = 0.1$ (blue). Energy is normalized by $m_e c^2$.}
\label{ilp2d03}
\end{center}
\end{figure}

\begin{figure}[htbp]
\begin{center}
\includegraphics[width=1\textwidth]{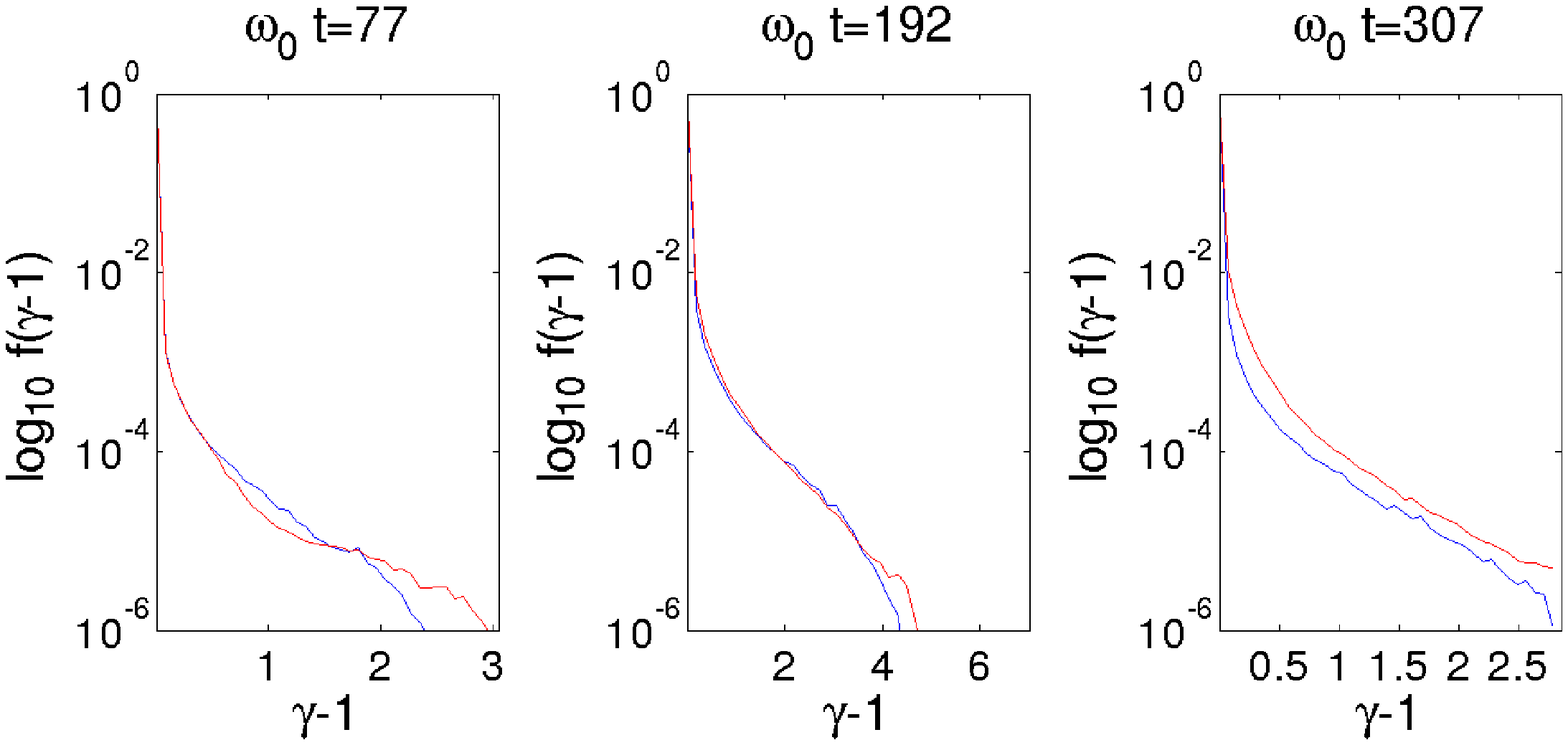}
\caption{Electron energy distribution at different times: explicit simulation (red) and implicit simulation with
$\theta_f = 0.5$ (blue). Energy is normalized by $m_e c^2$.}
\label{ilp2d03}
\end{center}
\end{figure}

\begin{figure}[htbp]
\begin{center}
\includegraphics[width=0.45\textwidth]{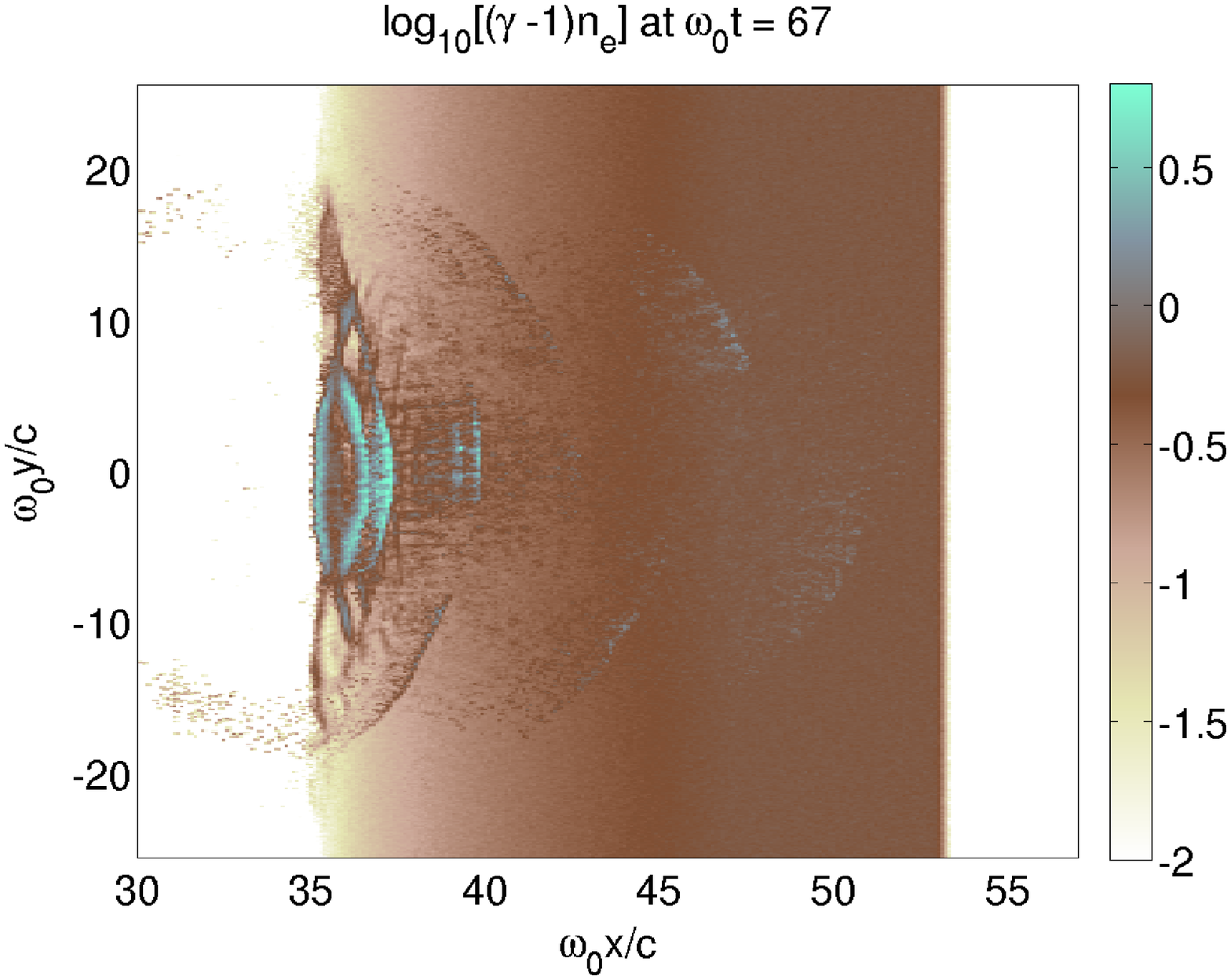}
\includegraphics[width=0.45\textwidth]{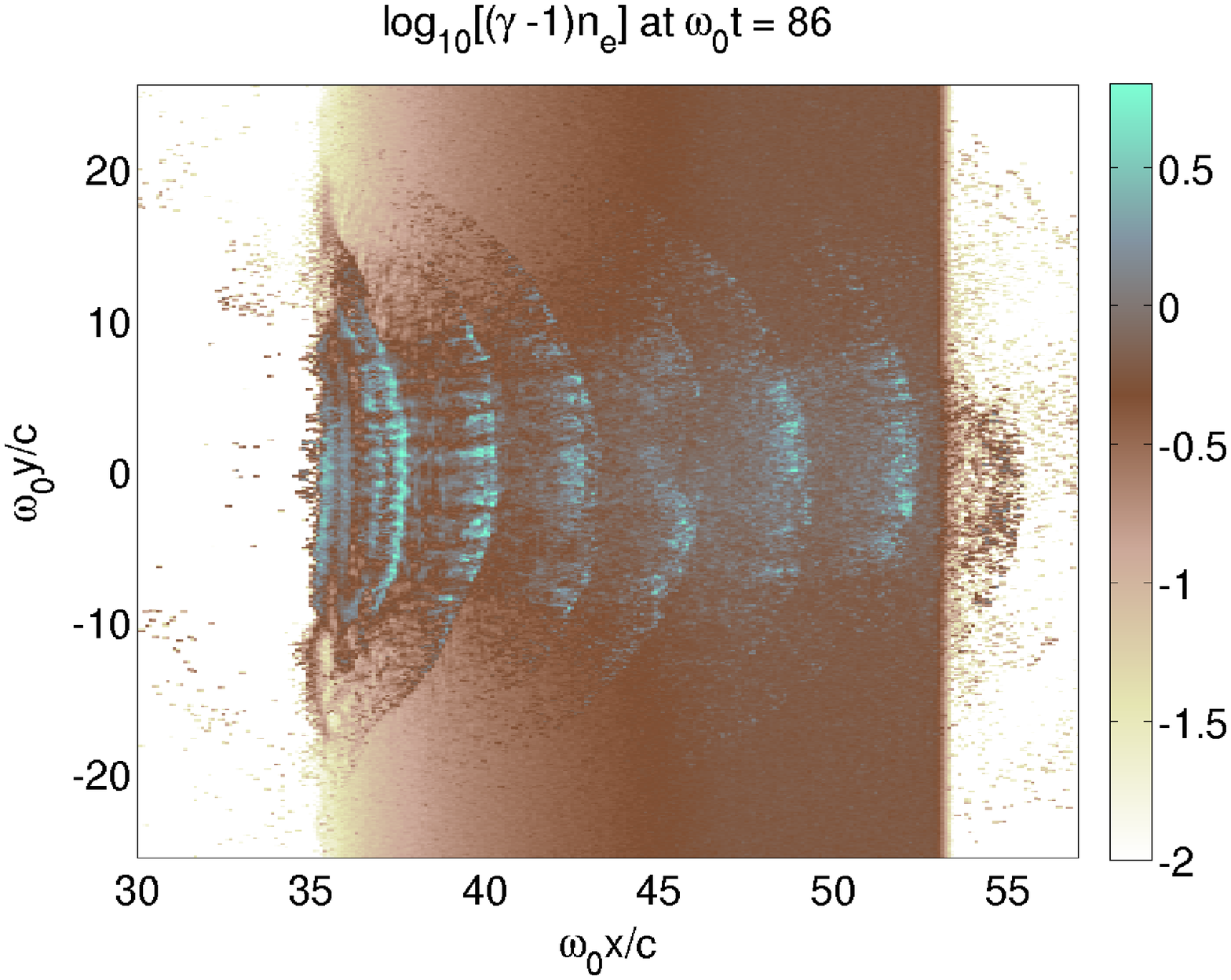} \\
\includegraphics[width=0.45\textwidth]{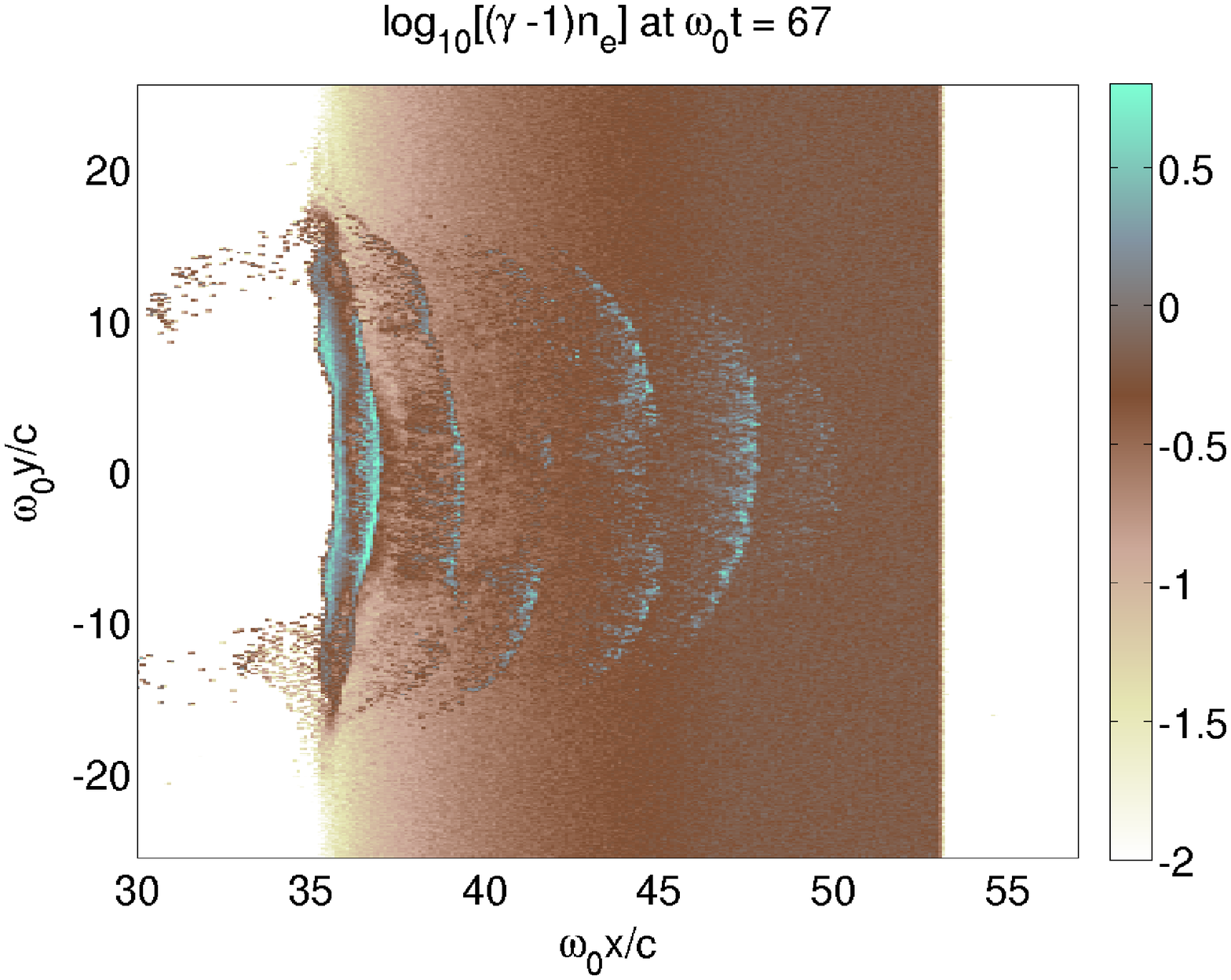}
\includegraphics[width=0.45\textwidth]{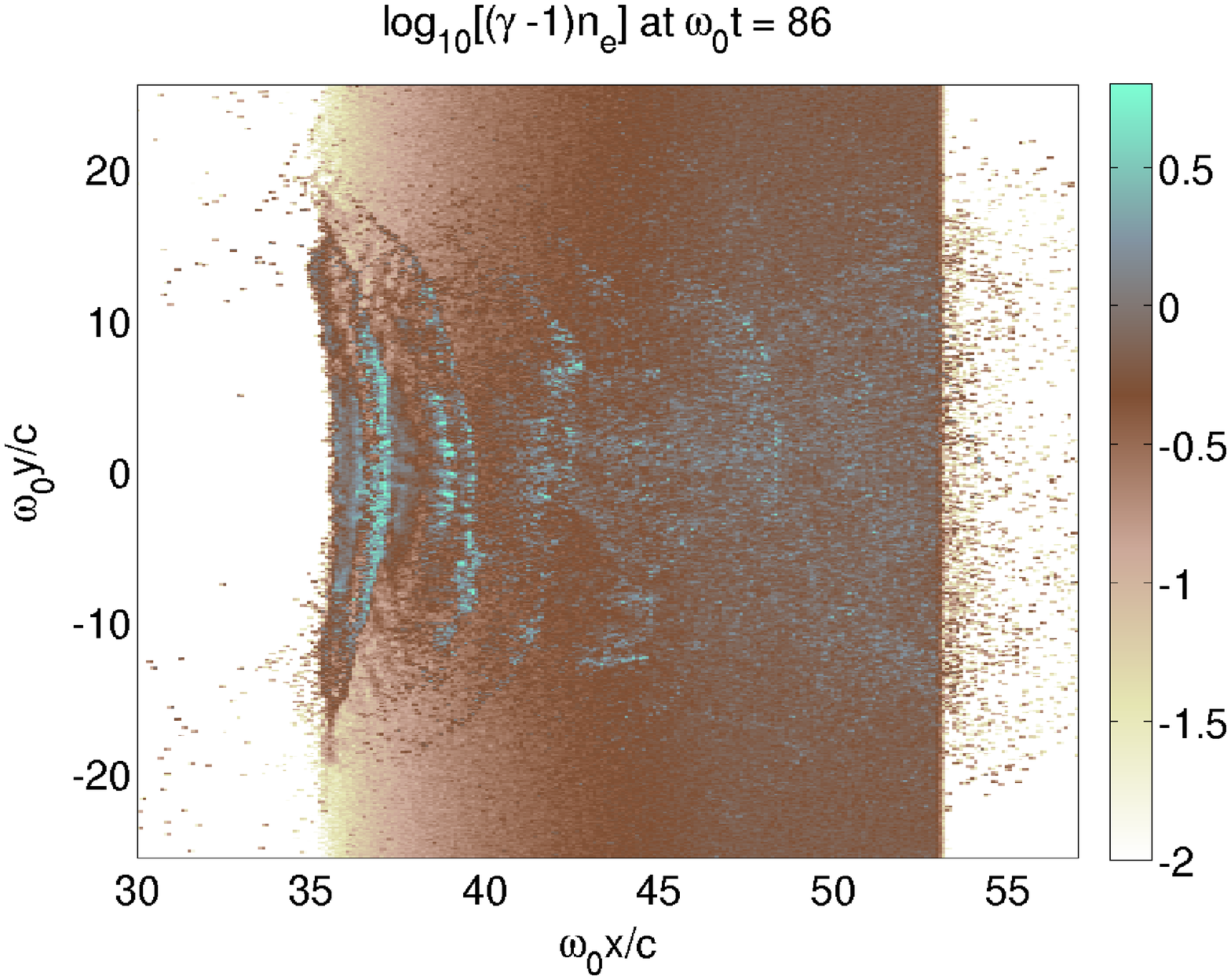} \\
\includegraphics[width=0.45\textwidth]{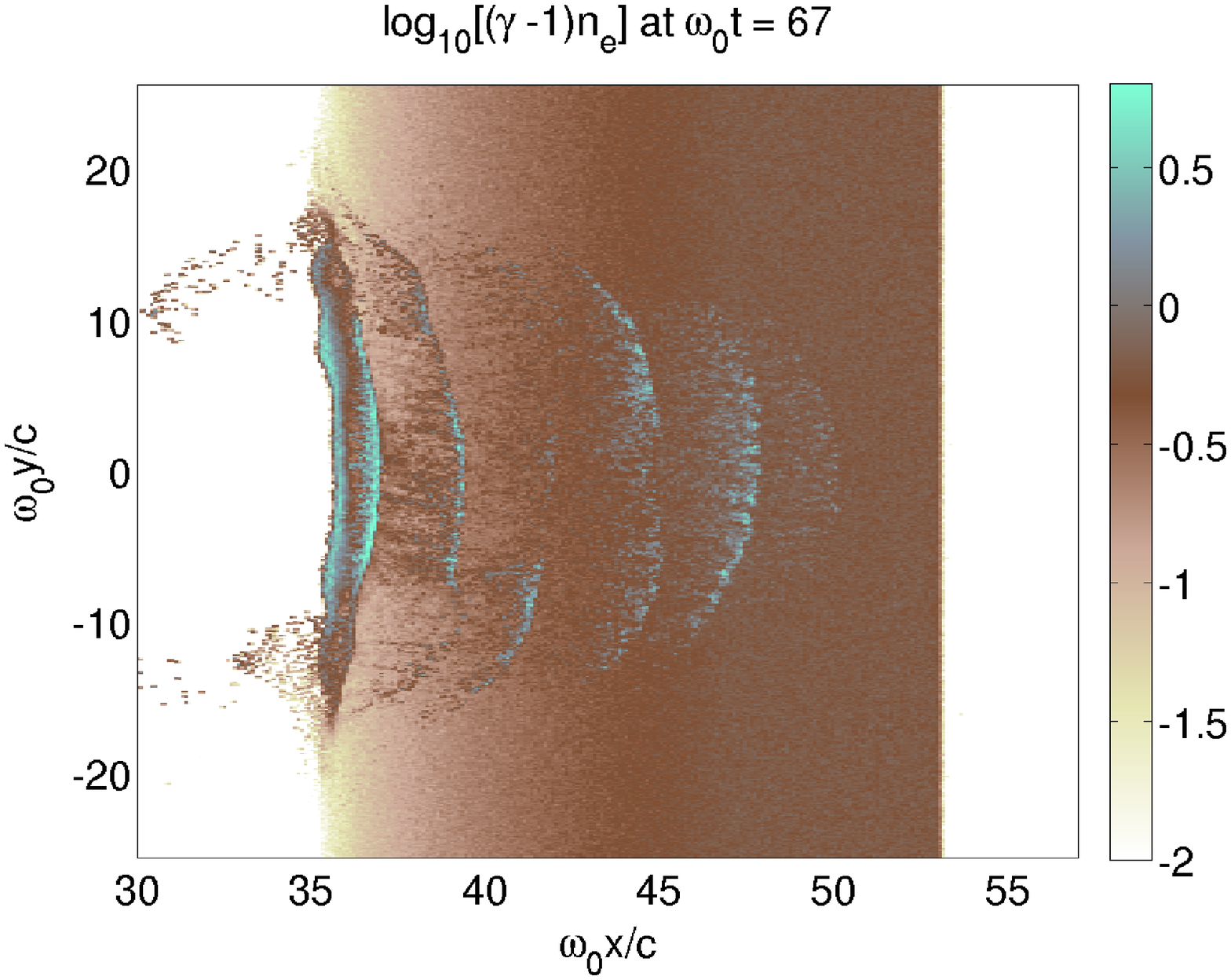}
\includegraphics[width=0.45\textwidth]{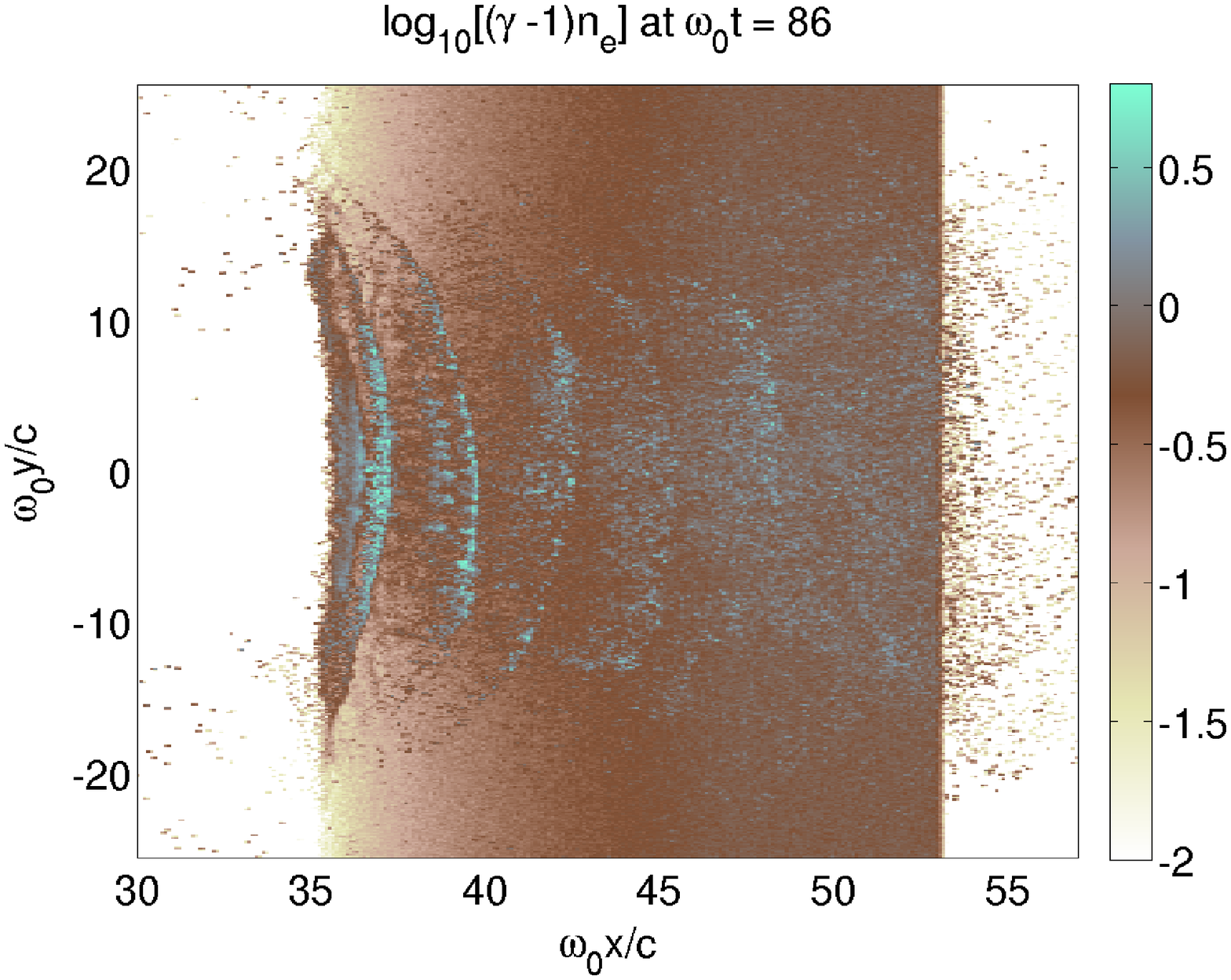}
\caption{Electron kinetic energy density (normalized by $m_ec^2n_c$) at $t=67 \omega_0^{-1}$ and $t=86 \omega_0^{-1}$:
explicit simulation (top) and implicit simulations with $\theta_f=0.1$ (center) and $\theta_f=0.5$ (bottom).}
\label{ilp2d04}
\end{center}
\end{figure}

\section{Conclusion}
\label{sec:conclusion}

This paper has been devoted to the application of the relativistic direct implicit method to the problem of laser-plasma interaction. In contrast to closely related works \cite{camp04, evans2006,weisolodov2008},
our scheme, implemented inside the 2Dx-3Dv code ELIXIRS, allows for high-order weight functions and adjustable damping of the high-frequency waves. The latter capability, which extends to electromagnetic
waves a method originally designed by Friedman \cite{friedman1990} for electrostatic waves, permits to manage within a unified algorithm the dissipation-free, Courant condition-free propagation
of the incident laser pulse through vacuum, while suppressing the need to resolve the high-frequency collective modes inside the dense plasma region. After having presented an original derivation
of the adjustable-damping, direct implicit method as a simplified, one-iteration Newton scheme, we have carried out a thorough analysis of its numerical properties regarding both electromagnetic
and electrostatic waves. The latter study, accounting for the effects of  finite $\Delta t$ and $\Delta x$, the weight factor order and the damping parameter is found to provide useful hints when
compared to the simulation results of the free evolution of a plasma slab. Several numerical tests have been presented and successfuly benchmarked against finely resolved explicit simulations.
In particular, we have demonstrated the ability of the code to capture the main features of the laser-plasma interaction despite cruder space-time resolution. Yet, our code being still sequential,
its increased stability domain remains insufficient to access the large space- and time-scales managed nowadays by massively parallel explicit codes. The parallelization of our code is therefore
required and will be the subject of a future work.

\section{Acknowledgments}
We gratefully acknowledged the work of U. Voss  on the application of the direct implicit method to the problem of
laser-plasma interaction. This study, which provided us with important guidelines, was carried out in 1998
at the CMAP/\'Ecole Polytechnique and supported by the EU TMR grant FMBICT972082.

\appendix

\section{Numerical implementation of the field equation}
\label{app:field_equation}
We detail here the numerical procedure to solve Eq. (\ref{bp28}) within a 2D geometry. The Concus and Golub iterative method \cite{golub1973} is applied to the three components of Eq. (\ref{bp28}).
The $x$-component writes
\begin{align}
\label{bp30a}
E_{x,i+1/2,j}^{n+1} & + \frac{c^2\Delta t^2}{2\Delta x\Delta y} \left(
E_{y,i+1,j+1/2}^{n+1} - E_{y,i+1,j-1/2}^{n+1}
 - E_{y,i,j+1/2}^{n+1} + E_{y,i,j-1/2}^{n+1} \right) \notag \\
 & - \frac{c^2\Delta t^2}{2 \Delta y^2} \left(E_{x,i+1/2,j+1}^{n+1} -2
E_{x,i+1/2,j}^{n+1} +E_{x,i+1/2,j-1}^{n+1}\right)
+ \chi^{11,0}_{i+1/2}E_{x,i+1/2,j}^{n+1} \notag \\
 & +\frac{1}{4}\left[ \chi^{12,0}_{i} E_{y,i,j+1/2}^{n+1} +
 \chi^{12,0}_{i} E_{y,i,j-1/2}^{n+1}  \right .
\left . + \chi^{12,0}_{i+1} E_{y,i+1,j-1/2}^{n+1}
+ \chi^{12,0}_{i+1} E_{y,i+1,j+1/2}^{n+1} \right] \notag \\
 & + \frac{1}{2} \chi^{13,0}_i E_{z,i,j}^{n+1} + \frac{1}{2}
\chi^{13,0}_{i+1}E_{z,i+1,j}^{n+1} - \frac{\Delta t}{2 \Delta
y}\left[ \zeta^{31,0}_{i+1/2}E_{x,i+1/2,j+1}^{n+1}
- \zeta^{31,0}_{i+1/2}E_{x,i+1/2,j-1}^{n+1} \right] \notag \\
 & - \frac{\Delta t}{2 \Delta y}\left[ \zeta^{32,0}_i
E_{y,i,j+1/2}^{n+1} +
 \zeta^{32,0}_{i+1} E_{y,i+1,j+1/2}^{n+1} - \zeta^{32,0}_i E_{y,i,j-1/2}^{n+1}
- \zeta^{32,0}_{i+1}E_{y,i+1,j-1/2}^{n+1} \right] \notag \\
 & - \frac{\Delta t}{4 \Delta y}\left[
\zeta^{33,0}_{i+1}E_{z,i+1,j+1}^{n+1} +
 \zeta^{33,0}_i E_{z,i,j+1}^{n+1} - \zeta^{33,0}_{i+1}E_{z,i+1,j-1}^{n+1}
- \zeta^{33,0}_i E_{z,i,j-1}^{n+1} \right] \notag \\
 & = \widetilde{Q}_{x,i+1/2,j}.
\end{align}
The $y$-component writes
\begin{align}
\label{bp30b}
E_{y,i,j+1/2}^{n+1} & - \frac{c^2\Delta t^2}{2\Delta x^2} \left(
E_{y,i+1,j+1/2}^{n+1} -2 E_{y,i,j+1/2}^{n+1}
 + E_{y,i-1,j+1/2}^{n+1} \right) \notag \\
 & + \frac{c^2\Delta t^2}{2\Delta x\Delta y}
 \left(E_{x,i+1/2,j+1}^{n+1}-E_{x,i-1/2,j+1}^{n+1}-E_{x,i+1/2,j}^{n+1}+E_{x,i-1/2,j}^{n+1}\right) \notag \\
 & + \frac{\chi^{21,0}_i}{4}
 \left(E_{x,i-1/2,j}^{n+1} + E_{x,i+1/2,j}^{n+1} + E_{x,i-1/2,j+1}^{n+1} +
 E_{x,i+1/2,j+1}^{n+1}\right) \notag \\
 & + \chi^{22,0}_i E_{y,i,j+1/2}^{n+1} + \frac{\chi^{23,0}_i}{2}(E_{z,i,j}^{n+1}+E_{z,i,j+1}^{n+1}) \notag \\
 & + \frac{\Delta t}{2\Delta x}\left[ \zeta^{31,0}_{i+1/2}(E_{x,i+1/2,j}^{n+1}+E_{x,i+1/2,j+1}^{n+1})
 - \zeta^{31,0}_{i-1/2}(E_{x,i-1/2,j}^{n+1}+E_{x,i-1/2,j+1}^{n+1}) \right] \notag \\
 & + \frac{\Delta t}{2 \Delta x}\left[
\zeta^{32,0}_{i+1}E_{y,i+1,j+1/2}^{n+1}
- \zeta^{32,0}_{i-1}E_{y,i-1,j+1/2}^{n+1} \right] \notag \\
 & + \frac{\Delta t}{4 \Delta x}\left[
\zeta^{33,0}_{i+1}(E_{z,i+1,j}^{n+1}+E_{z,i+1,j+1}^{n+1})
- \zeta^{33,0}_{i-1}(E_{z,i-1,j}^{n+1}+E_{z,i-1,j+1}^{n+1}) \right] \notag \\
 & = \widetilde{Q}_{y,i,j+1/2}.
\end{align}
The $z$-component writes
\begin{align}
\label{bp30c}
E_{z,i,j}^{n+1} & - \frac{c^2\Delta t^2}{2\Delta x^2} \left(
E_{z,i+1,j}^{n+1} -2 E_{z,i,j}^{n+1} + E_{z,i-1,j}^{n+1}\right) -
\frac{c^2\Delta t^2}{2\Delta y^2}
\left( E_{z,i,j+1}^{n+1} -2 E_{z,i,j}^{n+1} + E_{z,i,j-1}^{n+1} \right) \notag \\
 & + \frac{\chi^{31,0}_i}{2}\left( E_{x,i-1/2,j}^{n+1} +
E_{x,i+1/2,j}^{n+1} \right) + \frac{\chi^{32,0}_i}{2}\left(
E_{y,i,j-1/2}^{n+1} + E_{y,i,j+1/2}^{n+1} \right)
+ \chi^{33,0}_i E_{z,i,j}^{n+1} \notag \\
 & - \frac{\Delta t}{\Delta x}\left(
\zeta^{21,0}_{i+1/2}E_{x,i+1/2,j}^{n+1}
- \zeta^{21,0}_{i-1/2}E_{x,i-1/2,j}^{n+1}\right) \notag \\
 & - \frac{\Delta t}{4 \Delta x}\left[ \zeta^{22,0}_{i+1}
\left(E_{y,i+1,j-1/2}^{n+1} + E_{y,i+1,j+1/2}^{n+1} \right)
- \zeta^{22,0}_{i-1} \left( E_{y,i-1,j+1/2}^{n+1} + E_{y,i-1,j-1/2}^{n+1}\right)\right] \notag \\
 & - \frac{\Delta t}{2 \Delta x}\left( \zeta^{23,0}_{i+1}E_{z,i+1,j}^{n+1}
- \zeta^{23,0}_{i-1} E_{z,i-1,j}^{n+1} \right) \notag \\
 & + \frac{\Delta t}{4 \Delta y}\zeta^{11,0}_i \left(
E_{x,i+1/2,j+1}^{n+1} + E_{x,i-1/2,j+1}^{n+1} - E_{x,i+1/2,j-1}^{n+1} - E_{x,i-1/2,j-1}^{n+1}\right) \notag \\
 & + \frac{\Delta t}{\Delta y}\zeta^{12,0}_i\left( E_{y,i,j+1/2}^{n+1}
- E_{y,i,j-1/2}^{n+1} \right) + \frac{\Delta t}{2\Delta
y}\zeta^{13,0}_i\left( E_{z,i,j+1}^{n+1} - E_{z,i,j-1}^{n+1}\right) \notag \\
 & = \widetilde{Q}_{z,i,j}.
\end{align}
The right-hand sides of Eqs. (\ref{bp30a})-(\ref{bp30c}) are given by
\begin{align}
 \tilde{Q}_{x,i+1/2,j}^{(m)} & = Q_{x,i+1/2,j} -(\chi_{i+1/2,j}^{11} - \chi_{i+1/2,j}^{11,0})E_{x,i+1/2,j}^{(m)} \notag \\
 & -\frac{1}{4}\left[(\chi_{i,j}^{12} - \chi_{i}^{12,0})\left(E_{y,i,j+1/2}^{(m)}
   + E_{y,i,j-1/2}^{(m)} \right) \right. \notag \\
 & \left. +(\chi_{i+1,j}^{12} - \chi_{i+1}^{12,0})\left(E_{y,i+1,j-1/2}^{(m)} + E_{y,i+1,j+1/2}^{(m)} \right) \right]
 -(\chi_{i,j}^{13} - \chi_{i,j}^{13,0})E_{z,i,j}^{(m)} \notag \\
 & +\frac{\Delta t}{2\Delta y} \left[ \left( \zeta_{i+1/2,j+1}^{31}-\zeta_{i+1/2}^{31,0}\right)E_{x,i+1/2,j+1}^{(m)}
 - \left( \zeta_{i+1/2,j-1}^{31}-\zeta_{i+1/2}^{31,0}\right)E_{x,i+1/2,j-1}^{(m)} \right] \notag \\
 & + \frac{\Delta t}{2\Delta y} \left[ \left( \zeta_{i,j+1/2}^{32}-\zeta_{i}^{32,0}\right)E_{y,i,j+1/2}^{(m)}
 + \left( \zeta_{i+1,j+1/2}^{32}-\zeta_{i+1}^{32,0}\right)E_{y,i+1,j+1/2}^{(m)} \right. \notag \\
 & \left. - \left( \zeta_{i,j-1/2}^{32}-\zeta_{i}^{32,0}\right)E_{y,i,j-1/2}^{(m)}
 - \left( \zeta_{i+1,j-1/2}^{32}-\zeta_{i+1}^{32,0}\right)E_{y,i+1,j-1/2}^{(m)} \right] \notag \\
 & + \frac{\Delta t}{4\Delta y} \left[\left( \zeta_{i+1,j+1}^{33}-\zeta_{i+1}^{33,0}\right)E_{z,i+1,j+1}^{(m)}
 + \left( \zeta_{i,j+1}^{33}-\zeta_{i}^{33,0}\right)E_{z,i,j+1}^{(m)} \right. \notag \\
 & \left. - \left( \zeta_{i+1,j-1}^{33}-\zeta_{i+1}^{33,0}\right)E_{z,i+1,j-1}^{(m)}
 - \left( \zeta_{i,j-1}^{33}-\zeta_{i}^{33,0}\right)E_{z,i,j-1}^{(m)} \right]\, ,
\end{align}
\begin{align}
  \tilde{Q}_{y,i,j+1/2}^{(m)}  = & Q_{y,i,j+1/2} -\frac{1}{4}\left[
  \left(\chi_{i,j}^{21} - \chi_{i}^{21,0}\right)\left(E_{x,i-1/2,j}^{(m)} + E_{x,i+1/2,j}^{(m)}\right) \right.\notag \\
 & \left.+ \left( \chi_{i,j+1}^{21}-\chi_{i}^{21,0}\right)\left(E_{x,i-1/2,j+1}^{(m)}+E_{x,i+1/2,j+1}^{(m)} \right) \right]
  -(\chi_{i,j+1/2}^{22} - \chi_{i}^{22,0})E_{y,i,j+1/2}^{(m)} \notag \\
 & -\frac{1}{2}\left[\left( \chi_{i,j}^{23}-\chi_{i}^{23,0} \right)E_{z,i,j}^{(m)}
 + \left( \chi_{i,j+1}^{23}-\chi_{i}^{23,0} \right)E_{z,i,j+1}^{(m)} \right] \notag \\
 & -\frac{\Delta t}{2\Delta x}
 \left[ \left( \zeta_{i+1/2,j}^{31}-\zeta_{i+1/2}^{31,0}\right)E_{x,i+1/2,j}^{(m)}\right. \notag
 + \left( \zeta_{i+1/2,j+1}^{31}-\zeta_{i+1/2}^{31,0}\right)E_{x,i+1/2,j+1}^{(m)} \notag \\
 & - \left( \zeta_{i-1/2,j}^{31}-\zeta_{i-1/2}^{31,0}\right)E_{x,i-1/2,j}^{(m)} \notag
 \left. - \left( \zeta_{i-1/2,j+1}^{31}-\zeta_{i-1/2}^{31,0}\right)E_{x,i-1/2,j+1}^{(m)} \right] \notag \\
 & -\frac{\Delta t}{2\Delta x}
 \left[ \left( \zeta_{i+1,j+1/2}^{32}-\zeta_{i+1}^{32,0} \right)E_{y,i+1,j+1/2}^{(m)}\right. \notag
 \left. - \left( \zeta_{i-1,j+1/2}^{32}-\zeta_{i-1}^{32,0} \right)E_{y,i-1,j+1/2}^{(m)}\right] \notag \\
 & -\frac{\Delta t}{4\Delta x}\left[
 \left( \zeta_{i+1,j}^{33}-\zeta_{i+1}^{33,0} \right)E_{z,i+1,j}^{(m)} \right. \notag
 + \left( \zeta_{i+1,j+1}^{33}-\zeta_{i+1}^{33,0} \right)E_{z,i+1,j+1}^{(m)} \notag \\
 & - \left( \zeta_{i-1,j}^{33}-\zeta_{i-1}^{33,0} \right)E_{z,i-1,j}^{(m)} \notag
 \left. - \left( \zeta_{i-1,j+1}^{33}-\zeta_{i-1}^{33,0} \right)E_{z,i-1,j+1}^{(m)} \right]\, ,
 \end{align}
 \begin{align}
 \tilde{Q}_{z,i,j}^{(m)}  = & Q_{z,i,j}
 -\frac{1}{2} (\chi_{i,j}^{31} - \chi_{i}^{31,0})\left(E_{x,i-1/2,j}^{(m)}+ E_{x,i+1/2,j}^{(m)} \right) \notag \\
 & -\frac{1}{2}\left( \chi_{i,j}^{32}-\chi_{i}^{32,0} \right)\left( E_{y,i,j-1/2}^{(m)}+E_{y,i,j+1/2}^{(m)} \right)
 - \left( \chi_{i,j}^{33} - \chi_{i}^{33,0}\right) E_{z,i,j}^{(m)} \notag \\
 & + \frac{\Delta t}{\Delta x}\left[
 \left( \zeta_{i+1/2,j}^{21}-\zeta_{i+1/2}^{21,0}\right)E_{x,i+1/2,j}^{(m)} \right. \notag
 \left. - \left( \zeta_{i-1/2,j}^{21}-\zeta_{i-1/2}^{21,0}\right)E_{x,i-1/2,j}^{(m)} \right] \notag \\
 & + \frac{\Delta t}{4\Delta x}\left[
 \left( \zeta_{i+1,j-1/2}^{22}-\zeta_{i+1}^{22,0}\right)E_{y,i+1,j-1/2}^{(m)} \right. \notag
 +  \left( \zeta_{i+1,j+1/2}^{22}-\zeta_{i+1}^{22,0}\right)E_{y,i+1,j+1/2}^{(m)} \notag \\
 & -  \left( \zeta_{i-1,j-1/2}^{22}-\zeta_{i-1}^{22,0}\right)E_{y,i-1,j-1/2}^{(m)} \notag
 \left. -  \left( \zeta_{i-1,j+1/2}^{22}-\zeta_{i-1}^{22,0}\right)E_{y,i-1,j+1/2}^{(m)} \right]\notag \\
 & + \frac{\Delta t}{2\Delta x}\left[
 \left( \zeta_{i+1,j}^{23}-\zeta_{i+1}^{23,0}\right)E_{z,i+1,j}^{(m)}\right. \notag
 \left. - \left( \zeta_{i-1,j}^{23}-\zeta_{i-1}^{23,0}\right)E_{z,i-1,j}^{(m)} \right] \notag \\
 & - \frac{\Delta t}{4\Delta y}\left[
 \left( \zeta_{i+1/2,j+1}^{11}-\zeta_{i+1/2}^{11,0}\right) E_{x,i+1/2,j+1}^{(m)} \right. \notag
 + \left( \zeta_{i-1/2,j+1}^{11}-\zeta_{i-1/2}^{11,0}\right) E_{x,i-1/2,j+1}^{(m)} \notag \\
 & - \left( \zeta_{i+1/2,j-1}^{11}-\zeta_{i+1/2}^{11,0}\right) E_{x,i+1/2,j-1}^{(m)} \notag
 \left. - \left( \zeta_{i-1/2,j-1}^{11}-\zeta_{i-1/2}^{11,0}\right) E_{x,i-1/2,j-1}^{(m)}\right] \notag \\
 & - \frac{\Delta t}{\Delta y}\left[
 \left( \zeta_{i,j+1/2}^{12}-\zeta_{i}^{12,0} \right)E_{y,i,j+1/2}^{(m)} \right. \notag
 \left. - \left( \zeta_{i,j-1/2}^{12}-\zeta_{i}^{12,0} \right)E_{y,i,j-1/2}^{(m)} \right] \notag \\
 & - \frac{\Delta t}{2\Delta y}\left[
  \left( \zeta_{i,j+1}^{13}-\zeta_{i}^{13,0} \right)E_{z,i,j+1}^{(m)} \right. \notag
 \left. - \left( \zeta_{i,j-1}^{13}-\zeta_{i}^{13,0} \right)E_{z,i,j-1}^{(m)} \right].
\end{align}
Assuming periodicity of the electric field along the $y$ direction, we Fourier transform
Eqs. (\ref{bp30a})-(\ref{bp30c}) in this direction.
We introduce $E_k^R$ and $E_k^I$ the real and imaginary parts of the Fourier transformed electric field.
For notational simplicity, the index $k$ will be omitted in the following.
The real part of the Fourier transform of Eq. (\ref{bp30a}) reads
\begin{align}
\label{apAA01}
& \left( E_y^R \right)_i \left\{\frac{-c^2\Delta t^2}{2\Delta x\Delta y}\left( \cos(\tilde{k}\Delta y)-1\right)
+ \frac{\chi_i^{12,0}}{4}\left( \cos(\tilde{k}\Delta y)+1\right)
 -\frac{\Delta t}{2\Delta y}\zeta_i^{32,0}\left( \cos(\tilde{k}\Delta y)-1\right)\right\} \notag\\
& +\left( E_y^I \right)_i \left\{ \frac{c^2\Delta t^2}{2\Delta x\Delta y}-\frac{\chi_i^{12,0}}{4}
 +\frac{\Delta t}{2\Delta y}\zeta_i^{32,0}\right\}\sin(\tilde{k}\Delta y) \notag\\
& + \left( E_z^R \right)_i \left\{\frac{\chi_i^{13,0}}{2}\right\} +
\left( E_z^I \right)_i \left\{\frac{\Delta t}{2\Delta y} \zeta_i^{33,0}\sin(\tilde{k}\Delta y) \right\} \notag\\
& +\left( E_x^R \right)_{i+1/2} \left\{1-\frac{c^2\Delta t^2}{\Delta y^2}\left( \cos(\tilde{k}\Delta y)-1\right)  +\chi_{i+1/2}^{11,0} \right\}+
\left( E_x^I \right)_{i+1/2} \left\{ \frac{\Delta t}{\Delta y}\zeta_{i+1/2}^{31,0}\sin(\tilde{k}\Delta y) \right\} \notag\\
& +\left( E_y^R \right)_{i+1} \left\{\frac{c^2\Delta t^2}{2\Delta x\Delta y}\left( \cos(\tilde{k}\Delta y)-1\right)+
\frac{\chi_{i+1}^{12,0}}{4}\left( \cos(\tilde{k}\Delta y)+1\right)-
\frac{\Delta t}{2\Delta y}\zeta_{i+1}^{32,0}\left( \cos(\tilde{k}\Delta y)-1\right)\right\} \notag\\
& +\left( E_y^I \right)_{i+1} \left\{\frac{-c^2\Delta t^2}{2\Delta x\Delta y}-
\frac{\chi_{i+1}^{12,0}}{4}+
\frac{\Delta t}{2\Delta y}\zeta_{i+1}^{32,0}\right\}\sin(\tilde{k}\Delta y) \notag\\
& +\left( E_z^R \right)_{i+1} \left\{ \frac{\chi_{i+1}^{13,0}}{2} \right\}+
\left( E_z^I \right)_{i+1} \left\{\frac{\Delta t}{2\Delta y}\zeta_{i+1}^{33,0}\sin(\tilde{k}\Delta y)\right\}=
\left( \widetilde{Q}_x^R \right)_{i+1/2}. 
\end{align}
The imaginary part of the Fourier transform of Eq. (\ref{bp30a}) reads
\begin{align}
\label{apAA02}
& \left( E_y^R \right)_i \left\{ -\frac{c^2\Delta t^2}{2\Delta x\Delta y}
+\frac{\chi_i^{12,0}}{4} -\frac{\Delta t}{2\Delta y}\zeta_i^{32,0} \right\}\sin(\tilde{k}\Delta y) \notag\\
& + \left( E_y^I \right)_i \left\{ -\frac{c^2\Delta t^2}{2\Delta x\Delta y}\left( \cos(\tilde{k}\Delta y)-1\right)
+\frac{\chi_i^{12,0}}{4}\left( \cos(\tilde{k}\Delta y)+1\right)
-\frac{\Delta t}{2\Delta y}\zeta_i^{32,0}\left( \cos(\tilde{k}\Delta y)-1\right) \right\} \notag\\
& + \left( E_z^R \right)_i \left\{ -\frac{\Delta t}{2\Delta y}\zeta_i^{33,0}\sin(\tilde{k}\Delta y) \right\} +
\left( E_z^I \right)_i \left\{ \frac{\chi_i^{13,0}}{2} \right\} \notag\\
& + \left( E_x^R \right)_{i+1/2}\left\{ -\frac{\Delta t}{\Delta y}\zeta_{i+1/2}^{31,0}\sin(\tilde{k}\Delta y)\right\} +
\left( E_x^I \right)_{i+1/2}\left\{ 1-\frac{c^2\Delta t^2}{\Delta y^2}\left( \cos(\tilde{k}\Delta y)-1\right)
+\chi_{i+1/2}^{11,0} \right\} \notag\\
& + \left( E_y^R \right)_{i+1} \left\{ \frac{c^2\Delta t^2}{2\Delta x\Delta y} + \frac{\chi_{i+1}^{12,0}}{4}
-\frac{\Delta t}{2\Delta y}\zeta_{i+1}^{32,0} \right\}\sin(\tilde{k}\Delta y) \notag\\
& + \left( E_y^I \right)_{i+1}\left\{ \frac{c^2\Delta t^2}{2\Delta x\Delta y}\left( \cos(\tilde{k}\Delta y)-1\right)
+\frac{\chi_{i+1}^{12,0}}{4}\left( \cos(\tilde{k}\Delta y)+1\right)
-\frac{\Delta t}{2 \Delta y}\zeta_{i+1}^{32,0}\left( \cos(\tilde{k}\Delta y)-1\right) \right\} \notag\\
& + \left( E_z^R \right)_{i+1}\left\{ -\frac{\Delta t}{2 \Delta y}\zeta_{i+1}^{33,0}\sin(\tilde{k}\Delta y)\right\} +
\left( E_z^I \right)_{i+1}\left\{ \frac{\chi_{i+1}^{13,0}}{2}\right\}=
\left( \widetilde{Q}_x^I \right)_{i+1/2}.
\end{align}
The real part of the Fourier transform of Eq. (\ref{bp30b}) reads
\begin{align}
\label{apAA03}
& \left( E_y^R \right)_{i-1}\left\{ -\frac{c^2\Delta t^2}{2\Delta x^2}
-\frac{\Delta t}{2\Delta x}\zeta_{i-1}^{32,0} \right\} \notag \\
& + \left( E_z^R \right)_{i-1}\left\{ -\frac{\Delta t}{4\Delta x}\zeta_{i-1}^{33,0}
\left( \cos(\tilde{k}\Delta y)+1\right)\right\} +
\left( E_z^I \right)_{i-1} \left\{ -\frac{\Delta t}{4\Delta x}\zeta_{i-1}^{33,0}
\sin(\tilde{k}\Delta y) \right\} \notag \\
& + \left( E_x^R \right)_{i-1/2}\left\{ \frac{c^2\Delta t^2}{2\Delta x\Delta y}\left( \cos(\tilde{k}\Delta y)-1\right)
+\frac{\chi_{i}^{21,0}}{4}\left( \cos(\tilde{k}\Delta y)+1\right)
-\frac{\Delta t}{2\Delta x}\zeta_{i-1/2}^{31,0}\left( \cos(\tilde{k}\Delta y)+1\right) \right\} \notag \\
& + \left( E_x^I \right)_{i-1/2}\left\{ \frac{c^2\Delta t^2}{2\Delta x\Delta y} +\frac{\chi_i^{21,0}}{4}
-\frac{\Delta t}{2\Delta x}\zeta_{i-1/2}^{31,0} \right\} \sin(\tilde{k}\Delta y) \notag \\
& + \left( E_y^R \right)_{i} \left\{ 1+\frac{c^2 \Delta t^2}{\Delta x^2} +\chi_i^{22,0} \right\} \notag \\
& + \left( E_z^R \right)_{i} \left\{ \frac{\chi_i^{23,0}}{2}\left(\cos(\tilde{k}\Delta y)+1 \right) \right\}+
\left( E_z^I \right)_{i} \left\{ \frac{\chi_{i}^{23,0}}{2}\sin(\tilde{k}\Delta y)\right\} \notag \\
& + \left( E_x^R \right)_{i+1/2} \left\{-\frac{c^2\Delta t^2}{2\Delta x\Delta y}\left( \cos(\tilde{k}\Delta y)-1\right)
+ \frac{\chi_i^{21,0}}{4}\left( \cos(\tilde{k}\Delta y)+1\right)
+ \frac{\Delta t}{2\Delta x}\zeta_{i+1/2}^{31,0}\left( \cos(\tilde{k}\Delta y)+1\right) \right\} \notag \\
& + \left( E_x^I \right)_{i+1/2}\left\{-\frac{c^2\Delta t^2}{2\Delta x\Delta y} +\frac{\chi_i^{21,0}}{4}
+ \frac{\Delta t}{2\Delta x}\zeta_{i+1/2}^{31,0} \right\}\sin(\tilde{k}\Delta y) \notag \\
& + \left( E_y^R \right)_{i+1} \left\{ -\frac{c^2\Delta t^2}{2\Delta x^2}
+ \frac{\Delta t}{2\Delta x}\zeta_{i+1}^{32,0}\right\} \notag \\
& + \left( E_z^R \right)_{i+1} \left\{ \frac{\Delta t}{4\Delta x}\zeta_{i+1}^{33,0}
\left( \cos(\tilde{k}\Delta y)+1 \right)\right\}+
\left( E_z^I \right)_{i+1}\left\{ \frac{\Delta t}{4\Delta x}\zeta_{i+1}^{33,0}\sin(\tilde{k}\Delta y)\right\}=
\left( \widetilde{Q}_y^R \right)_i.
\end{align}
The imaginary part of the Fourier transform of Eq. (\ref{bp30b}) reads
\begin{align}
\label{apAA04}
& \left( E_y^I \right)_{i-1} \left\{-\frac{c^2\Delta t^2}{2\Delta x^2} -\frac{\Delta t}{2\Delta x}
\zeta_{i-1}^{32,0}\right\} \notag\\
& + \left( E_z^R \right)_{i-1} \left\{ \frac{\Delta t}{4\Delta x}\zeta_{i-1}^{33,0}\sin(\tilde{k}\Delta y)\right\} +
\left( E_z^I \right)_{i-1} \left\{ -\frac{\Delta t}{4\Delta x}\zeta_{i-1}^{33,0}
\left( \cos(\tilde{k}\Delta y)+1\right)\right\} \notag\\
& + \left( E_x^R \right)_{i-1/2} \left\{ -\frac{c^2\Delta t^2}{2\Delta x\Delta y}\sin(\tilde{k}\Delta y)
-\frac{\chi_i^{21,0}}{4}\sin(\tilde{k}\Delta y) +\frac{\Delta t}{2\Delta x}\zeta_{i-1/2}^{31,0}\sin(\tilde{k}\Delta y)\right\} \notag\\
& + \left( E_x^I \right)_{i-1/2} \left\{ \frac{c^2\Delta t^2}{2\Delta x\Delta y}\left( \cos(\tilde{k}\Delta y)-1\right)+
\frac{\chi_i^{21,0}}{4}\left( \cos(\tilde{k}\Delta y)+1\right)
-\frac{\Delta t}{2\Delta x}\zeta_{i-1/2}^{31,0}\left( \cos(\tilde{k}\Delta y)+1\right)\right\} \notag\\
& + \left( E_y^I \right)_{i} \left\{1+\frac{c^2\Delta t^2}{\Delta x^2} +\chi_{i}^{22,0} \right\} \notag\\
& + \left( E_z^R \right)_{i} \left\{ -\frac{\chi_i^{23,0}}{2}\sin(\tilde{k}\Delta y) \right\} +
\left( E_z^I \right)_{i} \left\{ \frac{\chi_i^{23,0}}{2}\left( \cos(\tilde{k}\Delta y)+1 \right) \right\} \notag\\
& + \left( E_x^R \right)_{i+1/2} \left\{ \frac{c^2\Delta t^2}{2\Delta x\Delta y}\sin(\tilde{k}\Delta y)
-\frac{\chi_i^{21,0}}{4}\sin(\tilde{k}\Delta y) -\frac{\Delta t}{2\Delta x}\zeta_{i+1/2}^{31,0}
\sin(\tilde{k}\Delta y)\right\} \notag\\
& + \left( E_x^I \right)_{i+1/2} \left\{-\frac{c^2\Delta t^2}{2\Delta x\Delta y}\left( \cos(\tilde{k}\Delta y)-1\right)
+\frac{\chi_{i}^{21,0}}{4}\left( \cos(\tilde{k}\Delta y)+1\right)
+\frac{\Delta t}{2\Delta x}\zeta_{i+1/2}^{31,0}\left( \cos(\tilde{k}\Delta y)+1\right)\right\} \notag\\
& + \left( E_y^I \right)_{i+1} \left\{ -\frac{c^2\Delta t^2}{2\Delta x^2} +\frac{\Delta t}{2\Delta x}
\zeta_{i+1}^{32,0} \right\} \notag\\
& + \left( E_z^R \right)_{i+1} \left\{-\frac{\Delta t}{4\Delta x}\zeta_{i+1}^{33,0}\sin(\tilde{k}\Delta y)\right\} +
\left( E_z^I \right)_{i+1} \left\{ \frac{\Delta t}{4\Delta x}\zeta_{i+1}^{33,0}\left( \cos(\tilde{k}\Delta y)+1\right)\right\}
= \left( \widetilde{Q}_y^I\right)_i.
\end{align}
The real part of the Fourier transform of Eq. (\ref{bp30c}) reads
\begin{align}
\label{apAA05}
& \left( E_y^R \right)_{i-1}\left\{ \frac{\Delta t}{4\Delta x}
\zeta_{i-1}^{22,0}\left( \cos(\tilde{k}\Delta y)+1 \right)\right\} +
\left( E_y^I \right)_{i-1}\left\{ -\frac{\Delta t}{4\Delta x}\zeta_{i-1}^{22,0}\sin(\tilde{k}\Delta y)\right\} \notag\\
& + \left( E_z^R \right)_{i-1}\left\{ -\frac{c^2\Delta t^2}{2\Delta x^2}
+ \frac{\Delta t}{2\Delta x}\zeta_{i-1}^{23,0}\right\} \notag\\
& + \left( E_x^R \right)_{i-1/2}\left\{ \frac{\chi_{i}^{31,0}}{2}
+\frac{\Delta t}{\Delta x}\zeta_{i-1/2}^{21,0} \right\} +
\left( E_x^I \right)_{i-1/2}\left\{ -\frac{\Delta t}{2\Delta y}\zeta_{i}^{11,0}\sin(\tilde{k}\Delta y)\right\} \notag\\
& + \left( E_y^R \right)_{i}\left\{ \frac{\chi_i^{32,0}}{2}\left( \cos(\tilde{k}\Delta y)+1 \right)
+\frac{\Delta t}{\Delta y}\zeta_{i}^{12,0}\left( \cos(\tilde{k}\Delta y)-1 \right)\right\} \notag\\
& + \left( E_y^I \right)_{i}\left\{-\frac{\chi_{i}^{32,0}}{2}\sin(\tilde{k}\Delta y)
-\frac{\Delta t}{\Delta y}\zeta_i^{12,0}\sin(\tilde{k}\Delta y) \right\} \notag\\
& + \left( E_z^R \right)_{i}\left\{ 1+\frac{c^2\Delta t^2}{\Delta x^2}
+\frac{c^2\Delta t^2}{\Delta y^2}\left(1-\cos(\tilde{k}\Delta y) \right) +\chi_i^{33,0}\right\} \notag\\
& + \left( E_z^I \right)_{i}\left\{ -\frac{\Delta t}{\Delta y}\zeta_i^{13,0}\sin(\tilde{k}\Delta y)\right\} \notag\\
& + \left( E_x^R \right)_{i+1/2}\left\{ \frac{\chi_i^{31,0}}{2}-\frac{\Delta t}{\Delta x}\zeta_{i+1/2}^{21,0}\right\} +
\left( E_x^I \right)_{i+1/2}\left\{ -\frac{\Delta t}{2 \Delta y}\zeta_i^{11,0}\sin(\tilde{k}\Delta y)\right\} \notag\\
& + \left( E_y^R \right)_{i+1}\left\{ -\frac{\Delta t}{4\Delta x}\zeta_{i+1}^{22,0}
\left( \cos(\tilde{k}\Delta y)+1 \right)\right\} +
\left( E_y^I \right)_{i+1}\left\{ \frac{\Delta t}{4\Delta x}\zeta_{i+1}^{22,0}\sin(\tilde{k}\Delta y)\right\} \notag\\
& + \left( E_z^R \right)_{i+1}\left\{ -\frac{c^2\Delta t^2}{2\Delta x^2}
-\frac{\Delta t}{2 \Delta x}\zeta_{i+1}^{23,0}\right\} = \left( \widetilde{Q}_z^R \right)_i.
\end{align}
The imaginary part of the Fourier transform of Eq. (\ref{bp30c}) reads
\begin{align}
\label{apAA06}
& \left( E_y^R \right)_{i-1}\left\{ \frac{\Delta t}{4\Delta x}\zeta_{i-1}^{22,0}\sin(\tilde{k}\Delta y)\right\} +
\left( E_y^I \right)_{i-1}\left\{ \frac{\Delta t}{4\Delta x}\zeta_{i-1}^{22,0}
\left( \cos(\tilde{k}\Delta y)+1\right)\right\} \notag\\
& + \left( E_z^I \right)_{i-1}\left\{ -\frac{c^2\Delta t^2}{2\Delta x^2}
+\frac{\Delta t}{2\Delta x}\zeta_{i-1}^{23,0}\right\} \notag\\
& + \left( E_x^R \right)_{i-1/2}\left\{ \frac{\Delta t}{2\Delta y}\zeta_i^{11,0} \sin(\tilde{k}\Delta y)\right\} +
\left( E_x^I \right)_{i-1/2}\left\{ \frac{\chi_i^{31,0}}{2}
+ \frac{\Delta t}{\Delta x}\zeta_{i-1/2}^{21,0}\right\} \notag\\
& + \left( E_y^R \right)_{i}\left\{ \frac{\chi_{i}^{32,0}}{2}\sin(\tilde{k}\Delta y)
+ \frac{\Delta t}{\Delta y}\zeta_{i}^{12,0}\sin(\tilde{k}\Delta y) \right\} \notag\\
& + \left( E_y^I \right)_{i}\left\{ \frac{\chi_i^{32,0}}{2}\left( \cos(\tilde{k}\Delta y)+1\right)
+\frac{\Delta t}{\Delta y}\zeta_i^{12,0}\left( \cos(\tilde{k}\Delta y)-1 \right) \right\} \notag\\
& + \left( E_z^R \right)_{i}\left\{ \frac{\Delta t}{\Delta y}\zeta_i^{13,0}\sin(\tilde{k}\Delta y)\right\} \notag\\
& + \left( E_z^I \right)_{i}\left\{ 1+\frac{c^2\Delta t^2}{\Delta x^2}
+\frac{c^2\Delta t^2}{\Delta y^2}\left( 1-\cos(\tilde{k}\Delta y)\right) +\chi_i^{33,0}\right\} \notag\\
& + \left( E_x^R \right)_{i+1/2}\left\{ \frac{\Delta t}{2\Delta y}\zeta_i^{11,0}\sin(\tilde{k}\Delta y)\right\} +
\left( E_x^I \right)_{i+1/2}\left\{ \frac{\chi_i^{31,0}}{2}-\frac{\Delta t}{\Delta x}\zeta_{i+1/2}^{21,0}\right\} \notag\\
& + \left( E_y^R \right)_{i+1}\left\{ -\frac{\Delta t}{4\Delta x}\zeta_{i+1}^{22,0}\sin(\tilde{k}\Delta y)\right\} +
\left( E_y^I \right)_{i+1}\left\{ -\frac{\Delta t}{4\Delta x}\zeta_{i+1}^{22,0}
\left( \cos(\tilde{k}\Delta y)+1\right)\right\} \notag\\
& + \left( E_z^I \right)_{i+1}\left\{ -\frac{c^2\Delta t^2}{2\Delta x^2}
-\frac{\Delta t}{2 \Delta x}\zeta_{i+1}^{23,0}\right\} = \left( \widetilde{Q}_z^I \right)_i.
\end{align}
Considering $N_x$ grid points along $x$-direction Eqs. (\ref{apAA01})-(\ref{apAA06}) can be formulated as
a band-diagonal system of equations, which we solve using a LU technique \cite{numrecipf90} 
for each of the $N_y$ modes of the discrete Fourier transform. Then we compute the
field solution in real space by inverse Fourier transformation.

\section{Numerical implementation of the charge correction step}
\label{app:charge_correction}

We detail here the numerical procedure to solve Eq. (\ref{eqcharge03}) within a 2D geometry. As for the wave equation, we make use of the Concus and Golub iterative method \cite{golub1973}, which writes
in the present case
\begin{equation}
\label{apA01}
-\mathbf{\nabla}\cdot\left[ (1+\chi^0)\mathbf{\nabla}\psi^{(m+1)} \right]=
\rho-\mathbf{\nabla}\cdot\left[ (1+\chi)\mathbf{E}_{n+1} \right]
+\mathbf{\nabla}\cdot\left[ (\chi-\chi^0)\mathbf{\nabla}\psi^{(m)} \right]
\end{equation}
where $\chi^0=\left[ \chi^{kl,0}\right]_{1\le k,l\le 3}$ denotes the $y$-averaged $\chi$
susceptibility tensor with $\chi^{kl,0}=<\chi^{kl}>_y$. $\mathbf{E}_{n+1}$ is the solution of the wave
equation (\ref{bp26}) and $m$ denotes the iteration index. Omitting the latter, we discretize the above equation in the form
\begin{align}
\label{apA02}
 -\frac{1}{\Delta x} & \left[
\left( 1 + \chi_{i+1/2,j}^{11,0}\right) \frac{1}{\Delta x}(\psi_{i+1,j}-\psi_{i,j}) -
\left( 1 + \chi_{i-1/2,j}^{11,0}\right) \frac{1}{\Delta x}(\psi_{i,j}-\psi_{i-1,j})\right] \notag \\
 -\frac{1}{2\Delta x} & \left[
 \chi_{i+1,j}^{12,0} \frac{1}{2\Delta y}(\psi_{i+1,j+1}-\psi_{i+1,j-1}) -
 \chi_{i-1,j}^{12,0} \frac{1}{2\Delta y}(\psi_{i-1,j+1}-\psi_{i-1,j-1})\right] \notag \\
 -\frac{1}{2\Delta y} & \left[
 \chi_{i,j+1}^{21,0} \frac{1}{2\Delta x}(\psi_{i+1,j+1}-\psi_{i-1,j+1}) -
 \chi_{i,j-1}^{21,0} \frac{1}{2\Delta x}(\psi_{i+1,j-1}-\psi_{i-1,j-1})\right] \notag \\
 -\frac{1}{\Delta y} & \left[
\left( 1 + \chi_{i,j+1/2}^{22,0}\right) \frac{1}{\Delta y}(\psi_{i,j+1}-\psi_{i,j}) -
\left( 1 + \chi_{i,j-1/2}^{22,0}\right) \frac{1}{\Delta y}(\psi_{i,j}-\psi_{i,j-1})\right] \notag \\
 & = S_{i,j} \, ,
\end{align}
where we have defined the source term
\begin{align}
\label{apA03}
S= & \partial_x \left[
(\chi^{11}-\chi^{11,0})\partial_x\psi +
(\chi^{12}-\chi^{12,0})\partial_y\psi \right] \notag \\
 + & \partial_y \left[
(\chi^{21}-\chi^{21,0})\partial_x\psi +
(\chi^{22}-\chi^{22,0})\partial_y\psi \right] + \rho \notag \\
 - & \partial_x\left[ (1+\chi^{11})E_x \right]
-\partial_x\left( \chi^{12}E_y \right)
-\partial_x\left( \chi^{13}E_z \right) \notag \\
 - & \partial_y\left( \chi^{21}E_x \right)
-\partial_y\left[ (1+\chi^{22})E_y \right]
-\partial_y\left( \chi^{23}E_z \right)
\end{align}
A centered spatial discretization of Eq. (\ref{apA03}) is given by
\begin{align}
S_{i,j} = &
+\frac{1}{\Delta x}\left[
(\chi_{i+1/2,j}^{11}-\chi_{i+1/2}^{11,0})\frac{1}{\Delta x}(\psi_{i+1,j}-\psi_{i,j})-
(\chi_{i-1/2,j}^{11}-\chi_{i-1/2}^{11,0})\frac{1}{\Delta x}(\psi_{i,j}-\psi_{i-1,j})\right] \notag \\
 & +\frac{1}{2\Delta x}\left[
(\chi_{i+1,j}^{12}-\chi_{i+1}^{12,0})\frac{1}{2\Delta y}(\psi_{i+1,j+1}-\psi_{i+1,j-1}) \right.\notag \\
 & \left.- (\chi_{i-1,j}^{12}-\chi_{i-1}^{12,0})\frac{1}{2\Delta y}(\psi_{i-1,j+1}-\psi_{i-1,j-1})\right] \notag\\
 & +\frac{1}{2\Delta y}\left[
(\chi_{i,j+1}^{21}-\chi_{i}^{21,0})\frac{1}{2\Delta x}(\psi_{i+1,j+1}-\psi_{i-1,j+1})\right. \notag \\
 & \left.- (\chi_{i,j-1}^{21}-\chi_{i}^{21,0})\frac{1}{2\Delta x}(\psi_{i+1,j-1}-\psi_{i-1,j-1})\right] \notag\\
 & +\frac{1}{\Delta y}\left[
(\chi_{i,j+1/2}^{22}-\chi_{i}^{22,0})\frac{1}{\Delta y}(\psi_{i,j+1}-\psi_{i,j})-
(\chi_{i,j-1/2}^{22}-\chi_{i}^{22,0})\frac{1}{\Delta y}(\psi_{i,j}-\psi_{i,j-1})\right] \notag\\
 & -\frac{1}{\Delta x}\left[
(1+\chi^{11}_{i+1/2,j})E_{x,i+1/2,j} - (1+\chi^{11}_{i-1/2,j})E_{x,i-1/2,j} \right] \notag\\
 & -\frac{1}{2\Delta x}\left[
\frac{\chi^{12}_{i+1,j}}{2}\left( E_{y,i+1,j+1/2} + E_{y,i+1,j-1/2} \right)-
\frac{\chi^{12}_{i-1,j}}{2}\left( E_{y,i-1,j+1/2} + E_{y,i-1,j-1/2} \right)\right] \notag\\
 & -\frac{1}{2\Delta x}\left[
\chi^{13}_{i+1,j}E_{z,i+1,j} - \chi^{13}_{i-1,j}E_{z,i-1,j}\right] \notag\\
 & -\frac{1}{2\Delta y}\left[
\frac{\chi_{i,j+1}^{21}}{2} \left( E_{x,i+1/2,j+1}+E_{x,i-1/2,j+1} \right) -
\frac{\chi_{i,j-1}^{21}}{2} \left( E_{x,i+1/2,j-1}+E_{x,i-1/2,j-1} \right)\right] \notag\\
 & -\frac{1}{\Delta y}\left[
\left( 1+\chi^{22}_{i,j+1/2}\right) E_{y,i,j+1/2} -
\left( 1+\chi^{22}_{i,j-1/2}\right) E_{y,i,j-1/2} \right] \notag\\
 & -\frac{1}{2\Delta y}\left[
\chi^{23}_{i,j+1}E_{z,i,j+1} - \chi^{23}_{i,j-1}E_{z,i,j-1}\right] \notag\\
 & + \rho_{i,j}
\end{align}
The above equations are Fourier transformed along the $y$ direction. Considering 
$N_y$ grid cells we have to solve $N_y$ one-dimensional equations. Assuming $N_x$ grid cells in the 
$x$ direction, each equation turns out into a $2N_x$ system of equations.
These systems have a band-diagonal structure and are solved with a LU technique \cite{numrecipf90}.

\section{Derivation of the dispersion relation of electron plasma waves with finite $\Delta x$ and $\Delta t$}
\label{app:dispersion_relation}

We restrict our analysis to a one-dimensional, nonrelativistic electrostatic plasma with immobile ions. In the following, we adopt the methodology and notations of Ref.  \cite{bird85}.
For a single macro-particle, the adjustable-damping scheme (\ref{bp02a})-(\ref{bp02d}) can be formulated as
\begin{align}
&x_{n+1}-2x_n+x_{n-1} = \frac{\Delta t^2}{2}\left\{
a_{n+1}+\frac{a_n}{2} + \frac{a_{n-1}}{2^2} + \frac{a_{n-2}}{2^3} +\dots \right\} \notag \\
& = \frac{\Delta t^2}{2}\left\{
a_{n+1}+\frac{\theta_f}{2}a_n + \left(1-\frac{\theta_f}{2}\right)^2 \left[ a_{n-1} + \frac{\theta_f}{2} a_{n-2}+ \left(\frac{\theta_f}{2}\right)^2 a_{n-3}+ \dots \right] \right\}
\end{align}
where $n$ stands for the time step index. We now assume a harmonic form for the interpolated electric force $F^{(1)}=F(k)e^{i(kx-\omega t)}$. As a direct consequence of the PIC
interpolation scheme, we have the relation \cite{bird85}
\begin{equation}
F(k)=qE(k)S(-k)
\end{equation}
where $E(k)$ and $S(k)$ are the discrete Fourier transforms of the electric field and the $m$-order weight function, respectively. The latter reads
\begin{equation}
S(k) = \left[ \frac{\sin \left( k\Delta x/2 \right)}{k\Delta x/2}\right]^{m+1} \, .
\end{equation}
The first-order acceleration term can then be expressed as
\begin{align}
\label{apxC01}
a_n & = \frac{F(k)}{m}\exp  \left[ i (kx_n^{(0)}-\omega t_n) \right] \notag \\
& = \frac{F(k)}{m}\exp  \left [i k(x_0+v^{(0)}t_n)- i \omega t_n \right] \notag \\
& = \frac{F(k)}{m}\exp  i k x_0 \exp \left[ i(kv-\omega)n\Delta t \right] \, .
\end{align}
Defining $A(k)=\frac{F(k)}{m}e^{ikx_0}$ and $z=e^{i(kv-\omega)\Delta t}$, Eq. (\ref{apxC01}) reads
\begin{align}
x_{n+1}-2x_n+x_{n-1} & = \frac{\Delta t^2}{2}A(k)
\left\{ z^{n+1} + \frac{1}{2}z^n +\left(\frac{1}{2}\right)^2 z^{n-1}
+\left(\frac{1}{2}\right)^3 z^{n-2} + \dots \right\} \notag \\
x_{n+1}-2x_n+x_{n-1} & = \frac{\Delta t^2}{2}A(k)z^n
\Bigg\{ z^{-1} \left[ \left(1-\frac{\theta_f}{2} \right)^2 +\frac{\theta_f}{2}z +z^2\right] \notag \\
& + \left( 1-\frac{\theta_f}{2} \right)^2 \frac{\theta_f}{2}z^{-2}
\left( 1+\frac{\theta_f}{2}z^{-1}+\left(\frac{\theta_f}{2}\right)^2 z^{-2}+ \dots \right) \Bigg\} \, .
\end{align}
This equation can be further simplified as
\begin{equation}
x_{n+1}-2x_n+x_{n-1}= \frac{\Delta t^2}{2}A(k) 2z^n \frac{\left[ (1-\theta_f)+z^2\right]}{2z-\theta_f} \, .
\end{equation}
We linearize $x_n= x_n^{(0)}+ x_n^{(1)}$ where $x_n^{(0)}=x_0^{(0)}+v_0^{(0)}t_n$
\begin{equation}
 \label{apxC03}
x_{n+1}^{(1)}-2x_n^{(1)}+x_{n-1}^{(1)}=\frac{\Delta t^2}{2}A(k)\mathcal{P}(k)
\end{equation}
Where the polynomial $\mathcal{P}$ reads
\begin{align}
\mathcal{P}(k) & = 2z^n \frac{\left[(1-\theta_f)+z^2\right]}{2z-\theta_f}  \\
\end{align}
We deduce that $x_n^{(1)}(x_0,v_0,t_n)$ varies as $e^{i(kv-\omega)n\Delta t}=z^n$. 
Hence we find the solution 
\begin{align}
x_n^{(1)} & = \frac{\Delta t^2}{m} F(k)e^{i(kx -\omega t)}\left[
\frac{z}{(z-1)^2} + \frac{z}{2z-\theta_f} \right]
\end{align}
To evaluate the charge density, we introduce the dipole density
\begin{align}
P(x,t) & = n_0 q \int dv f_0(v) x_n^{(1)}(x,v,t) \notag\\
       & = -\frac{n_0 q}{m} F(k) e^{i(kx-\omega t)}
\int dv f_0(v) \frac{1}{\left( \frac{2}{\Delta t} \sin(\omega -kv)\frac{\Delta t}{2}\right)^2} \notag\\
 & + \frac{n_0q\Delta t^2}{2m}F(k) e^{i(kx-\omega t)} \int dv f_0(v)
 \sum_{s=0}^{\infty}\frac{e^{i(\omega-kv)s\Delta t}}{(2/\theta_f)^s}
\end{align}
The first and second terms of the right-hand side correspond to the explicit leapfrog scheme and 
the implicit correction, respectively.
Assuming a Maxwellian distribution $f_0(v)=\frac{1}{v_t\sqrt{2\pi}}exp\left[-\left( \frac{v}{\sqrt{2}v_t}\right)^2\right]$, 
the latter can be written
\begin{align}
\int dv f_0(v) \sum_{s=0}^{\infty}\frac{e^{i(\omega-kv)s\Delta t}}{(2/\theta_f)^s} &
  = \sum_{s=0}^{\infty}\frac{e^{i\omega s \Delta t}}{(2/\theta_f)^s} \int dv f_0(v) e^{-ikvs\Delta t}\notag\\
& = \sum_{s=0}^{\infty}\frac{e^{i\omega s \Delta t}}{(2/\theta_f)^s} \mathcal{F}(f_0)(ks\Delta t)\notag\\
& = \sum_{s=0}^{\infty}\frac{e^{i\omega s \Delta t}}{(2/\theta_f)^s}
e^{-\frac{v_t^2}{2}(sk\Delta t)^2}
\end{align}
where $\mathcal{F}$ denotes the Fourier transform. Thus the polarisation becomes
\begin{align}
P(x,t) & = \frac{n_0 q}{m}F(k)e^{i(kx-\omega t)}\frac{\Delta t^2}{4}
\int f^{\prime}_0(v)\frac{2}{k\Delta t}\mathop{\mathrm{cotan}}\left[ (\omega-kv)\frac{\Delta t}{2}\right] dv \notag \\
& + \frac{n_0 q \Delta t^2}{2 m}F(k)e^{i(kx-\omega t)} \sum_{s=0}^{\infty}\frac{e^{i\omega s \Delta t}}{(2/\theta_f)^s} e^{-\frac{v_t^2}{2}(sk\Delta t)^2}
\end{align}
We can develop $\mathop{\mathrm{cotan}}$ as a series in the form
\begin{equation}
\mathop{\mathrm{cotan}}\left[ (\omega -kv)\frac{\Delta t}{2}\right] =
\frac{2}{\Delta t} \sum_{q=-\infty}^{+\infty} \frac{1}{\omega -kv -q\omega_g}
\end{equation}
The continuous charge density is given by $\rho_p = -\bold{\nabla}\cdot\bold{P}$, which writes in Fourier space $\rho_p(k)=-ikP(k)$.
The discrete charge density is then given by
\begin{align}
\label{apxC033}
\rho(k) = & \sum_p S(k_p)\rho_p(k_p) \notag\\
 = & -i \sum_p k_p S(k_p)P(k_p) \notag\\
 = & -i \sum_p |S(k_p)|^2 \frac{n_0q^2}{m}E(k_p)\sum_{q=-\infty}^{+\infty}
 \int dv \frac{\partial f_0(v)/\partial v}{\omega -k_pv -q\omega_g}\notag\\
 & - i \sum_p k_p |S(k_p)|^2 \frac{n_0q^2\Delta t^2}{2 m} E(k_p)
 \sum_{s=0}^{\infty}\frac{e^{i\omega s \Delta t}}{(2/\theta)^s} e^{-\frac{v_t^2}{2}(sk\Delta t)^2} \, .
\end{align}
Using centered space-differencing, discrete Fourier transform of the relation $E=-\partial \phi/\partial x$ gives
\begin{equation}
E(k)=-iK(k)\phi(k)=-iK(k)\phi(k) \, ,
\end{equation}
where
\begin{equation}
K(k)=k\frac{\sin(k\Delta x)}{k\Delta x} \, .
\end{equation}
The Poisson equation as modified by the direct implicit method reads
\begin{equation}
\mathbf{\nabla}\cdot\left( \mathbf{\nabla}\phi_{n+1} \right)=-\frac{\rho_{n+1}}{\epsilon_0}
\end{equation}
Centered space-differencing followed by a Fourier transformation gives
\begin{equation}
\kappa^2(k)\phi(k)=\frac{\rho(k)}{\epsilon_0} \,
\end{equation}
where we have defined
\begin{equation}
\label{apxC038}
\kappa^2(k)=k^2\left[\frac{\sin\left(k \Delta x/2\right)}{k\Delta x/2} \right]^2 \, .
\end{equation}
Combining Eqs. (\ref{apxC033})-(\ref{apxC038}), 
we obtain the dispersion relation for an infinite electrostatic one dimensional plasma taking into account 
both spatial and temporal discretizations
\begin{align}
\epsilon(\omega,k) = 1 & +\frac{\omega_p^2}{\kappa^2(k)}\sum_p |S(k_p)|^2 K(k_p)
\sum_{q=-\infty}^{+\infty} \int dv \frac{\partial f_0(v)/\partial v}{\omega - k_pv -q\omega_g}  \notag \\
& + \frac{\omega_p^2}{\kappa^2(k)}\frac{\Delta t^2}{2} \sum_p k_p |S(k_p)|^2 K(k_p)
\sum_{s=0}^{+\infty} \frac{e^{i\omega s \Delta t}}{(2/\theta)^s} e^{-\frac{1}{2}v_t^2(sk\Delta t)^2} =0\, ,
\label{apxC02}
\end{align}
where $k_g=2\pi/\Delta x$, $\omega_g=2\pi/\Delta t$, $\omega_q=\omega -q\omega_g$ and $k_p=k-pk_g$.

Exploiting the Maxwellian form of $f_0$, we have 
\begin{equation}
\int dv\frac{\partial f_0/\partial v}{\omega_q - k_pv}  = 
\frac{1}{k_pv_t^2} \left[1+\xi_q \mathcal{Z}(\xi_q) \right] \, ,
\label{apxC04}
\end{equation}
where $\xi_q=\frac{\omega_q}{\sqrt{2}k_p v_t}$ and $\mathcal{Z}$ denotes the Fried and Conte plasma dispersion function $\mathcal{Z}$ \cite{frie61}, defined by
\begin{equation}
\mathcal{Z}(\xi)=\pi^{-1/2}\int_{-\infty}^{\infty}du \frac{e^{-u^2}}{u-\xi} \textrm{ with } \Im(\xi)>0 \, .
\end{equation}
Finally, substituting Eq. (\ref{apxC04}) into Eq. \ref{apxC02} yields Eq. (\ref{heat03}).

\bibliography{direct_ilp}
\bibliographystyle{elsarticle-num}

\end{document}